# Comparative Studies of 10 Programming Languages within 10 Diverse Criteria

Revision 1.0

Rana Naim*
Concordia University Montreal,
Quebec, Canada
ra_hass@cse.concordia.ca

Mohammad Fahim Nizam†
Concordia University Montreal,
Quebec, Canada
m_niza@cse.concordia.ca

Sheetal Hanamasagar‡
Concordia University Montreal,
Quebec, Canada
s_hanama@cse.concordia.ca

Jalal Noureddine§
Concordia University Montreal,
Quebec, Canada
j_noure@cse.concordia.ca

Marinela Miladinova¶
Concordia University Montreal,
Quebec, Canada
marin_mi@cse.concordia.ca

Abstract

*This is a survey on the programming languages: C++, JavaScript, AspectJ, C#, Haskell, Java, PHP, Scala, Scheme, and BPEL. Our survey work involves a comparative study of these ten programming languages with respect to the following criteria: secure programming practices, web application development, web service composition, OOP-based abstractions, reflection, aspect orientation, functional programming, declarative programming, batch scripting, and UI prototyping. We study these languages in the context of the above mentioned criteria and the level of support they provide for each one of them.*

**Keywords:** programming languages, programming paradigms, language features, language design and implementation

## 1 Introduction

Choosing the best language that would satisfy all requirements for the given problem domain can be a difficult task. Some languages are better suited for specific applications than others. In order to select the proper one for the specific problem domain, one has to know what features it provides to support the requirements. Different languages support different paradigms, provide different abstractions, and have different levels of expressive power. Some are better suited to express algorithms and others are targeting the non-technical users. The question is then what is the best tool for a particular problem. Aspects, like security and language safety, UI prototyping capabilities, language support for building distributed systems, and support for automating existing processes, and portability are also important issues to consider when choosing the programming language. In our analysis we discussed the suitability of the selected languages for the specified criteria.

---

* Primarily focused on Programming Languages C++ and JavaScript
† Primarily focused on Programming Languages AspectJ and C#
‡ Primarily focused on Programming Languages Haskell and Java
§ Primarily focused on Programming Languages PHP and Scala
¶ Primarily focused on Programming Languages Scheme and BPEL

## 1.1 Related Work

The work of [6], [7], [15], [32], [54], [55], [67], [69], [84], [87], [95], [96], [147], [148] was a major influence of how this paper is written and the approach taken to compare the programming languages. In [6] the author emphasises the difference between design and implementation of languages and how the decisions in one influences the other. In [7] the author performs comparative study of model transformation languages in the context of MDE. The author in [15] compares different Web service technologies, paradigms, and their evolution. The author in [32] discusses technologies and architectures in the context of distributed systems. In [54] the author studies fundamental concepts of programming languages. In [55] study of language concepts of varying degree of abstraction is performed in the context of C++. Several studies have been performed previously among several programming languages and several programming paradigms. But, from our observation, AspectJ has not yet compared with C#, such way in any previous research work. We have found some great research work like [69], [84], [87], [95], [96] which were really interesting and motivated us to do some extensive research. Finally, [67] provided all the guidance necessary to perform this study.

## 1.2 Overview

The rest of this paper is organized as follows. First we provide an overview of all the languages we have studied, beginning from Section 1.3 to Section 1.12. In Section 2, we have performed a detailed analysis of the studied languages in the context of the specified criteria. In Section 3, we consolidated our results, into a table supported by brief description. In Section 4 we conclude our work and incorporated some statistics about programming languages.

### 1.3 C++ Language Overview

Different platform such as windows, Linux and UNIX has different instruction set. Any executable file that produced from compiling C++ code under X platform can't run under Y platform, e.g. a windows-executable file that comes from C++ code can only run under windows compatible environment.[199]

### 1.4 JavaScript Language Overview

Java compiler in the other side converts the source code into intermediate code, each platform has a specific java virtual machine that run any intermediate code was been produced under any platform [199]. JavaScript language is able to use java objects and call their public methods, this facility earn the JavaScript to use what is exist in java and not exist in JavaScript. JavaScript code is compiled through interpreters been implemented in java language, these interpreters have external interface used to communicate with other external component. Script code that runs in the user side used to control the behavior of the end user [250].

### 1.5 AspectJ Language Overview

AspectJ is a general purpose programming language, which is simple and a practical Aspect-oriented extension to Java. AspectJ extends the Java language with keywords for writing aspects, pointcuts, advice code, and intertype declarations. Gregor Kiczales and his team, who has created a new programming paradigm AOP at the Palo Alto Research Center (PARC) [68], has also developed AspectJ—which is now the leading tool for Aspect-oriented Programming. Using this, it is possible to create clean modular implementations of crosscutting concerns such as tracing, login, user session management, synchronization, consistency checking, protocol management etc. By creating, just a few new constructs, AspectJ provides support for modular implementation of a range of crosscutting concerns. Here in dynamic join point model, join points are well-defined points of the program where the advice code will be executed; pointcuts are collections of join points; advice are special method-like constructs that can be attached to pointcuts; and aspects are modular units of crosscutting implementation, comprising pointcuts, advice, and ordinary Java member declarations [70].

## 1.6 C# Language Overview

C# is a general-purpose, multi-paradigm, object-oriented programming language. The goal of the language is the increased programmer productivity. The language is becoming very popular because of its perfect balances of simplicity, expressiveness, and performance. The language is developed by Microsoft Corporation within the .Net initiative. The chief architect of the language since its first version is Anders Hejlsberg [75] who is also the original author of Turbo Pascal and the chief architect of Delphi.

C# is a rich implementation of the object-orientation paradigm, which includes Data abstraction, Encapsulation, Inheritance, and Polymorphism. It is typically used for writing code that runs on Windows platforms. Although Microsoft standardized the C# language and the CLR through ECMA, the total amount of resources to support C# on other platforms than Windows is relatively small. The purpose of the Common Language Runtime (CLR) is to provide a language-neutral platform for application development and execution, including functions for exception handling, garbage collection, security, and interoperability [81].

## 1.7 Haskell Language Overview

Haskell is a Functional Programming Language. Haskell is named after Haskell B Curry who was one of the pioneers of the lambda calculus, which is a mathematical theory of functions and has been an inspiration to designers of number of functional languages [99]. Haskell is a Higher-Order Functions, Lazy Evaluation, light Syntax for data type, Purity- Not having side effects makes it possible to safely abstract any part of a function and replace it by parameter. The paradigm treats computation of mathematical functions and avoids state and mutable data. It is strongly typed, eliminating a huge class of easy to make errors at compile time. No possibility core dumps [100]. For example, in Java one can write a function that accepts two arguments of any possible type. However, Haskell goes further by allowing a function to accept two arguments of any type so long as they are both the same type [101].

## 1.8 Java Language Overview

Java is a programming language originally developed by James Gosling at Sun Microsystems. It is strongly typed object oriented that provides an excellent means of modularizing and structuring programs. It also supports concurrent execution of multiple threads as well as some key synchronization mechanisms. Integrated exception handling mechanism- Rich set of integer and floating-point data types, and Java.math package supports for user documentation. General coding style guidelines supports abstraction and information hiding, code portability [102].

## 1.9 PHP Language Overview

PHP is a powerful scripting language that can be run by itself in the command line of any computer with PHP installed [156]. PHP was originally created by Rasmus Lerdorf in 1995 and stood for "Personal Home Page" and was released as a free, open source project. In 1997, PHP was renamed to "PHP: Hypertext Preprocessor". PHP is especially well-suited for creating dynamic web pages with connectivity to various database systems (MySQL is the most widely used because PHP provides native support for it and the database is free and an open-source project). PHP runs on different platforms and is compatible with almost all servers used today. PHP is easy to learn and runs efficiently on the server side.

## 1.10 Scala Language Overview

Scala stands for “Scalable Language”. Scala is a general purpose programming language designed to express common programming patterns in a concise, elegant, and type-safe way. It smoothly integrates features of object-oriented and functional languages, enabling Java and other programmers to be more productive. Code sizes are typically reduced by a factor of two to three when compared to an equivalent Java application [157]. Scala runs on the standard Java platform and interoperates seamlessly with all Java libraries [158].

## 1.11 Scheme Language Overview

**General Characteristics**
Scheme is a multi-paradigm programming language supporting functional, procedural, object-oriented, meta, web applications, batch/CGI/Shell scripting application development. It has small language core and powerful tools to allow the language to be extended. The goal stated in the Scheme de facto standard is to provide a base from which compatible implementations can be built. A typical characteristic of all dialects of Lisp, including Scheme, is that they are homoiconic, i.e. self-representing. A list in Scheme can be interpreted as a data structure or a source listing to be compiled and executed. Type-safety is another characteristic targeted by the Scheme’s de facto standard. Although, the language is dynamically type (type checking, the binding of variable names to object types is done at run time) and supports implicit polymorphism (variables don’t have types and can be bound to any value that has a type), Scheme is type safe. The type system with its typing rules for type equivalence and type compatibility are strictly enforced by the implementation. Application of a function to objects with types that cannot support that function is not allowed. Scheme language is considered strict because its side-effect functions are always strict (the function is undefined when any if its arguments is undefined.) This is due to the applicative order of evaluation of arguments passed to functions [6]. Scheme is the first Lisp dialect to support lexical (static) scoping, i.e. the binding of the variable name usage to its declaration can be inferred from the program text, without the need to trace the execution history (dynamic scoping) to determine the name to declaration binding. This makes programs easier to write and understand, and prevents the introduction of many difficult to track programming errors [6]. Another characteristic of Scheme is that it is an expressive language in which algorithms can be represented concisely. Scheme’s homoiconic feature allows it to express uniformly the design, implementation, and semantics of the language in addition to the data that is handled by the programs expressed in the language. The architecture of the system, a Scheme program is implementing, can be expanded or created at runtime by compiling and interpreting a newly created source code. Scheme has a reach reflective system to support meta-programming. It also supports both applicative (a.k.a. eager) order of evaluation of arguments to functions and normal (a.k.a lazy) order of evaluation of special forms such as cond special form. Scheme uses Cambridge Polish (parenthesised) notation for expressions. The first element inside parenthesis is a function or a macro (special form) and the rest are its parameters. Anonymous functions are defined using lambda special form [6, 17].

**Functions and Control Flow**

During a typical function call scenario, the implementation restores the referencing environment conforming to the lexical scoping rules. At this time, arguments in the referencing environment are bound to the formal parameters of the function. Each expression inside the function is sequentially evaluated and the value of the last expression is the return value from the function, which can be another function [6].

Scheme provides multi-way conditional expression and special forms for assignment, sequencing, iteration (imperative constructs), delay and force for lazy evaluation. Arguments of special forms are passed unevaluated. The expression types (functions and special forms) in Scheme can be primitive (built-into the language implementation) and derived (defined in terms of primitives.) Lambda special form can be used to build derived functions. To build derived special forms (macros), syntax-rules special form is used. These forms can be bound to names using define-syntax and let-syntax. This mechanism allows for language expansion and creation of libraries [6].

**Scheme Implementations**

Implementations of Scheme languages include both commercial and free and run on different platforms. Some examples of graphical programming environment implementations follow: 3DScheme, 3DScheme Pro and EdScheme, Bee DrScheme, fluxus, a livecoding environment, SchemeWay, a Scheme environment built atop Eclipse, Pilo Visual Tools for Scheme [12, 13]. A rich list of Scheme implementations is given in [13].

## 1.12 BPEL Language Overview

**General Characteristics**

The goal of BPEL language is to allow for the smooth integration of heterogeneous systems in different application domains such as business-to-consumer, business-to-business, and enterprise application integration, for example. Typical interaction scenarios involve sequences of peer-to-peer, long-running, stateful message exchanges between two or more participants. Web services described by WSDL support stateless, one-way, uncorrelated message exchanges. BPEL is built to bridge this gap in the requirements [16].

The BPEL language automates business processes by a way of orchestrating existing functionality (services) that expose their interfaces as web services using WSDL as the standard language. BPEL processes are themselves web services described using WSDL. To achieve easy integration of heterogeneous web services, SOAP is used as the messaging protocol. It is an XML-based protocol which allows XML messages (data) to be transported in a platform neutral and programming model independent way [16].

The BPEL process consists of abstract and executable part. The abstract process is not executable and specifies the external message exchanges between partners and does not contain any internal details about the business process. The executable process defines both the external messages and the logic and it is executable [15].

Concepts relevant to BPEL processes are: (1) message flow to describe the order in which external web services are invoked; (2) control flow to specify the order in which activities within the process execute; (3) data flow of stateful interactions which include the content of the messages that are exchanged, the intermediate data used in internal computations, and the composed messages sent to partners [15].

**Functions and Control Flow**

The BPEL process specifies how multiple service interactions between partners are coordinated to achieve a business goal, as well as the state and the logic necessary for this coordination. It introduces exception handling mechanism and compensation mechanism to roll-back a failed unit of work transaction [16]. The computational model is imperative in nature and structured-programming constructs are used to specify the internal logic. The data model used by BPEL is defined using WSDL messages and XML Schema type [16]. Scopes are used as the encapsulation mechanisms and they are lexical (static) in nature. Scopes can be nested. Each control flow construct has its own scope. Scopes can exist without a control flow construct. Variables comply with the static rules that can be inferred from the program text. There are no subroutine or function constructs. The BPEL module itself can be considered as one function (in the imperative programming-model sense) or subroutine. Internal concurrency is afforded by the use of the Flow construct. Iteration is achieved using While, RepeatUntil, and ForEach constructs. Events handling is achieved thought the onMessage and onEvent that are encapsulated inside a Pick construct. Pick is blocking in nature and waits for a specific event to occur or a message to arrive. After the event is handled within the Pick scope, control is released to the rest of the process. Conditional logic is expressed using If construct. Although previously existed, the Switch construct is not part of BPEL version 2.0 OASIS standard [16]. Updates to variables are achieved using the Assign construct. BPEL uses Invoke to request a service or delegate a computation to another Web service. Since BPEL is a web service which exposes its capabilities using a WSDL interface, it can be invoked. BPEL is capable of receiving external requests using its Receive construct and Reply is used to return a response to the client that invoked the process.

**BPEL Implementations**

BPEL implementations include ActiveVOS, Open Source, The Eclipse STP BPMN Diagram Editor, Eclipse BPEL project, Orchestra Fully Open source, extensible and flexible BPEL Solution, The Open Source BPMS (Eclipse and Apache-based), Apache ODE, Open source BPEL server, NetBeans Enterprise Pack, and BPEL for Windows Workflow Foundation [14].

# 2 Analysis

## 2.1 C++vs. JavaScript

### 2.1.1 Default more secure programming practices

**C++**

The operating system is written in c language, therefore, C++ is close to the operating system, especially when it relates to pointers and memory access, this fact makes C/C++ unsafe [198]. In C++ it is easy to have illegal access to the memory object. There is no internal exception handling for out of memory boundary when dealing with arrays. Allocating and de-allocating memory space is error prone in C++ language, which easily may introduce privacy leak. Memory management with C++ in most cases needs to be done by the programmer, which complicates debugging task. C++ susceptible to buffer overflow if there is no checking to the input length of any function [198].

e.g the following code may lead to stack over flow see figure 2 and 3.

```
int main()
{
char *name[2];
name[0] = "/bin/sh";
name[1] = NULL;
execve(name[0], name, NULL);
return 0;
}
```

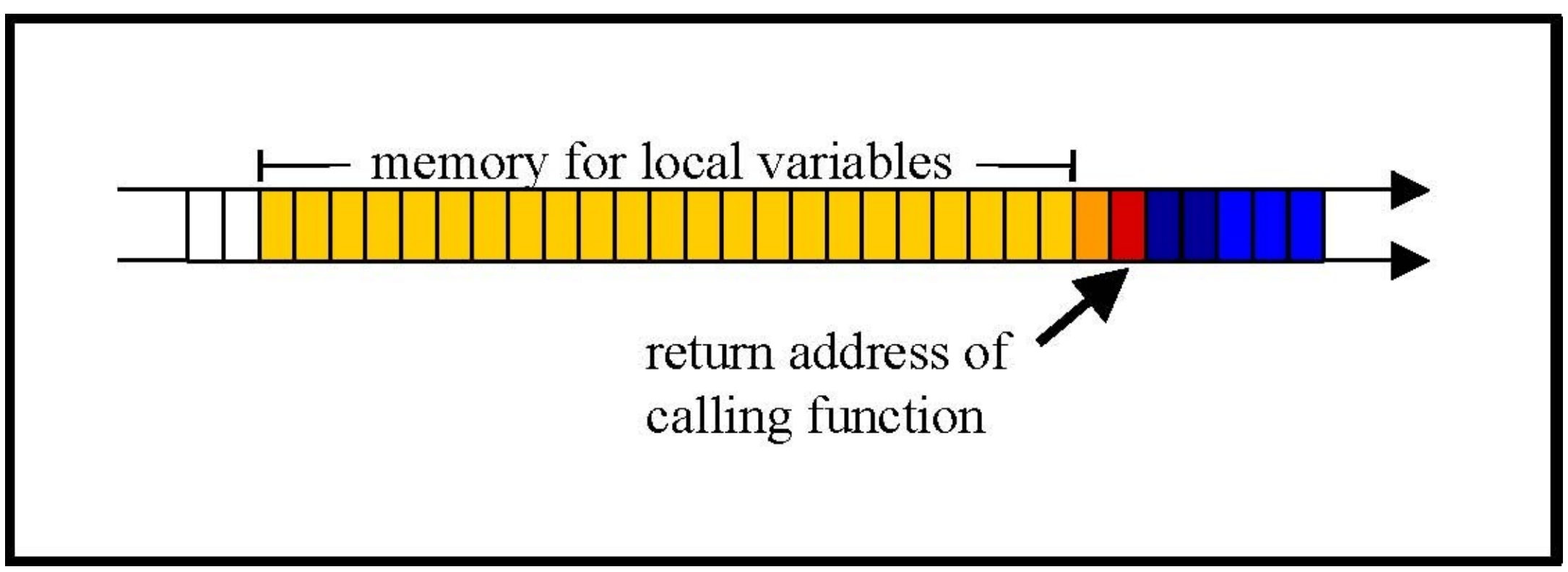

Fig. (Stack after some function has been called) [198]

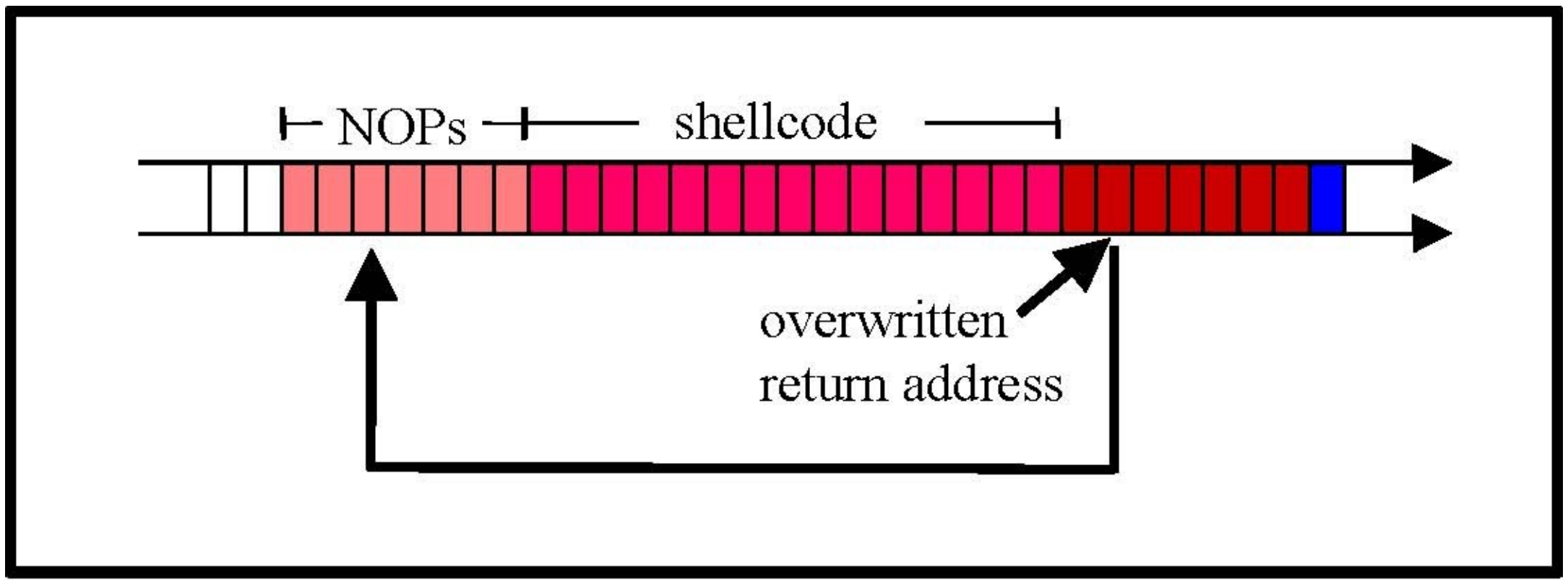

Fig 1. (Stack after buffer overflow) [198]

It is recommended to follow guidelines when programming in C++ [198], I will mention the following some:

- Enforce standard coding style guidelines
- Validate function's inputs
- Trace your code.
- Do not ignore warning prompts that is produced from code compilation.
- Use object oriented style that hides properties and encapsulates implementations.
- Prevent privacy leak.

There are more guidelines exists on [198].

### JavaScript

Most of the JavaScript features are inherited from Java language, Java language has a lot of build in security features [198], one of the security challenges of JavaScript is to provide a secure environment to run mobile code [200]. The Java language originally built to mitigate many programming mistakes, it is known that Java is type-safe and have the feature of memory management that checks arrays' boundary. Java virtual machine applies strict mechanism to prevent stack overflow from happening. Java and JavaScript faces the following threats [200]:

- A malicious subclass may clone or/and override random methods.
- If the code is operating with privileged access, then most probably there will be flaws. Privileged access need to be given to trusted entities. Running unprivileged code in a sandbox will solve the problem.
- Java library could be replaced with a malicious implementation.
- Exception handling may show sensitive knowledge. E.g, a java.io.FileNotFoundException shows the file path, since many attacks need to know the file path.
- Dynamic SQL creation along with untrusted input leads to injection flow.
- Code can be hidden in a script API, if that code is malicious, then a big threat may be introduced.
- Malicious code may use RMI and LDAP to inject himself to the system.
- JavaScript is dynamically typed [204].

- JavaScript does not check arrays boundary, so strings and arrays has variable length [204].

There are more threats exists on [200]. Most of previous threat may be isolated is we use isolate unrelated code using package.access security property.

```
private static final String PACKAGE_ACCESS_KEY = "package.access";

static {

  String packageAccess = java.security.Security.getProperty(

    PACKAGE_ACCESS_KEY

  );

  java.security.Security.setProperty(

    PACKAGE_ACCESS_KEY,

    (

      (packageAccess == null || packageAccess.trim().isEmpty()) ?

      "" :

      (packageAccess + ",")

    ) +

    "xx.example.product.implementation."

  );

}
```

**Guideline 1-1b Isolate unrelated code [200]**

## 2.1.2 Web Application Development

### C++

C++ can be used for creating web applications using Wt C++ Toolkit [210], [211], Wt C++ is C++ library developed for web applications [211]. Wt encapsulates many web details, e.g. Wt does not show details of embedded with the web form [211].

The components of WT C++ are ordered in hierarchical widget tree, The library provides Web Widget which is called WWebWidget it deals with server-side HTML DOM. Theses components are piece of text (WText), a table (WTable), a line edit (WLineEdit), and so on. The following example print hello world with quit button [211].

```
#include <WApplication>
#include <WText>
#include <WPushButton>

int wmain(int argc, char **argv)
{
   WApplication appl(argc, argv);

   // Widgets can be added to a parent
   // by calling addWidget() ...
   appl.root()->addWidget(new Wtext("<h1>Hello, World!</h1>"));

   // ... or by specifying a parent at
   // construction time
   WPushButton *Button = new WPushButton("Quit", appl.root());

   Button->clicked.connect(SLOT(&appl, Wapplication::quit));
   return appl.exec();
}
```

This code is taken from [211]

**JavaScript**

JavaScript is imperative scripting language that mainly used for internet applications, it helps to do computations associated with HTML and is interpreted by client browser, also it motivates the client to use websites, JavaScript brings the dynamic changing to the websites. JavaScript makes the computation done at client side, which increases the network bandwidth and improve computations performance [204], [227].

JavaScript in some cases used to check the validity of input data entered by user side, this will help to do the checking at client side, better than doing that computation at server side, otherwise the server has to send notification message and ask the user to redo data entry [227]. Moving the computation from server side into client side improves web computations and reduces the network bandwidth. Imagine some social web application like face book, a lot of updates happen in a second and needs to maintain the privacy of it's clients. If the server is responsible to control every client connected in any moments, then the face book servers is going to shutdown soon, due to the loads are considered huge. JavaScript script helps reducing this issue by doing this control at client side [242].

The following is an example of JavaScript code declares two integers, do some computations and print the result:

```
<html><body><script type="text/javascript">

var x = 8;
var y = 5;
var z = x*y;
```

```
document.write(z);

x++;
y--;
z=x*y;

document.write("<br />");

document.write(x*y);

</script>

<p>The script above declares variables,

assigns  values to them, do some computations,  displays the computation values,

and displays the value again.</p>

</body></html>
```

### 2.1.3 Web services design and composition

#### C++

C++ web services use XML remote procedure protocol (XML-RPC) or Simple Object Access Protocol (SOAP) as a transport layer to send a request and receive a response [212].

The following figures shows how to deploy a services as well as how to transport the request and receive the response [212].

XML-RPC helps the messaging and establishes the communication across different platforms.

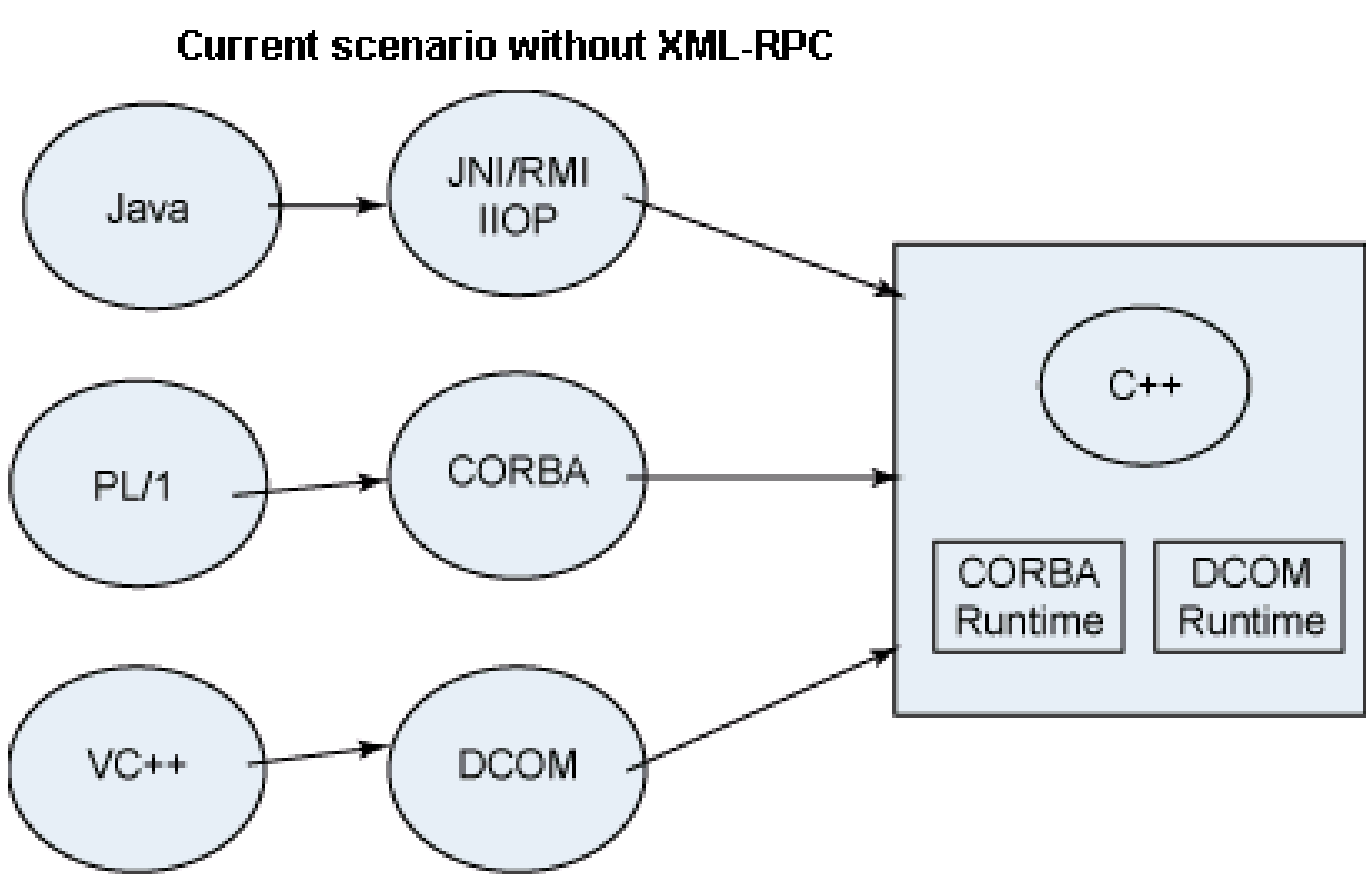

This picture is taken from [212]

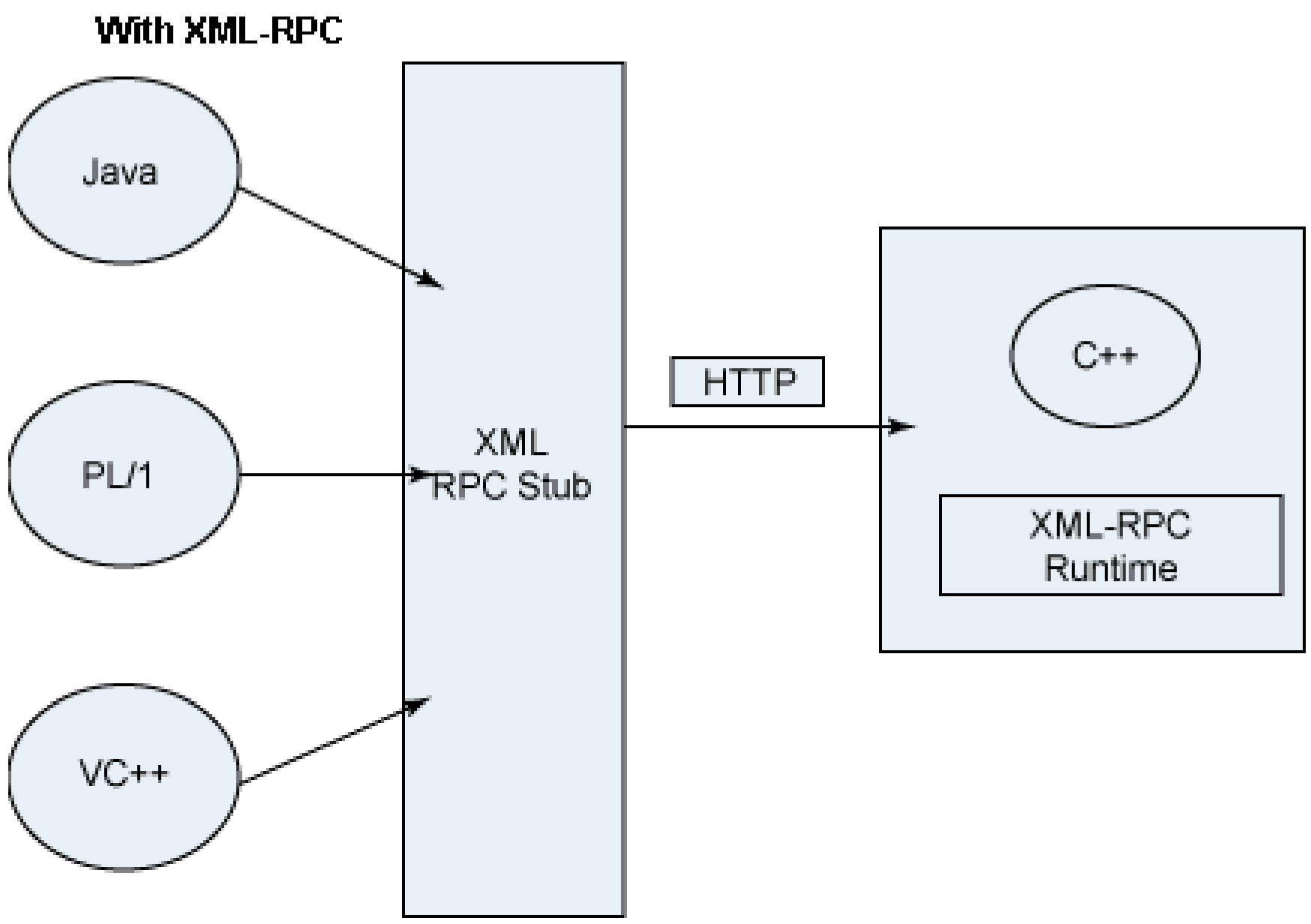

This picture is taken from [212]

There is a good example that shows how to deploy C++ web services and invoke those services at [212].

The following steps used to deploy a web service using SCA Service Component Architecture, the examples bellow is taken from [213]:

i. To declare services, we should declare an interface that abstracts super class, which only contains virtual functions [213]:

Service interface.

```
// LoanService interface
class LoanService {
public:
      virtual bool approveLoan(unsigned long customerNumber,
                               unsigned long loanAmount) = 0;
};
```

This code fragment is taken from [213].

ii. Then, the developer implements the interface already declared in the previous step.

Implementation declaration header file.

```
class LoanServiceImpl : public LoanService
{
public:
        LoanServiceImpl();
        virtual ~LoanServiceImpl();

        virtual bool approveLoan(unsigned long customerNumber,
                                 unsigned long loanAmount);
};
```

This code fragment is taken from [213]

Implementation.

```
#include "LoanServiceImpl.h"

LoanServiceImpl::LoanServiceImpl()
{
        …
}
LoanServiceImpl::~LoanServiceImpl()
{
        …
}

bool LoanServiceImpl::approveLoan(unsigned long customerNumber,
                                  unsigned long loanAmount)
{
        …
}
```

This code fragment is taken from [213]

iii. Use WSDL generator to generate a WSDL portType [213].

The following snippet shows the component type for this component implementation.

```
<?xml version="1.0" encoding="ASCII"?>
<componentType xmlns="http://www.osoa.org/xmlns/sca/1.0">
        <service name="LoanService">
                <interface.cpp header="LoanService.h"/>
        </service>
</componentType>
```

This code fragment is taken from [213]

The following figure shows the complete image and connection for the previous steps [213].

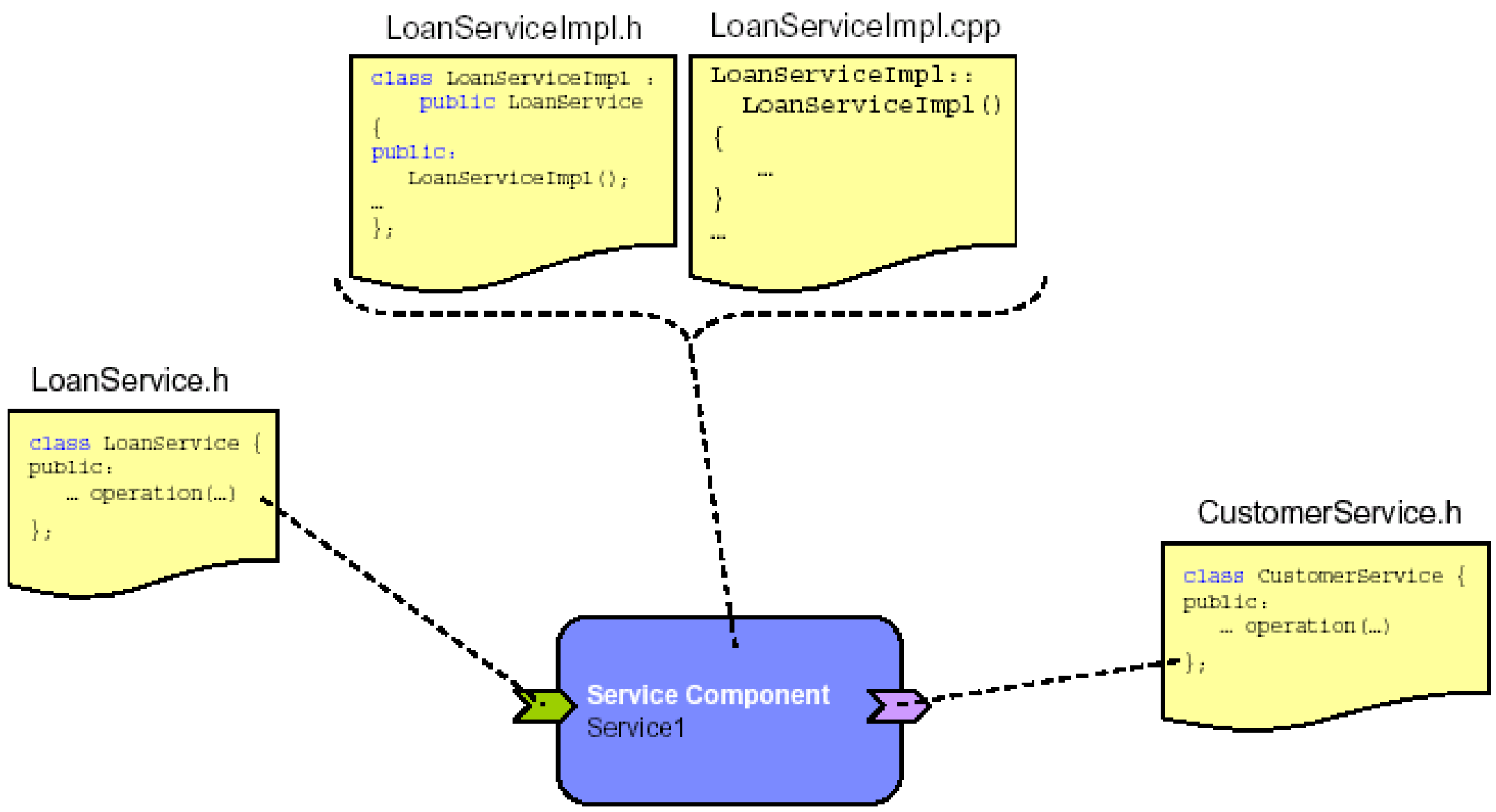

This code fragment is taken from [213]

iv. In the client side, we use a C++ code to call the method implemented in the server side. We need in this case to know the service name and the port address [213].

## JavaScript

We can call a web services from the client using script language such as JavaScript, JAX-WS is a Java API for XML-Based Web Services is a framework helps to do web service [243]

The following example defines a Web service with just one operation called hello that accepts one parameter of type string, you will notice the tag @ WebService to declare that the attached class is a web service interface. Webservice tag declare the service name and host address and the port number as well. If the port numer is not mentioned, the default port no is called. @Webmethod tag used to define a web service operation and its parameters as well as it implementation [243].

```
import javax.jws.WebService;
 import javax.jws.WebMethod;
 import javax.jws.WebParam;

 @WebService(targetNamespace="http://test.omii.ac.uk")
 public class Hello {

    @WebMethod
    public String hello(@WebParam(name = "name") String name) {
```

```
        return "Hello " + name + "!";
    }
}
```

The following code example is taken from [243]

Then we need to deploy it as a JAX-WS Web service, the following WSDL is generated by WSDL tool used to deploy previous declaration [243].

```
<definitions xmlns:soap="http://schemas.xmlsoap.org/wsdl/soap/"
          xmlns:tns="http://test.omii.ac.uk"
          xmlns:xsd="http://www.w3.org/2001/XMLSchema"
          xmlns="http://schemas.xmlsoap.org/wsdl/"
          targetNamespace="http://test.omii.ac.uk"
          name="HelloService">
  <types>
    <xs:schema xmlns:xs="http://www.w3.org/2001/XMLSchema" version="1.0"
            targetNamespace="http://test.omii.ac.uk">
      <xs:element name="hello" type="tns:hello" />
      <xs:element name="helloResponse" type="tns:helloResponse" />
      <xs:complexType name="hello">
        <xs:sequence>
          <xs:element name="name" type="xs:string" minOccurs="0" />
        </xs:sequence>
      </xs:complexType>
      <xs:complexType name="helloResponse">
        <xs:sequence>
          <xs:element name="return" type="xs:string" minOccurs="0" />
        </xs:sequence>
      </xs:complexType>
    </xs:schema>
  </types>
  <message name="hello">
    <part name="parameters" element="tns:hello" />
  </message>
  <message name="helloResponse">
    <part name="parameters" element="tns:helloResponse" />
  </message>
  <portType name="Hello">
    <operation name="hello">
      <input message="tns:hello" />
      <output message="tns:helloResponse" />
    </operation>
  </portType>
  <binding name="HelloPortBinding" type="tns:Hello">
    <soap:binding transport="http://schemas.xmlsoap.org/soap/http"
              style="document" />
    <operation name="hello">
      <soap:operation soapAction="" />
      <input>
        <soap:body use="literal" />
      </input>
      <output>
        <soap:body use="literal" />
      </output>
    </operation>
```

```
  </binding>
  <service name="HelloService">
    <port name="HelloPort" binding="tns:HelloPortBinding">
      <soap:address location="http://localhost:8080/context/HelloService" />
    </port>
  </service>
</definitions>
```

The following code example is taken from [243]

The following is SOAP invocation [243]

```
HelloService service = new HelloService();
Hello port = service.getHelloPort();

String response = port.hello(name);
```

The following code example is taken from [243]

The following the invocation for hello service:

```
import javax.xml.rpc.Call;
import javax.xml.rpc.Service;
import javax.xml.rpc.ServiceFactory;
[...]

String endPoint = "http://localhost:8080/context/HelloService";

Service service = ServiceFactory.newInstance().createService();

Call call = service.createCall();

call.setTargetEndpointAddress(endpoint);
call.setOperationName(new QName(OPERATION_NAMESPACE, "hello"));

String response = (String) call.invoke(new Object[] {name});
```

The following code example is taken from [243]

We can call the web services using JavaScript code, this will facilitate the design of web services, we do not need to create complex client design, all we need to implement JavaScript code which will be embedded in a HTML file and run this scrip remotely in client side, the following figure and JavaScript code shows how to call "hello" web service [214].

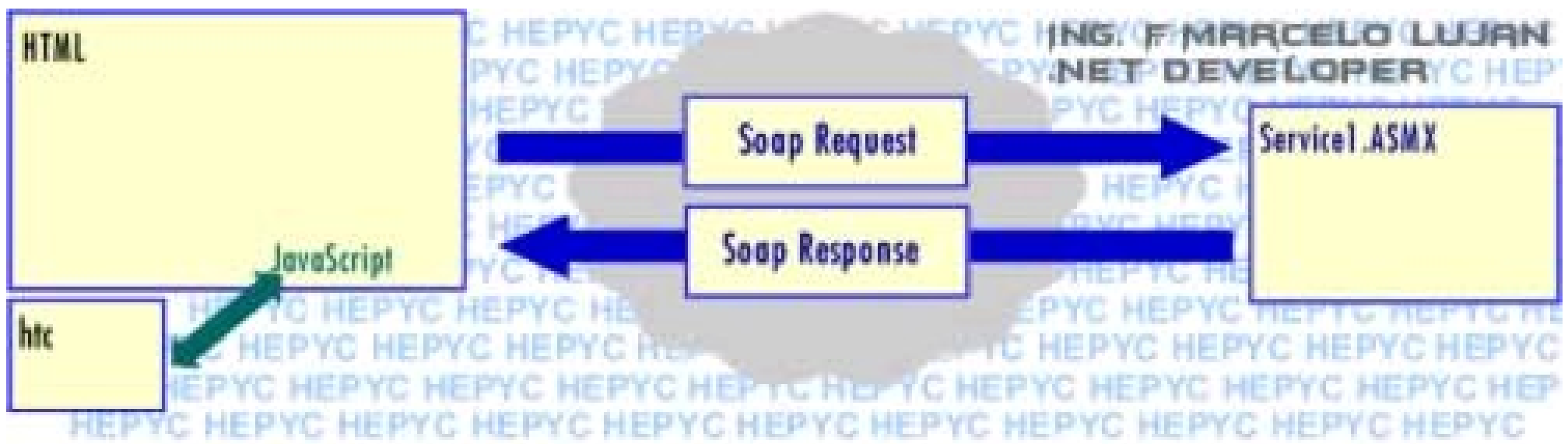

```
<script language="JavaScript">
init(){
 service.useService("http:// http://test.omii.ac.uk/hello.asmx?WSDL",_
                                                "hello");
}
function tst(){
  iCallID = service.Service1.callService("hello","Concordia");
}
function onmyresult(){
  service.innerHTML= "Resultado : " + event.result.value;
}
</script>
<body onLoad="init();">
  <button onclick="javascript:tst()" ID="Button1">Call Add Web Method</button>
  <div id="service" style="BEHAVIOR:url(webservice.htc)" onresult="onmyresult();">
  </div>
</body>
```

The following code example is taken from [214] after doing some modifications

Dealing with JavaScript in order to call web services leads to a simple and flexible solution, hence you don't need to be aware of all transportation mechanism details such as SOAP which is used as a transport layer for web service implementations [244].

### 2.1.4 OO-based abstraction

#### C++

C++ supports both imperative and object oriented based programming style, it provides stack and hap allocation, but it emphasizes the stack model. C++ provides multiple inheritance and exception handling as well as interface definition. C++ does not have garbage collection mechanism as Java did, therefore memory allocation that done by programmer, need to be returned by the programmer himself/herself. C++ support both static and dynamic binding [228].

CPP is built on C and enhance imperative properties and add new features to support object oriented programming. CPP support classes which consider the bases for object oriented programming, for example, to declare a String in CPP, the following class is useful:

```
class String{

private: char*  str;
public:
String(char [] str1)        {
Str=new char[strlen(str1)];
for (int i = 0; i<strlen(str1);i++)
   str[i] = str1[i];}
~String(){ delete [] str; } }
```

C++ support function overloading, also it supports namespaces, which is a methodology used to group multiple classes and namespaces as well as related properties. Also CPP supports Abstraction, encapsulation, inheritance, methods, virtual functions and polymorphism. The following example shows that [206]:

```
// file cats.h

class Felid {
public:
 virtual void meow() = 0;
};

class Cat : public Felid {
public:
 void meow() { std::cout << "Meowing like a regular cat! meow!\n"; }
};

class Tiger : public Felid {
public:
 void meow() { std::cout << "Meowing like a tiger! MREOWWW!\n"; }
};

class Ocelot : public Felid {
public:
 void meow() { std::cout << "Meowing like an ocelot! mews!\n"; }
};
```

This code is taken from [207]

In the previous example we have three types of animals: Cat, Tiger and Ocelot, each of which has different kind of sound "meow", now, we can create a pointer of base class which may point to an object of one of Cat, Tiger and Ocelot. If we call the method "meow" at run time, the implementation of meow method only determined at run time "dynamic binding" this is called polymorphism [206], see the run and the result below:

```
#include <iostream>
#include "cats.h"

void do_meowing(Felid *cat) {
 cat->meow();
}

int main() {
 Cat cat;
 Tiger tiger;
 Ocelot ocelot;

 do_meowing(&cat);
 do_meowing(&tiger);
 do_meowing(&ocelot);
}
```

```
Meowing like a regular cat! meow!
Meowing like a tiger! MREOWWW!
Meowing like an ocelot! mews!
```

This code and run is taken from [207]

## JavaScript

Java Language fully support object oriented concepts and features such as inheritance or polymorphism, however, JavaScript doesn't support inheritance or virtual functions. JavaScript can run within any browser that support JavaScript interpreter [204].

JavaScript support simple object-based paradigm, it allows the programmer to build complex data structure [245], [246]. JavaScript support class construct and it allows defining methods and properties as well as event handler, thus it allow to abstract different kind of objects [247], [248].

All JavaScript data types has instance methods, all data types in JavaScript are Inherited from Object, all objects in JavaScript have prototype [251].

JavaScript support namespaces, which is a methodology used to group multiple classes and namespaces as well as related properties [248]. JavaScript support static and dynamic binding, Example of dynamic binding [250].

```
importPackage(java.util);
var hashtable = new Hashtable();
$javadate = new Java(“java.util.Date”)
$socket = new Java(“java.net.Socket”, “java.sun.com”, 80)
```

JavaScript support function overloading, e.g. [250]:

```
void write(String s)
void write(int i)
```

JavaScript permits the developer to define classes and variable types as well as JavaScript has built-in JavaScript objects [205]. See the following example

```
<script type="text/javascript">
var str="Hello world!";
document.write(str.toUpperCase());
</script>
```

There is no class abstraction in JavaScript, but we can create a an abstract definition of an object in JavaScript. To start with, we need to declare a namespace, suppose that we want to declare a namespace Animal, the following line do that [248],[251]:

```
Type.registerNamespace("Animal");
```

Then, to declare a class Cat, we need to include it into a namespace, the following lines declares Cat constructor [248]:

```
Animal.Cat = function(catType, catName, catAge)

{
this._ catType = catType;
this._ catName = catName;
this._ catAge = catAge;
}
```

The following example shows how to define the class itself, prototyping the properties, methods [248]:

```
Animal.Cat.prototype =
{
getCatType: function()
{
return(this._catType);},
getCatName: function()
{
return(this._catName);},
getCatAge: function()
{
return(this._catAge);},
getCatInformation: function()
{
var str = this._catName + " is a "
+ this._catType + " Cat that is age: "
+ this._catAge;
return(strDetails);
```

```
},
dispose: function()
{alert("destroying " + this.getCatName());}}
```

The next step is to register the class [248]:

```
Animal.Cat.registerClass('Animal.Cat');
```

JavaScript support inheritance and it is based on prototypes concepts which works as a template for new objects [251], this mechanism helps the programmer to avoid declaring common tasks. JavaScript support interface definition and declaration, interfaces help to group common methods that distributed among different class definitions. JavaScript support reflection [248]. JavaScript support some basic collection and other objects like Date, Math, and Arrays [246].

The following example depicts how to deal with array's object

```
<html><body><script type="text/javascript">

var mycars = new Array();

mycars[0] = 1;

mycars[1] = 2;

mycars[2] = "Car";

for (i=0;i<mycars.length;i++)

{document.write((mycars[i]+1) + "<br />");}

</script></body></html>
```

You will figure out that to declare a variable in JavaScript, you need to use var keyword. This var keyword will help you to declare any type.

### 2.1.5 Reflection

#### C++

Reflection is a mechanism that helps the C++ program that is dealing with undefined object compile time, the program check itself and based on some conditions it going to create specific object and invoke its method. Reflection here in a programming context means to build a generic code that deals with undefined object. This mechanism facilitates remote method invocation and serialization. C++ doesn't support reflection [215].

Devadithya et al states that there are two types of reflections: compile-time reflection and run-time reflection, the decision for which code need to be bind at compile time, the decision depends on the metadata availability at compile time [217]. The run-time reflection depends on the program knows the type information at run time, in this case the program able to take the decision at run time, so the reflection behavior happens dynamically at execution time.

C++ doesn't support run-time reflection due to the unrecognizing detailed type information because of the C++ complier hides these information. Run-time reflection in C++ is limited to some features such as monitoring expression types and querying the type name [217].

The following fragment shows how to depicts reflection in C++ [217],

```
classType = ClassType::getClass(‘‘Service1’’);

obj = classType.createInstance();

obj.invoke(‘‘method1’’);
```

Previous code shows how to call any configurable methods. Here is another example [218]:

```
MemberFunction mf =

ct->getMemberFunction("m1");

mf.invoke<int>(&myClassObj);
```

This code to invoke member function m1 which returns int and does not accept any argument.

Remote method invocation is kind of reflection ability, which allow a program to run a code remotely, that code could change depends on the remote host, the remote host could decide to change the implementation of remote method. Also the local host is able to change which method to call by changing the calling argument m1, in the previous example, into different method name [217].

Devadithya et al demonstrated that reflection can be added to C++ language, for more details, see [217],[218].

## JavaScript

Java supports reflection as part of their standard specifications. Reflection in JavaScript context means the ability of a program to dynamically examine program structure and get current object information such as which instance of which class, or which inherited class and or which interface that it implements, as well as class properties and methods [217],[219], and[248].

JavaScript is build on top of Java, ExtendScript is a construct facilitates a reflection behavior; this construct helps to provide information about objects such as its name, a description and properties, as well as input parameters and return type [219].

JavaScript support reflection object, which is a construct that provides the program with reflected-objects' contents, short and long description, all method and defined interface, and all properties defined in the reflected-object's class, the following code is an example of this [219]:

```
var f = new File ("myfile");

var props = f.reflect.properties;

for (var i = 0; i < props.length; i++) {

        $.writeln('this property ' + props[i].name + ' is ' + f[props[i].name]);

}
```

The following example shows an example of object and retrieves it properties, methods information, and data type properties [219]

```
obj = new String ("hi");

obj.reflect.name; // => String

obj = new String ("hi");

obj.reflect.methods; //=> indexOf,slice,...

obj.reflect.find ("indexOf"); // => the method info

Math.reflect.properties; //=> PI,LOG10,... This code gets a list of properties:

Math.reflect.find ("PI").type; // => number   This code gets the data type of a property
```

### 2.1.6 Aspect-orientation

#### C++

Aspects in AspectC++ implement crosscutting concerns in a modular way. Pointcut is considered the major element in AspectC++, pointcut composes single or multiple joint points that need to be used across multiple functions or namespace. Joint point may be evaluated at compile time or at run time [195].

In AspectC++, there are two types of pointcuts: code pointcuts and name pointcuts, for more information, see aspectc++ manual [195]. The following code describes code joint point [195].

```
class Shape;
void draw(Shape&);

namespace Circle {
  typedef int PRECISION;

  class S_Circle : public Shape {
    PRECISION m_radius;
  public:
    ...
    void radius(PRECISION r) { m_radius=r; }
  };

  void draw(PRECISION r) {
    S_Circle circle;
    circle.radius(r);
    draw(circle);
  }
}

int main() {
  Circle::draw(10);
  return 0;
}
```

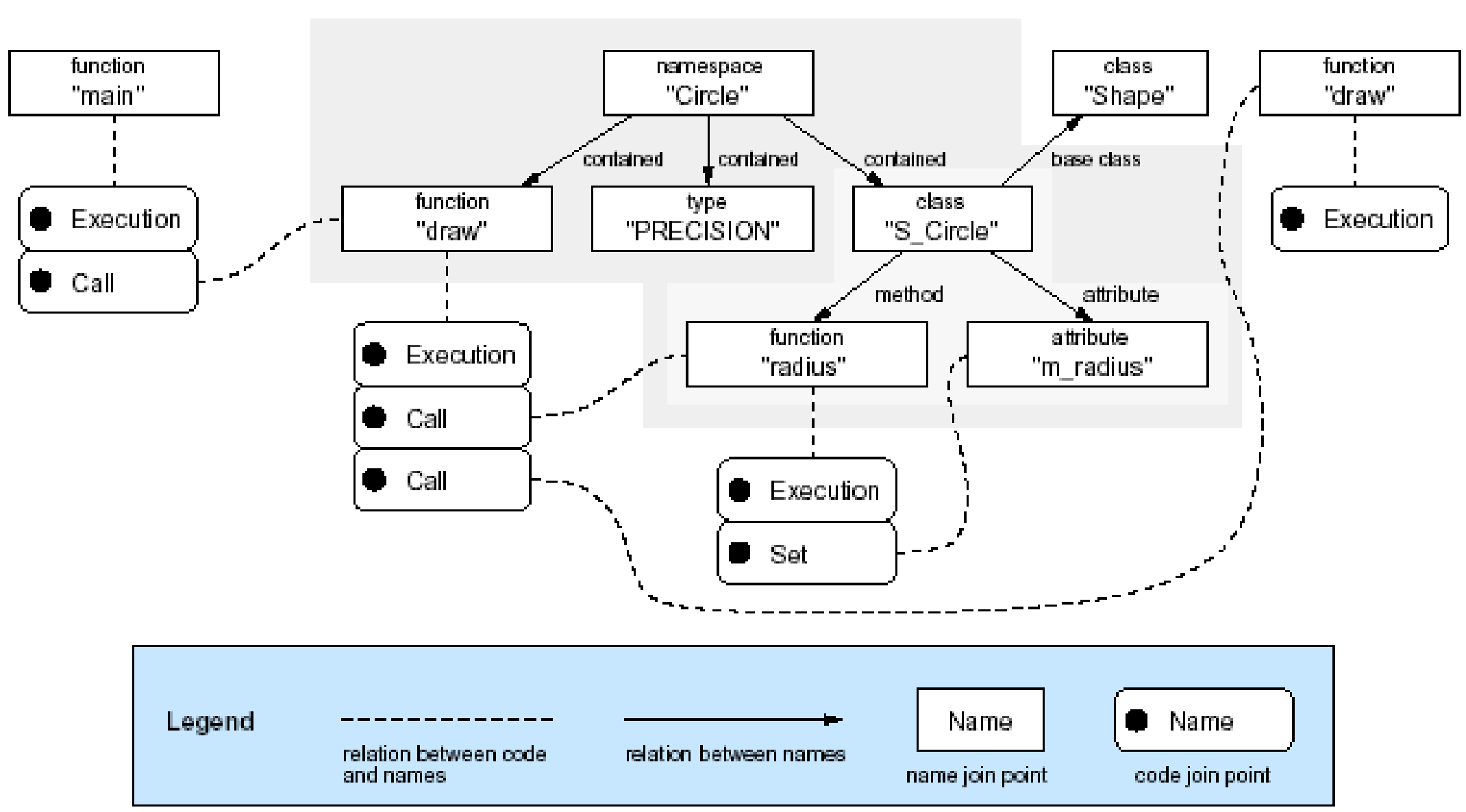

This figure is taken from [195]

The previous code example and figure is been taken from aspectC++ manual [195] to depicts the meaning of name joint point, code joint point.

In aspectC++, each join point could be a function, an attribute, a type or a variable. Depending on the kind of pointcuts, they are evaluated at compile time or at runtime. Name pointcuts may compose types, attributes, functions, variables, or namespaces. Code point cut is an execution point of a function that is repeated in multiple places ia a program [195]. It is possible to mix name pointcuts and code pointcuts within pointcut expression. Pointcut expression is a way for search pattern that uses special character as a wildcard used to aggregate and simplify describing collection of classes, cunfcting or names that share some characters [195].

## JavaScript

XHTML has the advantage to separate web form from its content, this helps to improve the separation of concern for aspect oriented language (AOP). AOP centers around the separation of concern approach, this means to create the main program that invoke the sub programs, each subprogram is concern to perform specific task or concern, such as logging, authentication and synchronization, which are usually repeated and used at many places in many programs [220].

Login and synchronization concerns called cross cutting, AOP helps to improve the memory usage and synchronize cross cutting modification by encapsulation these cross cutting in modular form [220].

JavaScript and C++, have the feature to weave advice code into destination code, this helps to change the behavior of the program without refractor the original code. The advice code is written in a separate file, and woven into the original code at execution time. In java platform, this integration could be in the source code level or byte cide level [220].

JavaScript support AOP utility, as example of advise idiom that have three kinds of aspect directions are seen in the following example [196].

```
Aspects = new Object();

Aspects._addIntroduction = function(intro, obj)
{ for (var m in intro.prototype)
    {
      obj.prototype[m] = intro.prototype[m];
    }
}

Aspects.addIntroduction = function(aspect, objs)
{
  var oType = typeof(objs);

  if (typeof(aspect) != 'function')
    throw(InvalidAspect);

  if (oType == 'function')
  {
    this._addIntroduction(aspect, objs);
  }
```

```
  else if (oType == 'object')
  {
    for (var n = 0; n < objs.length; n++)
    {
      this._addIntroduction(aspect, objs[n]);
    }
  }
  else
  {
    throw InvalidObject;
  }
}

Aspects.addBefore = function(aspect, obj, funcs)
{
  var fType = typeof(funcs);

  if (typeof(aspect) != 'function')
    throw(InvalidAspect);

  if (fType != 'object')
    funcs = Array(funcs);

  for (var n = 0; n < funcs.length; n++)
  {
    var fName = funcs[n];
    var old = obj.prototype[fName];

    if (!old)
      throw InvalidMethod;

    obj.prototype[fName] = function() {
      aspect.apply(this, arguments);
      return old.apply(this, arguments);
    }
  }
}

Aspects.addAfter = function(aspect, obj, funcs)
{
  if (typeof(aspect) != 'function')
    throw InvalidAspect;

  if (typeof(funcs) != 'object')
    funcs = Array(funcs);

  for (var n = 0; n < funcs.length; n++)
  {
    var fName = funcs[n];
    var old = obj.prototype[fName];

    if (!old)
      throw InvalidMethod;

    obj.prototype[fName] = function() {
      var args = old.apply(this, arguments);
```

```
      return ret = aspect.apply(this, Array(args, null));
    }
  }
}

Aspects._getLogic = function(func)
{
    var oSrc = new String(func);
    var nSrc = '';
    var n = 0;

    while (oSrc[n])
    {
      if (oSrc[n] == '\n' || oSrc[n] == '\r')
        nSrc[n++] += ' ';
      else
        nSrc += oSrc[n++];
    }

    n = 0;
    while (nSrc[n++] != '{');
    nSrc = nSrc.substring(n, nSrc.length - 1);
    return nSrc;
}

Aspects.addAround = function(aspect, obj, funcs)
{
  if (typeof(aspect) != 'function')
    throw InvalidAspect;

  if (typeof(funcs) != 'object')
    funcs = Array(funcs);

  var aSrc = this._getLogic(aspect);

  for (var n = 0; n < funcs.length; n++)
  {
    var fName = funcs[n];
    if (!obj.prototype[fName])
      throw InvalidMethod;

    var oSrc = 'var original = ' + obj.prototype[fName];
    var fSrc = oSrc + aSrc.replace('proceed();',
               'original.apply(this, arguments);');
    obj.prototype[fName] = Function(fSrc);
  }

  return true;
}
```

### 2.1.7 Functional programming

**C++**

FC++ is a rich library that supports functional programming in C++. FC++ library provides support to polymorphism using C++ type inference. Big part of c++ standards library is implemented in functional model. C++ language as a functional programming style has the advantage to direct mainplatation for the memory, and has primitive support for dealing with pointers [221].

C++ is object based programming language that support creating classes using the class or struct keyword, C++ support operator overloading. This overloading helps the developer to teach class of object how to deal with operator. The following example shows how to overload "()" operator [221]

```
struct Twice {
  int operator()( int x ) { return 2*x; }
} twice;

struct Inc {
  int operator()( int x ) { return x+1; }
} inc;

  twice(5)     // returns 10
  inc(5)       // returns 6
```

The figure above is taken from [255]

To declare a function in C++, the function should have the following syntax:

```
Return-type function-name(parameter-type parameter-list)
{|
  statement1
  statement2
}
```

The following in an example of function add that accepts two parameters and return the addition of the two parameters:

```
int add (int a, int b){

    return a+b;

}
```

## JavaScript

JavaScript support functional programming [222]. JavaScript support standard functions such as map reduce and select. See the following example [223]:

```
map('x*x', [1,2,3,4])
```

The result is [1, 4, 9, 16]

```
select('>2', [1,2,3,4])
```

The result is [3, 4]

```
reduce('x*2+y', 0, [1,0,1,0])
```

The result is 10

```
map(guard('2*', not('%2')), [1,2,3,4])
```

The result is [1, 4, 3, 8]

In JavaScript block does not have scope, only function have scope [224], To declare a function in JavaScript the key word "function" is used followed by function name and then parameters [225]:

```
function fname(value1,value2, ...)
{
  statement1
  statement2
  .
}
```

See the following example [225]:

```
<HTML>
<HEAD>
<TITLE>A function definition</TITLE>
<SCRIPT LANGUAGE="JavaScript">
<!-- Hiding JavaScript
function listItems(itemList)
```

```
{
  document.write("<UL>\n")
  for (i = 0;i < itemList.length;i++)
  {
    document.write("<LI>" + itemList[i] + "\n")
  }
  document.write("</UL>\n")
}
// End hiding JavaScript -->
</SCRIPT>
</HEAD>
<BODY>
<SCRIPT LANGUAGE="JavaScript">
<!--
days = new Array("Sunday","Monday","Tuesday","Wednesday","Thursday","Friday","Saturday")
listItems(days)
// -->
</SCRIPT>
</BODY>
</HTML>
```

Another example with function that have return value [225]:

```
function average()
{
  var items = average.arguments.length
  var sum = 0
  for (i = 0; i < items;i++)
  {
    sum += average.arguments[i]
  }
  return (sum/items)
}
```

The following is the call of the function average:

```
document.write(average(6,9,8,4,5,7,8,10))
```

### 2.1.8 Declarative programming

#### C++

C++ is an imperative, object based and functional programming paradigm. I didn't find enough references that depict C++ as supporting declarative language.

Pro*C mainly used for dealing with databases, it allows the developer to connect into a database and extract or manipulate SQL statements. Pro*C has some declarative statements mainly used to execute SQL statements, the following figure depicts eight declarative statements [233]:

**List of Embedded SQL Statements Supported by Pro*C**

| Declarative Statements | |
|---|---|
| EXEC SQL ARRAYLEN | To use host arrays with PL/SQL |
| EXEC SQL BEGIN DECLARE SECTION<br>EXEC SQL END DECLARE SECTION | To declare host variables |
| EXEC SQL DECLARE | To name Oracle objects |
| EXEC SQL INCLUDE | To copy in files |
| EXEC SQL TYPE | To equivalence datatypes |
| EXEC SQL VAR | To equivalence variables |
| EXEC SQL WHENEVER | To handle runtime errors |

The figure above is taken from [233]

Compiling Pro*C program is passé two phases, the first phase is to compile Pro*C pre-compiler and the second phase is to compile SQL statements and replaces them with suitable methods that perform the task [233].

## JavaScript

JavaFX is built on the top of Java platform and it supports web applications. JavaFX support desktops, browsers and mobile phones and TV set-top boxes. JavaFX Scrip language considered of declarative language [230].

Compiling JavaFX files will result in java byte code, this java byte code is protable and can execute in a plateform that has a suitable java virtual machine [230].

JavaFX software provide the ability to control playback videos, this ability comes from *javafx.scene.media API* package. The mentioned API allows the programmer to built new features in the elaborate video player. JavaFX technology allows the programmer to build customized graphical tool, to see more capabilities of JavaFX, see [229].

JavaFX script allows the programmer to declare web API application only by using predicates that describes desired GUI components as well as the relationship between them, and then JAVAFX script interprets these components [231],[232] the following example depects that and it is taken from [232].

```
import javafx.stage.Stage;
import javafx.scene.Scene;
import javafx.scene.shape.Rectangle;
import javafx.scene.paint.Color;
import javafx.scene.shape.Circle;

Stage {
    title: "Declaring Is Easy!"
    scene: Scene {
        width: 300
        height: 250
        content: [
            Rectangle {
                x: 25, y: 80 width: 250, height: 80
                arcWidth: 20 arcHeight: 20
```

```
            fill: Color.web("#6699ff")
            stroke: Color.web("#003399")
            strokeWidth: 5.0
          }, //Rectangle
          Circle {
            centerX: 150  centerY: 120 radius: 80
            fill: Color.MAROON
            stroke: Color.INDIANRED
            strokeWidth: 10.0
          } //Circle
        ] //Content
    } //Scene
} //Stage
```

The following picture shows the result of executing previous code [232]:

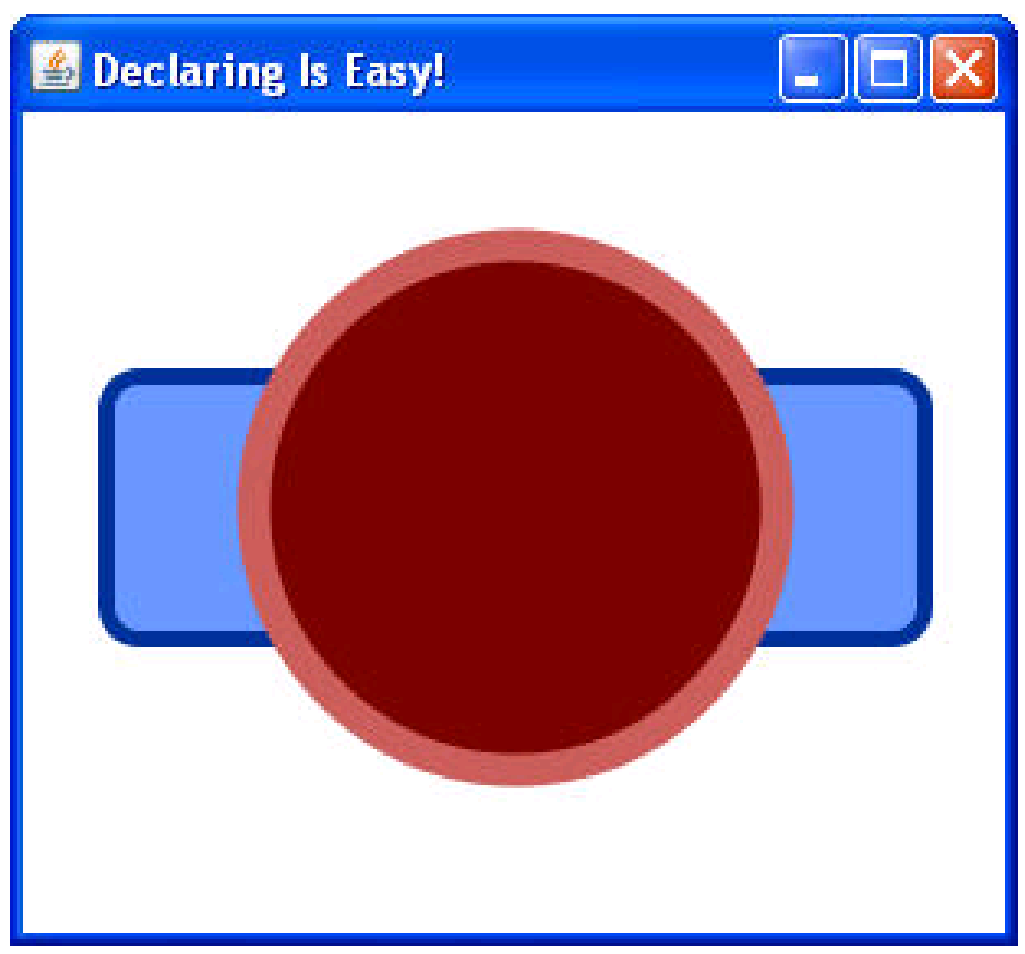

## 2.1.9 Batch scripting

### C++

I didn't find any resources talks about batch script with cpp, I found one cite [253] talks about C++ script.

C++Script uses C++ library and C++ header files, it uses a dynamic programming and cen be compiled with C++ compiler. With C++ scripting the developer can use dynamic typing by declaring a variable with "var" tag, and static typing by using C++ types. Dynamic typing may lead to runtime errors; the following code is in C++ script [253]:

```
#include <cppscript>
var script_main(var args)  // a function called script_main
{
   writeln("Hello world!");
   return 0;
}
```

Variable types in C++ script are : integers, doubles, strings, chars, Booleans containers (arrays, maps, lists etc), objects, methods, functions, iterators and null [253].

C++ Script support C++ operator, casting, comparison, scope, cloning, control flow, foreach statement, user defined functions, exception handling, throw exception, catch and finally statements [253].

### JavaScript

Javascript is able to detect, add, move, change, create, or delete, read and write remote files. The following is JScript Sample [236]:

```
// Instantiate a File System ActiveX Object:
var fso = new ActiveXObject("Scripting.FileSystemObject");
// Invoke the method:
var a = fso.CreateTextFile("c:\\testfile.txt", true);
// Do something with it:
a.WriteLine("This is a test.");
// Close the connection:
a.Close();
```

## 2.1.10 UI prototype design

### C++

Prototype means analyzing the problem, mapping to existing solutions and design, producing a simple design that simulate the solution and the improving the design until we reach to that final result, this will defiantly leads to save money and resources. Prototyping helps the developer to Proof of concept, exploring the design and choose the suitable language to implement the design [239].

C++ provides many compilers for GUI designs and prototypes, such as [238]:

- **ActiveX**: allows the developer to use existing components, such as Internet ActiveX controls, which helps to use the Interent as virtual environment.
- **ATL**: Active Template Library. A collection of template-based C++ classes that works with MFC components.
- **COM:** Component Object Model. A technique used for interprocess communications like remote method invocation in Java.
- **DCOM:** Distributed COM**.**
- **GLUT:** OpenGL Utility Toolkit. An independent library used for OpenGL programs.
- **GUI:** graphical user interface.

More information about these compilers and other can be extracted from [238].

MFC C++ provides the help for developers to develop windows applications and user interface applications this is considered fast technology to create UI prototypes in windows platform. The Web browser has the capabilities to reuse HTML UI with windows' components UI-prototype-windows-style [238], [239].

## JavaScript

JavaScript support prototyping, which helps the developer to use this concept to simulate inheritance in JavaScript programs, all objects have their own prototypes, and all of them inherent the base object prototype. This fact makes adding a new method instance a simple task, the following example shows how to add method push to object of type array [251]:

```
Array.prototype.double = function() {

        var newArray = [];

        // inside this function, "this" refers to the array

        for (var i = 0, length = this.length; i < length; i++) {

newArray.push(this[i]);
newArray.push(this[i]);

        }

        return newArray;

};



var exampleArray = new Array(1, 2, 3);

// inherits all the properties of Array.prototype

exampleArray.double();

//-> [1, 1, 2, 2, 3, 3]
```

Because exampleArray is a instance of Array object, then it inherent the new instance method push and works as depicted in the previous implementation [251].

There is a javascript library called Prototype, this library used with dynamic web applications. This library aims to reduce JavaScript coding [252].

Prototype Window Class (PWC) is a JavaScript class that provides the developer the ability to easily generate windows-style API within an HTML file [240],[241]. The following lines shows an example of windows-style API [241]:

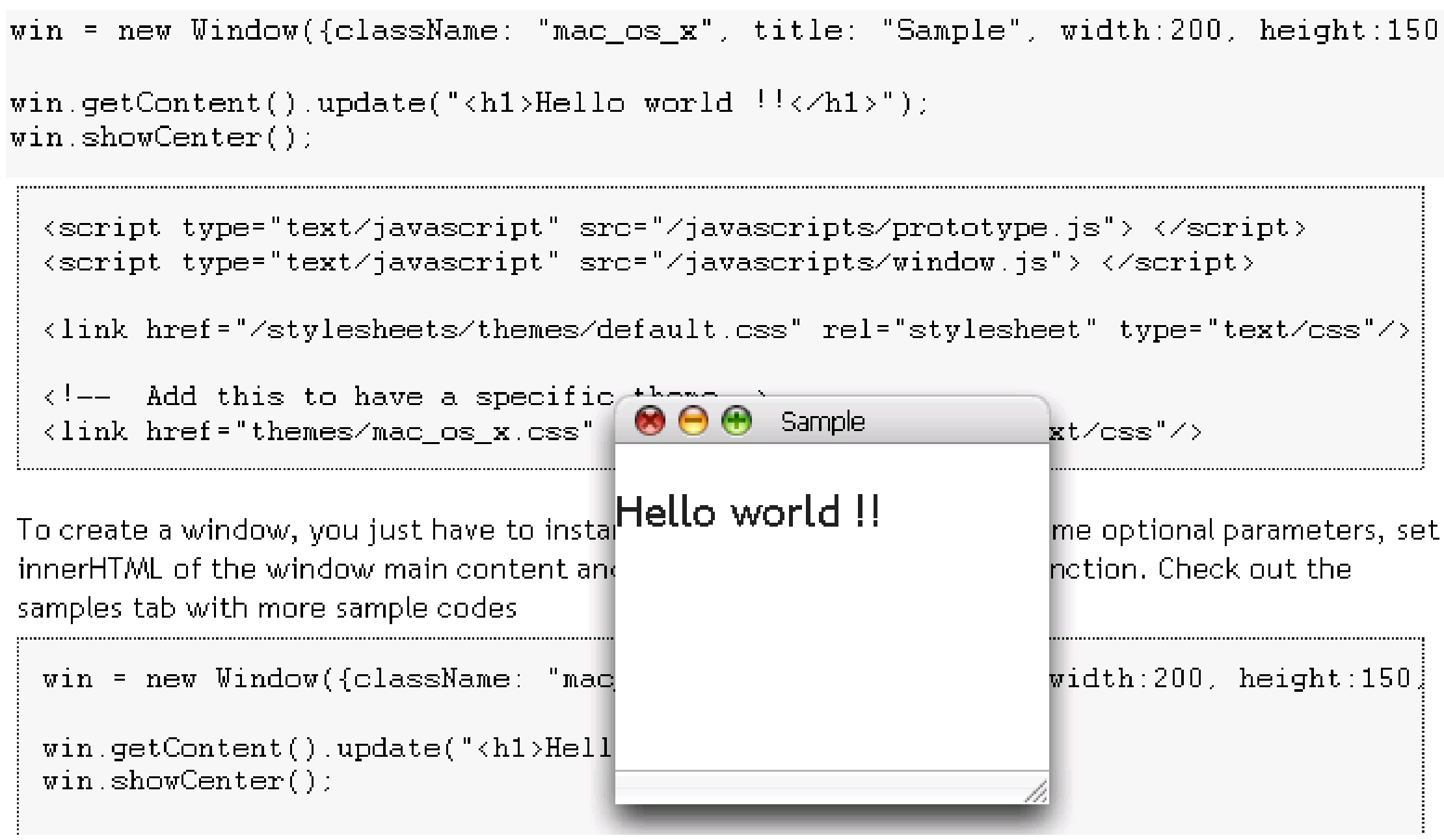

The generated windows inherent all features from windows GUI components. Means, you can maximize, minimize, move generated windows. The following figure shows Prototype Window Class properties and methods [241].

| Window Class | Dialog Module | Windows Module | Windows Add-ons |
|---|---|---|---|
| - keepMultiModalWindow | - confirm | - overlayShowEffectOptions | - WindowsStore.init |
| - initialize | - alert | - overlayHideEffectOptions | - WindowCloseKey.init |
| - destroy | - info | - addObserver | - TooltipManager |
| - getId | - setInfoMessage | - removeObserver | |
| - setDestroyOnClose | - closeInfo | - closeAll | |
| - setCloseCallback | | - getFocusedWindow | |
| - setContent | | - focus | |
| - setHTMLContent | | - blur | |
| - setURL | | | |
| - getURL | | | |
| - refresh | | | |
| - setAjaxContent | | | |
| - getContent | | | |
| - setCookie | | | |
| - setLocation | | | |
| - getSize | | | |
| - setSize | | | |
| - updateWidth | | | |
| - updateHeight | | | |
| - toFront | | | |
| - show | | | |
| - showCenter | | | |
| - minimize | | | |
| - maximize | | | |
| - isMinimized | | | |

JavaScript uses The Scripting API javax.script package which includes the following interfaces: Bindings, Compilable, , Invocable, ScriptContext, ScriptEngine and ScriptEngineFactory. And classes: AbstractScriptEngine, CompiledScript, ScriptEngineManager and SimpleBindings, and one exception class: ScriptException. The detailed information about these classes, interfaces are found in the following cite [249].

## 2.2 AspectJ vs. C#

### 2.2.1 Default more secure programming practices

Secure programming or safe programming is primarily based on two important properties of programming languages: type safety and memory safety. The fundamental characteristic guaranteed by type safety (strong typing) is type soundness, because ensuring type-safety is one of the key issues to protect the system against various security threats. Also as a requirement to be type-safe, a language should have garbage collection or otherwise it has to restrict the allocation and de-allocation of memory [74].

C# is primarily a type-safe language, i.e., types can interact only according to the rules defined and not violated by program semantics, thereby ensuring each type's internal consistency. More specifically C# supports static typing, which allows it to perform the type checking at compile time. Moreover, C# has one distinguishable feature to make the certain part of a program dynamic typed via the *dynamic* keyword.

C#'s CLR has a garbage collector that executes as part of the program, frees the memory for managed objects automatically that are no longer referenced. This technique relieves programmers from explicitly de-allocating the memory for an object. In Listing 1, when Test method executes, an array to hold 500 bytes is allocated on the memory heap. When the method exits, this local variable *test_arr* pops out of scope, meaning that nothing is left to reference the array on the memory heap. The array then becomes eligible to be reclaimed in garbage collection to free the memory. However, it is possible to force to work Garbage Collector by calling GC.Collect.

```
public void Test()
{
        byte[] test_arr = new byte[500];
        ..........
}
```

Listing 1: Example C# memory allocation

In addition to this C# supports the disposal pattern explicitly by implementing the *IDisposal* interface. The *Dispose* method released the unmanaged resources like database connections, handles, files etc. As shown in the Listing 2, MyClass has implemented the *IDisposable* interfaces to release the unmanaged resources like file object.

```
class MyClass : IDisposable
{
  FileStream myStream; //large object

  public MyClass (string filePath)
  {
    myStream = new FileStream
      (filePath, FileAccess.ReadWrite);
  }

  public void Dispose()
  {
```

```
        myStream.Dispose();
        //remove myStream from the GC finalize queue
        GC.SuppressFinalize(myStream);
    }
}
```

Listing 2: Example C# Dispose method for unmanaged resource [80]

On the other hand, AspectJ is also a static-typing programming language and also considered as type-safe like its base class Java. However, researches [95] have revealed that, unlike Java, AspectJ does not have a safe type system, a binding between a pointcut and an advice can rise to type errors at runtime. Also, AspectJ's typing rules severely restrict the definition of certain generic advice behavior. In AspectJ, a cross-cutting concern i.e., memory monitoring and management can be applied at a pointcut of the program for better memory management. This approach can be used for both managed and unmanaged resources (files, handles, DB connections etc.).

C# is a static typing and type-safe programming language, whereas AspectJ is not really a type-safe language in some sense. But a cross-cutting concern can applied for better memory management in AspectJ, which could be an important application of this.

### 2.2.2 Web Application Development

A web application is a kind of application that is accessed over a network such as the internet or an intranet [75]. It is usually hosted in a web server and from the client side operated by thin client. Web applications are getting very popular due to the ubiquity of web browsers, and the convenience of using a web browser as a client. Nowadays, various platforms and frameworks have been implemented for web development. Primarily, web application development required coding and development in two specific areas, i.e. server sided scripting and client side scripting. For client side development, programming/scripting languages like JavaScript, Flash - Action Script, Ajax etc. are very popular. For server side development, C#, AspectJ can be used in their own framework or platform.

C# language of .Net platform is used for web application development in ASP .Net technology, where every ASPX file can have a C# class. The C# file that contains this class used for ASP .Net programming implementation is called *Code Behind File.* This class may contain initializes, event handlers, supporting methods and codes etc. The developed web applications are needed to host in the Microsoft IIS server. The web development using C# is performed in the Visual Studio IDE, which provides extensive and rich support for development like web forms designer, various built in web server controls, framework libraries etc. Development using C# is flexible, faster and convenient for the programmers.

In Web application development, AspectJ separates the cross-cutting concern from the core concern and perform several activities. In a research of IBM, Ron Bodkin has implemented several monitoring, checking and error handling technique using AspectJ [87]. These can be implemented in Web application development:

- Monitor multiple applications
- Monitor & detect for common application failure
- Error handling & repair
- Database request monitoring
- Authentication, verification and user session management

AspectJ provides an improved technique to implement the cross-cutting concern in a better way without hampering the existing code. But, comparatively C# is better for rapid and convenient web development, whereas AspectJ is better for modular coding.

### 2.2.3 Web services design and composition

A Web service is a software system designed to support interoperable machine-to-machine interaction over a network. It has an interface described in a machine-process able format (specifically WSDL). Other systems interact with the Web service by its description using SOAP messages, typically conveyed using HTTP with an XML serialization in conjunction with other Web-related standards [76]. So, basically Web services are based on a core set of standards that describe the syntax and semantics of communication. Here, XML provides the common syntax for representing data; the SOAP (Simple Object Access Protocol) provides the semantics for data exchange; and the WSDL (Web Services Description Language) provides a mechanism to describe the capabilities of a Web service. Unlike traditional client-server systems, such as a web page, Web service do not provide the users with GUI rather share business logic and various processing through a interfaces across the network.

The .NET Framework provides extensive support for interoperability through Web services. In C#, using the .NET Framework, Visual Studio, and ASP.NET, creating a Web service is as simple as just create a Web service project and add a public attribute *WebMethod,* which is to expose [81].

Creating a Web Service is simpler in C# using .NET Framework and the Visual Studio IDE. Compiling and deployment of the Web service is easy and faster. A simple Web service example is given here, which receives a message from the client and returns the same.

```
//services.asmx – C# Web Service
using System;
using System.Web;
using System.Web.Services;
using System.Web.Services.Protocols;

[WebService(Namespace = "http://tempuri.org/")]
[WebServiceBinding(ConformsTo = WsiProfiles.BasicProfile1_1)]
public class Service : System.Web.Services.WebService
{
   public Service () {

   }

   [WebMethod]
   public string echo(string message) {
      return message;
   }
}
```

Listing 3: A simple Web service developed in C#

A simple aspect-oriented Apache Axis Web service (equivalent to the previous one created in ASP .Net) that has been created using AspectJ is mentioned here. Apache Axis is rapidly becoming popular web service implementations for Java developers. It also has a high scale of success in interoperability with other web services and frameworks such as .NET. When Apache Axis running within Tomcat server in a machine, we can use the following code to compile and deploy as a simple aspect-oriented web service:

```
//MyWebService.java

package com.oreilly.aspectjcookbook;
public class MyWebService
{
  public String echo(String message)
  {
    return message;
  }
}
```

Listing 4: Web service Class – Web Method

```
//AddMessageHeaderAspect.java

package com.oreilly.aspectjcookbook;

public aspect AddMessageHeaderAspect
{
  public pointcut captureEcho(String message) :
    execution(public void MyWebService.echo(String)) && args(message);

  Object around(String message) : captureEcho(message)
  {
    return "Your original message was: " + message;
  }
}
```

Listing 5: Web service Aspect

```
//deploy.wsdd

<deployment xmlns="http://xml.apache.org/axis/wsdd/"
         xmlns:java="http://xml.apache.org/axis/wsdd/providers/java">

 <service name="MyWebService" provider="java:RPC">
  <parameter name="className" value="com.oreilly.aspectjcookbook.MyWebService"/>
  <parameter name="allowedMethods" value="*"/>
 </service>
</deployment>
```

Listing 6: deploy.wsdd for automatic deployment

This example is taken from the “AspectJ Cookbook” by Russell Miles. Basically, AspectJ can perform various functionalities as a Web service [87]:

- It can monitor multiple services hosted on a single server
- It can monitor web service call
- Monitor for service failure and repairmen in the service pipeline

Creating a web service in AspectJ is comparatively more tedious and complicated than C#. It required to compile the source file to class file, than creating of *deploy.wsdd* is required for automatic deployment of the service and also some additional work to make the service live in the tomcat server. Whereas, in the case of C#, it is very fast and easy to deploy a service in the IIS server. But, in AspectJ the separation of cross-cutting concern can make the service more modular. AspectJ can separate the cross-cutting concern from the core concern and perform monitoring, verification, authentication or providing notification without tangling or scattering the code.

### 2.2.4 OO-based abstraction

Object-oriented programming (OOP) is a programming paradigm that uses “objects" data structures consisting of data fields and methods together with their interactions to design applications and computer programs [75]. In OOP, programming techniques may include features such as data abstraction, encapsulation, modularity, polymorphism, and inheritance. Object-Oriented Development Methodology has basically four principles. Its *entities* and the *relationships* between them must satisfy the following four principles [77].

**Abstraction of data and functions principle:** This means that the computation is separated specifically into entities, each consisted of data and functions. These are the only functions permitted to manipulate the data of the entity directly. The whole entity can be manipulated as a unit. This is one of the most important principles of Object-Oriented paradigm.

**Information encapsulation principle:** This means that the implementation information and the mechanism of a computational entity are hidden from other computational entities. Each entity is only required to provide a well defined interface for the other entities to interact with each other. The purpose of this principle is to shield a service using entity like class from the implementation details of the service provider entity.

**Inheritance principle:** This means that we can create a child entity based on a parent entity. So the child entity can obtain all the properties from its parent entity and the new entity can be changed in data and functionality to perform the specific new task. This capability provides different ways of utilizing and reusing previously developed entities.

**Polymorphism of methods:** The polymorphism principle is the ability to create methods which have similar but some entity specific functionality. The principle is implemented in conjunction with the inheritance principle.

```
//Example Simple Object-oriented approach by C#

class MyClass
{
   private int x; // private access
   private int y; // private access
   public int z; // public access
   // Methods to access x and y.
   //a member of a class can access the private members of the same class.
   public void Setx(int a)
   {
      x = a;
   }
   public int Getx()
   {
      return x;
   }
   public void Sety(int a)
   {
      y = a;
   }
   public int Gety()
   {
      return y;
   }
}

//Using the class from main
class Program
{
   static void Main(string[] args)
   {
      MyClass ob = new MyClass();
      // Access to x and y is allowed only through setx() and getx() methods.
      ob.Setx(-99);
      ob.Sety(19);
      ob.z = 5;

      Console.WriteLine("ob.x is " + ob.Getx());
      Console.WriteLine("ob.y is " + ob.Gety());

      // It cannot be accessed for x or y like this:
      // ob.x = 10; // Wrong! x is private!
```

```
            // It is OK ob.z
            Console.WriteLine("ob.z is " + ob.z);

            Console.ReadKey();
        }
    }
```

Listing 7: Simple Object-oriented program by C#

From the Listing 7, here we see a simple C# program that satisfies the basic principles and properties of OO paradigm like class, method, message passing, abstraction, message passing, encapsulation etc.

On the other hand, in AspectJ an aspect can encapsulates the implementations of cross-cutting functionalities. Aspects are similar to classes in many ways. The similarities [82] are discussed below in detail…

**Aspects Can Include Data Members And Methods:** The data members and methods in aspects have the same role as in classes. For instance, an aspect can manage its state using data members, whereas methods can implement behavior that supports the crosscutting concern's implementation or can be utility methods. Aspects may also have constructors included.

**Aspects Can Have Access Specifications**: An aspect's access specifier provides its visibility like the same as classes and interfaces. Top-level aspects can have only public or package access. Moreover, nested aspects are also similar to nested classes, which can have a public, private, protected access specifier.

**Aspects Can Be Abstract:** Like a class, an aspect that contains abstract pointcuts or methods must declare itself as an abstract aspect. An abstract aspect can mark any pointcut or method as abstract and refer to it from other constructs. Any subaspect of an abstract aspect that doesn't define every abstract pointcut and method in the base aspect, or that adds additional abstract pointcuts or methods, must also declare itself abstract. The following example shows an abstract aspect that contains an abstract pointcut and an abstract method:

**Aspects Can Extend Classes And Abstract Aspects, as well as Implement Interfaces:** Aspect has the inheritance properties, which is one of the basic principles of OOP. In the example, below, the following concrete aspect extends the AbstractTracing aspect. So that, it is able to provide the definitions for its abstract pointcut and method by matching the requirements of tracing the banking system:

```
public aspect BankingTracing extends AbstractTracing
{
        public pointcut traced()
        : execution(* banking..*.*(..));
        public Logger getLogger()
        {
                return Logger.getLogger("banking");
```

```
    }
}
```

**Aspects Can Be Embedded In Classes And Interfaces As Nested Aspects:** Aspects can be embedded into classes and interfaces when the aspect's implementation is closely tied to its enclosed class or interface. .

Though C# and AspectJ are two languages from two different programming paradigms (object-oriented and aspect-oriented respectively), their key properties like classes & aspects have some significant similarities. An aspect supports the object-oriented properties like abstraction, encapsulation, inheritance etc (partially in some cases). Moreover, AspectJ is developed as extension of Java, which is often called a pure object-oriented language and undoubtedly it supports all the four major principles of OO programming [72]. Basically, an Aspect is required for separating the cross-cutting concerns, but the core concerns can be implemented using object-oriented methodology. And, eventually AspectJ comes up with the support of both the programming paradigm OOP & AOP.

### 2.2.5 Reflection

Reflection-oriented programming, or reflective programming, is a functional extension to the object-oriented programming paradigm. Reflection-oriented programming includes self-examination, self-modification, and self-replication. However, the emphasis of the reflection-oriented paradigm is dynamic program modification, which can be determined and executed at runtime. Reflection in a programming language can be used to observe and dynamically modify or change the program execution at runtime [75].

In C#, many of the classes that support reflection are part of the .NET Reflection API, which is in the System.Reflection namespace. Thus, any program that will use reflection will have to include the System.Reflection namespace. The reflection namespace has very rich and powerful support for reflective programming, which has extended C# for reflection-oriented programming paradigm. A static binding example below is represented dynamic use of reflection.

```
// Without Reflection
string s = "Programming";
int length = s.Length;

// With Reflection
object s = "Programming";
PropertyInfo prop = s.GetType().GetProperty("Length");
// GetValue can get the value of the PropertyInfo
 int length = (int)prop.GetValue(s, null);
```

Listing 8: Reflection example of C#

In AO languages such as AspectJ offer very powerful but, controlled mechanisms to modify the execution flow. AspectJ offers an alternative way to access the static and dynamic context associated with the join points through a reflection API. For example, through this API, we can access the name of the currently advised method as well as the argument objects to that method. The most common use of this reflective information is to implement tracing and similar aspects [82].

Both C# and AspectJ supports reflection in their area with their powerful and controllable API/namespace. C#'s Reflection namespace is based on its OO paradigm; whereas AspectJ's Reflection API is on AO paradigm.

### 2.2.6 Aspect-orientation

Aspect-oriented programming (AOP) is a programming paradigm in which secondary or supporting functions are isolated from the main program's business logic [67]. It aims to increase modularity by allowing the separation of cross-cutting concerns and forming a basis for aspect-oriented software development. All AOP implementations have some crosscutting expressions that encapsulate each concern in one place.

Till the release of C# 4.0, there are no extensions or implementation was carried out by Microsoft Corporation for Aspect-oriented programming paradigm. But, there are currently several AOP Frameworks available for the .NET space implemented by third party, each with their own approach and having their own positive and negative attributes. Among those, Spring.Net [83] is the most renowned one, which provides comprehensive infrastructural support for developing enterprise.NET applications, with AOP implementation. Spring.NET AOP is implemented in C# language. There is no need for a special compilation process - all weaving is done at runtime. Spring.NET AOP does not need to control or modify the way in which assemblies are loaded, and is thus suitable for use in any CLR environment. Since C# doesn't support the AOP yet, we'll see by example, an implementation of authentication procedure in C# and the response of AspectJ, on the other hand.

```
class Program
  {
    static void Main(string[] args)
    {
      SendMessage sm = new SendMessage();
      Authenticator auth = new Authenticator();
      try
      {
        auth.authenticate(); //check and authenticate
        sm.sendNow("This is an important message.");
      }
      catch(Exception ex)
      {
        //catch the exception generated for verification failure
        Console.WriteLine("User/password didn't match");
      };

      Console.ReadKey();
    }
  }

  //Sending the message
  public class SendMessage
  {
    public void sendNow(String message)
           {
                   Console.WriteLine(message);
           }
  }
```

```
// Authanticator.java
public class Authenticator
{

    public void authenticate()
            {
        string user = "";
                    string pass = "";
                    try
                    {
          //user input for user and password
                            Console.Write("Username: ");
                            user = Console.ReadLine().Trim();

                            Console.Write("Password: ");
                            pass = Console.ReadLine().Trim();

                    }
                    catch(Exception ex)
        {
          Console.WriteLine(ex);
        }

        //if user name and password is not same then throw exception
        if(user!=pass)
                            throw new Exception();
            }
}
```

Listing 9: Implementation of Authenticated Message Sending in C#

Before delivering a message, an authentication method required to invoke to check whether the user is authenticated or not. Each method that requires to be authenticated has to call authenticate method, which leads to code tangling. Similar code needed to be included in all the classes that require authentication.

```
// main.java

public class Main

{
        public static void main(String[] args)
        {
                SendMessage sm = new SendMessage();
                sm.sendNow("This is an important message.");
        }
}
// SendMessage.java
public class SendMessage

{
        public void sendNow(String message)
        {
                System.out.println(message);
        }
}
// SecurityAspect.aj
public aspect SecurityAspect
{
```

```
    private Authenticator authenticator = new Authenticator();

    pointcut secureAccess()
  : execution(* SendMessage.sendNow(..));

    before() : secureAccess() {
        System.out.println("Checking and authenticating user");
        authenticator.authenticate();
    }
}
// Authanticator.java
public class Authenticator
{

    public void authenticate()
    {
        String user = new String();
        String pass = new String();

        BufferedReader in = new BufferedReader(new InputStreamReader(System.in));

        try
        {
            System.out.print("Username: ");
            user = in.readLine().trim();

            System.out.print("Password: ");
            pass = in.readLine().trim();
            System.out.print(user + " " + pass);

        }
        catch(IOException ex)
        {}
        if(!user.equalsIgnoreCase(pass))
            throw new AuthenticationException("User/password didn't match");
    }
}
// AuthenticationException.java
public class AuthenticationException extends RuntimeException
{
    public AuthenticationException(String message)
    {
        super(message);
    }
}
```

Listing 10: Implementation of Authenticated Message Sending in AspectJ

Without changing a single line of code in the SendMessage class from listing 9, we can enhance its functionality by adding an aspect to the system, which is mentioned here as SecurityAspect. The Authenticator class asks for credentials (username and password) when the authenticate() method is called. The AspectJ example here is written by with help from the book AspectJ in Action by Ramnivas Laddad [82].

So, from two above examples, it is easily understandable that for cross-cutting concern like logging, authentication can make the OO code tangling or scattered, whereas AOP programming makes it more modular and readable.

### 2.2.7 Functional programming

Functional programming is a programming paradigm that treats computation as the evaluation of mathematical functions and avoids state changes and mutable data. It emphasizes the application of functions, in contrast to the imperative programming style, which emphasizes changes in state. Programming in a functional style can also be accomplished in languages that aren't specifically designed for functional programming. For example, in C# 3.0 or higher, lambda functions can be employed to write programs in a functional style [67].

The code example of functional implementation is taken from c-sharpconner.com [84]. With the release of the 3.5 framework, we have a completely different coding style available (functional programming). We could actually do functional programming in the 2.0 framework, but the resulting code was ugly and hard to understand at a glance and hard to maintain as well. Now, with lambda method syntax and extension methods, we can produce code written in a functional style as in the Listing below. Here, the convert method is used to trim the empty space from the ends of the strings by passing method defined using lambda syntax:

```
// Functional prgramming example with Lanbda expression
File.OpenText(str1[0]).Use(stream =>
{
    stream
        .ReadToEnd()
        .Split(' ')
        .Convert(str => str.Trim())
        .GetCounts((x, y) => x == y)
        .ForEach(kvp => String.Format("{0} count: {1}", kvp.Key, kvp.Value.ToString()));
});
```

Listing 11: Implementation Functional Style

Specifically in AspectJ, no functional programming implementation is observed. Rather in some research works, aspect-oriented functional language implementation or prototype implementations are found, such as Aspectual Caml [88], AspectFun [89] and AspectML [90]. In these papers, authors made an approach to build aspect-orient extension over a functional language to create a multi-paradigm or hybrid programming language, which is reverse in this occasion. Basically, Aspect-oriented programming paradigm is developed on top of Object-oriented programming paradigm. And AOP core concerns are primarily developed in object-oriented approach. Since AspectJ is developed as a superset of Java, it holds all the functionalities and properties of Java. Superficially, Java is not a functional language; however, using interfaces and inner classes it is fairly easy to mimic some of the important features of functional programming [92]. Using an interface for real functions, it is possible to write functions that take functions as parameters and construct and return new function. In addition to that, some open source implementations are available for functional programming in Java like FunctionalJ, Functional Java, JFun etc. In the next release of Java, which is JDK7, Lambda expression will be introduced with more functional programming concepts like closure, higher order function etc. [93].

So in comparison with AspecJ and its subset Java, C# provides more functional programming ability with its built in namespace and library.

### 2.2.8 Declarative programming

Declarative programming is a programming paradigm that expresses the logic of a computation without describing its control flow. Many languages applying this style attempt to minimize or eliminate side effects by describing what the program should accomplish, rather than describing how to go about accomplishing it. Declarative programming is often defined as any style of programming that is not imperative. A number of other common definitions exist that attempt to give the term a definition other than simply contrasting it with imperative programming[72]. For example:

- A program that describes what computation should be performed and not how to compute it
- Any programming language that lacks side effects (or more specifically, is referentially transparent)
- A language with a clear correspondence to mathematical logic.

C#'s ability to programming in declarative style can be observed through its two new implementation, those are LinQ and regular expression. LINQ, or Language Integrated Query, is a set of language and framework features for writing structured type-safe queries over local object collections and remote data sources. LINQ was introduced in C# 3.0 and Framework 3.5. It enables to query any collection implementing IEnumerable<T>, whether an array, list, or XML DOM, as well as remote data sources or database tables. Additionally, the regular expressions a notable implementation of C# language is able to identify complicated character patterns. [85].

```
//Example LINQ implementaion

var linqdata = row_line.AsQueryable(); //getting values from rowline collection

// Linq expression for getting the counts for each key/combination groups
var Found = from ad in linqdata
        group ad by ad into g
        select new { Names = g.Key, Counts = g.Count() };


//Example Regular Expression [85]

Regex bp = new Regex(@"\d{2,3}/\d{2,3}"); // exp. to find the blood pressure reading
Console.WriteLine(bp.Match("It used to be 160/110")); // 160/110
Console.WriteLine(bp.Match("Now it's only 115/75")); // 115/75
```

Listing 12: Declarative Programming in C#

Specifically in AspectJ, no declarative programming implementation is observed. But, for its base class Java, declarative programming can be implemented through annotations [97].

So in comparison with AspecJ and its subset Java, C# provides more declarative programming ability with it's built in library LinQ and Regex.

### 2.2.9 Batch scripting

Operating system's command line and its batch/shell scripting capabilities is a core support for systems administrators and power users but is relatively unknown to many PC users. The purpose of this is to make a comparative study of this powerful utility of the command line between C# and AspectJ.

Here a comprehensive analysis has been performed to identify the capabilities of various batch scripting techniques in C# from MSDN [81].

- C# has a rich and powerful implementation of System.Diagnostic namespaces, which can used for executing the external commands.

```
// Example implementation of external command

Process proc = new Process();
proc.StartInfo.WorkingDirectory = @"G:\My Documents\COMP 6411\";
proc.StartInfo.FileName = "notepad.exe";
proc.StartInfo.Arguments = "test.txt";
proc.StartInfo.UseShellExecute = false;
proc.StartInfo.RedirectStandardOutput = false;
proc.StartInfo.RedirectStandardError = true;
proc.Start();
proc.WaitForExit();
proc.Close();
```

Listing 13: Example Batch Script in C#

- t has extensive library support to work with Microsoft Office packages, like reading, writing, creating files, macros etc.
- Using the method and properties of diagnostics namespaces it can monitor the system performance and can send and receive external message/mail with the help of its internal library support.

Aspect-oriented programming (AOP) is a natural fit for solving the problems of system monitoring. AOP lets us define pointcuts that match the many join points where we want to monitor performance. We can then write advice that updates performance statistics/logs, which can be invoked automatically whenever the process enter or exit one of the join points. In a research of IBM named AOP@Work: Performance monitoring with AspectJ [87], a basic aspect-oriented performance monitoring system was developed by Ron Bodkin using AspectJ. This system captures the time and counts for different Servlets' processing incoming Web requests. Same approach can be applied to implement a batch script to monitor the system or any activities with the support of Java's API for external /internal command execution.

Unlike C#, a batch script in AspectJ, can be written for both Windows and Linux platforms. Moreover, its Java API support for external/internal command, automation, scheduling etc are more or less same like C#. Additionally, its AOP implementation can provide better modularity by separating the cross-cutting concern like security, authentication, statistics collection etc.

### 2.2.10 UI prototype design

In Visual C#, the most rapid and convenient way to create user interface (UI) is to do so visually, using the Windows Forms Designer and Toolbox. There are three basic steps in creating all user interfaces:

- Adding controls to the design surface and also dynamically when required.
- Setting initial properties for the controls.
- Writing handlers for specified events.
- It is possible implement thread in Windows Form program with the *BackgroundWorker* class. An intensive task needed, can be done on another thread so that it can be avoided for the UI from freezing or stop responding in this implementation.
- It has a portability problem, which makes it failed to acquire an important feature Platform Independence. UI designing is possible only in Microsoft Operating System, since it C# is a Microsoft proprietary language. But with the help of some other framework/tool it can be deployed on other platform.

Although it is possible to create UI by writing in the code, designers enable to do this work much more rapidly than is possible by manual coding [81].

Since, AspectJ is extended from Java, it came up with all the User Interface API provided by the Java. But, from the software development perspective, some distinguished properties & supports can be availed from aspect-oriented features of AspectJ:

- Some kind of policy, rules or regulation update can be provided by a notification or message without hampering the core concerns of the system, just by writing new aspect for a particular work [86].
- Memory management for GUI application (by calling the Garbage Collector on demand at certain pointcuts, as example) and performance can be monitored through the aspect implementation.
- Synchronizations of UI thread can be monitored and at the respective pointcut whenever the thread enters or exits the method [98].
- Java's extensive API can be used for implementing the user interface.

The IDE of C# is more convenient for designing user interface. Its extensive support of tools and controls provide more flexibility and faster development techniques for the programmers. Though, designing UI in AspectJ with Java API is little more tedious and less convenient than C#, but, it can provide some additional flexibility for the cross-cutting concerns by separating them from the core concern and making the code more modular than C#.

## 2.3 Haskell vs. Java

### 2.3.1 Default more secure programming practices

Security is a primary concern in language design & implementation. Well define & reliable program execution prevents attackers from circumventing security policies by exploiting weaknesses in language models. Unified modeling languages provide a way to express higher level of system abstraction. Language safety is a generalization of common notions such as type safety & memory safety. Advanced programming abstractions such as Threads & distributed message passing have become common in modern languages. Associated language features give programmers powerful, flexible control over various resources. Recent languages based security research & development seeks to add security behaviors to language execution models & features for programming security policies to language syntax [103].

### Security in Java

Java is a language mostly used for internet purpose hence security is important because networks provide an avenue of attack to any computer hooked to them.

- Robust- For controlling memory, I/O devices or other hardware.
- Type-safe-It is a strongly typed language helps reduce errors in programs at compile time enhances the integrity and security of software [102].
- Automatic memory management-
  Garbage collector- The correctness of the Garbage collector implementation is essential to the reliability and security. Java RMI collector is the most widely used distributed garbage collector [104].

**Drawback-**It is still difficult to implement system with rigorous safety and real time requirements, most of the overheads incurred by garbage collection. There has been much research on scheduling the garbage collector and improving the efficiency of code transformation, even though it has not proven particularly effective so far [102].

**Security in Haskell**

- Due to pure-function concept no memory or I/O side effects, if the result of any pure expression is not used, it can be removed without affecting other expressions.
- Thread safe [104]. It is strongly typed. Types may be polymorphic i.e. they may contain type variables which are universally quantified over all types. Haskell does not require explicit type declaration, the type interface system provides static type checking[153]
- Automatic memory management-
  Garbage Collector- Haskell has an internal Garbage collector [105].

### 2.3.2 Web Application Development

A web application is any application that uses a web browser as a client. Most web applications are based on the client-server architecture where the client enters information while the server stores and retrieves information.

The Applications are broken into different layers are known as tires.

1. Presentation layer -Client side program-This Layer creates a visual gateway for the consumer to interact with the application. This can range from basic HTML and DHTML to complex COM components and Java applets.

2. Application Logic/Business Logic- This tier can range from Web scripting in ASP/PHP/JSP/HSP to server side programming such as TCL, CORBA and PERL, that allows the user to perform complex actions through a Web interface.

3. Database Layer- Databases allow developers to store, retrieve, add to, and update categorical information in a systematic and organized fashion.

For more complex application a 3-tier solution may fall short so n tiered approach is benefited to break the business logic, which resides on the application tier into a more fine grained model [106].

### Web Application in Java

Java builds high quality web application. Java has APIs to create these applications.

- The servlet component is used as a controller, the Java bean component is a model and JSP is used as a view template. The Enterprise Java Beans can be used as a model which can be located in distributed environment.
- Java EE platform provides support for security, authentication, authorization, transaction.
- The Database Connection management with JNDI API provides flexibility [107].
- Java API for XML Processing (JAXP), part of the Java SE platform, supports the processing of XML documents using the Document Object Model (DOM). JAXP enables applications to parse and transform XML documents independent of a particular XML-processing implementation.
- Java Persistence API is a Java technology standards-based solution for persistence. Persistence uses an object-relational mapping approach to bridge the gap between an object-oriented model and a relational database[108]

#### Web Application in Haskell

Haskell does not have inbuilt components like Servlets in JAVA to create web application it relies on certain libraries –

- CGI & XHtml libraries the monad transformers in Haskell are used to add application specific functionality [109].
- Database connectivity system HDBC. This library allows to write a code & access data stored in almost any SQL database. [104]

## 2.3.3 Web services design and composition

Web services can be considered as one kind of service, which is useful as it offers syntactical interoperability with remote services in a platform independent way [110].

Using Web service technologies like SOAP (Simple Object Access Protocol), WSDL (Web Service Description language)-to provide a web service document, UDDI (Universal Description Discovery & Integration)-web service registry [111].

A web service can interact with any other Web service, no matter which platform the Web service is developed and run on. A Web service client can be of many types, such as another Web service, a client written in a scripting language, a C# Client, a java client etc [112].

#### Web Service composition in Java

The development of java technology standards occurs through Java Specification Requests being submitted to Java Community Process.

Java Remote Method Invocation enables the programmer to create distributed java technology, which invoke objects from other Java Virtual machines [145].
Two popular java web services architecture

1.Java API for XML based RPC (JAX-RPC) – This is a standard way for J2SE clients & are invoked with a simple java commands, using JAX-RPC Service Factory for creation of instances of services access points

2. Implementing Enterprise web services – This specifies the web services for Java 2 Enterprise Edition (J2EE) builds on SOAP & WEDL to cover the use of JAX-RPC. It uses Java Naming & Directory interface (JNDI) to obtain service interface with two steps- * Instantiate a local JNDI Context. *Do a JNDI lookup for the Web Service name in this context. [114]

**Web Service composition in Haskell**

1. Web services can be performed by integrating with XML Serializer for producing and consuming message representations & HTTP- based client-side executor for services [115].

2. The HaskellXML toolbox is a validating XML-parser. It introduces a more general approach for processing XML with Haskell. This toolbox uses generic data model for representing XML documents, including the DTD subset & the document subset, in Haskell. This data model makes it possible to use filter functions as a uniform design of XML processing applications [104].

3. Hayoo! provides a JSON-based web service API, which can be used to retrieve search result in a structured format [144].

### 2.3.4 OO-based abstraction

Abstraction is the mechanism and practice which reduces and factors out details so that one can focus on a few concepts at a time. Abstraction can apply to control or to data: Control abstraction is the abstraction of actions while data abstraction is that of data structures.

- Control abstraction involves the use of subprograms and related concepts control flows
- Data abstraction allows handling data bits in meaningful ways. For example, it is the basic motivation behind data types[116]

**Abstraction in Java**

- Java programs are organized as object that normally consists of visible and non visible data fields and methods.
- Java offers abstraction by means of the abstract class type and interface, where no implementation details are allowed. Information hiding is naturally supported.
- Abstract classes can be used to abstract data structure as well as functions, and behavior, to make cohesion of data structure with related behavior.

Disadvantage- The hierarchy of classes built on these bases one can't choose to reuse just behavior code or just data structure because they are bound to each other, and the behavior is built for its associated data structure and not for properties in it. [117]

Program which contains an abstract method named check in an abstract class compare.- Self written Code

```
import java.lang.*;                                                                      abstract class
compare    // Class Has to be declared abstract
{                                                                                        abstract void
check(int x, int y);       // abstract method                                                          }
//close abstract class
class child extends compare        //Abstarct class should have a subclass
{                                                                                          void
check(int x,int y)            //Defined abstract method
{                                                                                          if(x==y)
System.out.println("True"); //output on output console

else

System.out.println("false");

}// close check method
}//class child                                                                             class
demoabstract    //Main program in which the abstract function is call
{                                                                                        public static
void main(String[] args)                                                          {
int a,b;
a=Integer.parseInt(args[0]);                                         b=Integer.parseInt(args[1]);
child ob=new child();     // creating an object for child class
ob.check(a,b);       //calling abstract method
}                                                                                          }

The program has to be executed  at command prompt as  Java abs

Command line inputs 5 6

The output of this program- false
```

**Abstraction in Haskell**

- The MonadLab DsL embedded in Template Haskell-Monads have become a central programming abstraction in Haskell with benefits including modularity and maintainability, effective programming and mathematical precision. Monads can be used to support imperative features within a purely functional language. What they really provide is modularity. That is, by defining an operation monadically, we can hide underlying machinery in a way that allows new features to be incorporated into the monad in a transparent manner [118]. Abstraction can be performed with the help of Type Classes. Which provide a very clear separation between data abstraction and function abstraction

1. Type Classes allow a default implementation which enables implementation generalization.
2. Type Classes instance declaration can be done outside the type declaration site, or even outside the whole module that contains the type.
3. **Type class substitute for class and type for object** [152].
4. One powerful abstraction mechanism available is higher order function. In Haskell a function is a first-class citizen, it can freely be passed to other functions, retuned as the result of a function, stored in a data structure [146].

The Program **illustrates the Typeclass** in Haskell.

```
--The Decleration of function isEqual in a class

class BasicEq a where

   isEqual :: a -> a -> Bool



--Defination of function in an instance of above class

instance BasicEq Bool where

   isEqual True  True  = True

   isEqual False False = True




INPUT of program

ghci:>. is Equal False False



Output of program

 True
```

**Comparison: Java and Haskell**

In Java**-** An abstract function is declared in abstract class which has to be defined in a subclass which extends abstract class which is mandatory. It is 24 line of code. Then the program has to be compiled using command prompt where class path & path has to be set as per the complete syntax. The command line arguments have to be converted to **primitive data** type as the prototype of the function which takes integer nos.

In Haskell **–** which may look like the objects of Object Oriented programming [104].As Haskell does not have subtypes [109]. We will refer to instance types with the letter a. An instance type of this typeclass is any type that implements the functions defined in the typeclass. This typeclass defines one function. That function takes two parameters- both corresponding to instance types- and returns a Boolean.

## 2.3.5 Reflection

In computer science, reflection is the process by which a computer program can observe and modify its own structure and behavior.
Normally, instructions are executed and data is processed; however, in some languages, programs can also treat instructions as data and therefore make reflective modifications.
Reflection is most commonly used in high-level virtual machine programming languages like and scripting languages, and less commonly used in manifestly typed and/or statically typed programming languages.
Reflection can be used for observing and/or modifying program execution at runtime. A reflection-oriented program component can monitor the execution of an enclosure of code and can modify itself according to a desired goal related to that enclosure.
Reflection can also be used to adapt a given program to different situations dynamically [116].

**Reflection in Java**

Reflection is commonly used by programs which require the ability to examine or modify the runtime behaviour of applications running in the Java virtual machine.

- The two main aspects of self-manipulation are *introspection* and *intercession*, which are the abilities of a program to observe and modify (respectively) its own state and behavior.
- Java reflection API- Distinguishes between Static & Dynamic objects. Meta-objects cannot be modified. They can only be created, read & used. A mechanism provides Partial Evaluation. Ability to examine itself and its environment [122].
- Java.lang.reflection package provide access to the representation of Java Classes [121].
- One tangible use of reflection is in JavaBeans, where software components can be manipulated visually via a builder tool. The tool uses reflection to obtain the properties of Java components (classes) as they are dynamically loaded [122].

**Reflection in Haskell**

Type-safe reflection in Haskell-

- The Type-safe reflection is the ability for a program to generate new code and incorporate this into its own execution. The main use of this would be to allow further abstraction over types than provided by parametric polymorphism & type classes.
- The structure of types can be observed, including names of constructors, fields and types. **This is similar to the reflection API of JAVA** where attributes and method signatures can be observed, and objects can be constructed from class name.
- This approach handles regular data types, nested data types, mutually-recursive data types constructor parameterised in additional types & it handles single and multi-parameter term traversal. It supports higher order Generic programming, reusable definitions of traversal strategies & overriding of generic functions at specified types. Generic functions are directly defined on Haskell data types[123]
- One way to provide this would be to extend the language with new constructs but this would require changes to existing compilers and interpreters and would cause compatibility problems for existing code & the other way to use separate tool.
- To provide linguistic reflection Haskell uses separate tool **Derive** is pre-processor written entirely in Haskell. Derive can only perform **compile time Reflection**. [123]

### 2.3.6 Aspect-orientation

In computing, aspect-oriented programming (AOP) is a programming paradigm which isolates secondary or supporting functions from the main program's business logic. It aims to increase modularity by allowing the separation of cross-cutting concerns, forming a basis for aspect-oriented software development.

AOP includes programming techniques and tools that support the modularization of concerns at the level of the source code, while "aspect-oriented software development" refers to a whole engineering discipline [124].

**Aspect Orientation in Java**

AOP is a concept and as such it is not bound to a certain programming language or programming paradigm. It can help with the shortcomings of all languages that use single, hierarchical decomposition. This may be procedural, object oriented, or functional. The Java implementation of AOP is called AspectJ (TM) and has been created at Xerox PARC.

Like any other aspect-oriented compiler, the AspectJ compiler includes a weaving phase that unambiguously resolves the cross-cutting between classes and aspects. AspectJ implements this additional phase by first weaving aspects with regular code and then calling the standard Java compiler. [125]

**Aspect Oriented Programming in Haskell**

- GHC can model an AOP style of programming via a simple syntax-directed transformation scheme where AOP programming idioms are mapped to type classes. But we cannot easily advise functions in programs which carry type annotations [126].
- Embedding of an aspect-oriented programming style in Haskell provides a structuring syntax directed compilers implemented as attribute grammars which is a convenient notation for specifying the functions that deal with each of the production rules in the abstract syntax.
- Attribute grammar systems offer decomposition by aspect, but only at a syntactic level, not at the semantic level [128].
- Trex extension of Haskell,provides a rich set of records TRex stands for "typed records with extensibility". These extensible records are a key component of modularity approach to define attribute grammar.
- The standard strategy for writing an attribute grammar consists of three steps, namely the definition of semantic **domains, semantic functions, and translators**.

### 2.3.7 Functional programming

The fundamental approach in functional languages is the definition and application of functions. They allow functions to be treated as values and support higher order functions i.e. functions which can take functions as parameters and which can combine functions to create new functions [129].

**Functional Programming in JAVA**

- Java is a "true" object-oriented language, in the sense that it cannot be executed without having a class.
- Superficially, Java has no functions; however, using interfaces and inner classes it is fairly easy to mimic some of the important features of functional programming.
- By using interfaces and inner classes it is possible to do much more than just pass functions as parameters. It is fairly easy to mimic many more of the features of a functional programming language. [130]

Program example which reads & transfers the data from one text file to another text file which uses input output buffer stream

```
import java.lang.*;

import java.io.*;

public class demofilejava

{
```

```
public static void main(String[] args) throws IOException
{
String str;
File fi;
fi= new File("c:/program/input.txt");
if(fi.length()<0)
  System.out.println("data does not exist");
else
{
   FileInputStream fin=new FileInputStream(fi);
   byte b;
   do
    {
     b=(byte)fin.read();
char c[]=Character.toUpperCase((char)b))
str=new String
System.out.print((char)b);
    }
    while(b!=-1);
    fin.close();
 }//else


File fo=new file("c:/program/output.txt");
FileOutputStream fop=new FileOutputStream(fo);


      if(fo.exists())
```

```
    {
    fop.write(str.getBytes());
    fop.flush();
    fop.close();
    System.out.println("The data has been written");
    }//if

      else
        System.out.println("This file is not exist");

}
}

Output of this file:
The data has been written- It will copy the data from input.txt to output.txt
Check the file input.txt & output.txt
```

The program can be compiled in same manner as mentioned in section 1.5A

**Functional Programming in Haskell**

- It is a pure functional language, which is side-effect free.
- We can replace any expression in a program with its resulting value without changing the program's meaning (referential transparency).
- This makes it possible to reason about programs and their correctness, similar to the way we would reason about mathematical formulas.
- Every variable is defined exactly once and can't be modified later[131]

The program which read & writes data from & to the file without side effects. This is an Example which illustrates the combination of monad (System IO), lazy evaluation, & function purity[104], As I have mentioned in section 1.4B the abstraction like Java can be implemented in Haskell in this way [118].

```
-- file: ch07/toupper-lazy1.hs

import System.IO

import Data.Char(toUpper)


main = do

    inh <- openFile "input.txt" ReadMode

    outh <- openFile "output.txt" WriteMode

    inpStr <- hGetContents inh

--using hGetContents methods you are not required to ever consume all the data from the input ----file.

    hPutStr outh (map toUpper inpStr)

    hClose inh

    hClose outh

--hClose close the instance of the file
```

The above program **map** is a function defined by Haskell's prelude in order to apply it to all elements of the list.

**Comparison**

Java- The program code consists of 23 lines. We call a function within a function but contains impure function as the functions have to be enclosed within the class so contains some side effects. Java has sequential evaluation.
Haskell- The code is small. Function does not contain side-effects which returns the same result when it is called. Haskell supports Lazy Evaluation. You are not required to ever consume all the data from the input file when using hGetContents. Whenever the Haskell system determines that the entire string hGetContents returned can be garbage collected.

### 2.3.8 Declarative programming

Declarative programming is a way of specifying what a program should do, rather that specifying how to do it. It's not needed to prescribe the computer which steps to take and in what order, it can rearrange your program much more freely, maybe even execute some tasks in parallel.

A declarative programming is often defined as

1. A program that describes what computation should be performed & not how to compute it

2. Any programming language that lacks side effects

3. A language with a clear correspondence to mathematical logic [132].

**Declarative Programming in Java**

Java is an Object Oriented language but it can be integrated with facilities for supporting declarative programming, by providing them as a library called JSetL. A programmers cant not restrict themselves to constraints, but they can try to provide a more comprehensive collection of facilities to support a real declarative style of programming [133].

**Declarative Programming in Haskell**

- Haskell is based on models of computation that are fundamentally different from the state-machine model underlying imperative programming languages.
- Haskell is a functional programming language based on the Lambda Calculus and it makes extensive use of pattern matching, which encourages a new, and extremely useful way of programming.
- The Haskell compiler automatically infers the type of each expression, thereby enabling it to catch type errors without the programmer having to explicitly specify types for each datum. [135]

### 2.3.9 Batch scripting

Batch script allow several commands that would be entered manually at a command line interface to be executed automatically and without having to wait for a user to trigger each stage of the sequence[137].

**Batch Script in Java**

- Java has a java.lang.**ProcessBuilder** class used to create operating system process. Each PeocessBuilder instance manages a collection of process attributes [136].
- Java can either load a DLL (Direct linking library) that contain any external program implementation of the required native methods or some other DLL that can dynamically register the native methods using the JRegister Natives JNI entry [138].

**Batch Script in Haskell**

GHC package contains GHC API which can be used in module to access an external commands [139]. It is possible to invoke external commands from Haskell. The **raw System** from the System.cmd**[104]**. This will invoke a specified program with the specified arguments, and return the exit code from that program.

The important process is termination of an external program or exit an external command after invoking. Haskell automatically indicates a non-successful exit whenever a program is aborted by an exception [104]. Haskell Foreign Function Interface is the means by which Haskell code can use & be used by, code written in other language. In order to call a foreign function from Haskell, we import externally defined functions into Haskell, either by static linking or by dynamic linking.

### 2.3.10 UI prototype design

Prototyping means exploring ideas before you invest them. Most often they are created early in the project, during the planning and specification phase, before developers write any production code. That's when the need for exploration is greatest, and when the time investment needed is most viable Software & web designers create mock-ups of how users will interact with their design. The real product prototypes are easy and inexpensive to create. The conventional purpose of prototype is to allow users of the software to evaluate developers proposals for the design of the eventual product by actually trying them out, rather than having to interpret & evaluate the design based on descriptions. The prototyping can also be used by end users to describe & prove requirements that developers have not considered. Commonly used UI prototypes are Windows based GUI [140].

**GUI development in Java**

User interface libraries which are used in Java are:

- The (heavyweight, or native) Abstract Window Toolkit (AWT), which provides GUI components, the means for laying out those components and the means for handling events from those components.
- The (lightweight) Swing libraries, which are built on AWT but provide (non-native) implementations of the AWT widgetry
- APIs for audio capture, processing, and playback [141].

**GUI development in Haskell**

There are several toolkits are available for Haskell but 2 are most popular & commonly used-

- Gtk2hs is a GUI library for Haskell based on Gtk+. Gtk is an extensive multiplatform toolkit for creating GUI interfaces. The fundamental thing of Gtk+ is widget. A widget represents any part of the GUI & may contain other widgets. Some examples of widgets include a window, dialog box, button, and text within the button [104].
- wxHaskell is a portable and native GUI library for Haskell.
- wxHaskell is built on top of wxWidgets – a comprehensive C++ library that is portable across all major GUI platforms; including GTK, Windows, X11, and MacOS X.
- It is a mature library (in development since 1992) that supports a wide range of widgets with the native look-and-feel [142]
- wxHaskell consists of two libraries - *WXCore* and *WX*. The *WXCore* library provides the core interface to wxWidgets functionality. Using this library is just like programming wxWidgets in C++ and provides the raw functionality of wxWidgets [143].

## 2.4 PHP vs. Scala

### 2.4.1 Default more secure programming practices

Evert program is insecure. Security has been an important factor in the way programming languages are designed and maintained. Every programmer should know how to avoid critical security mistakes by performing reviews, testing for security bugs, and so on. "Security is a measurement, not a characteristic. Security must be balanced with expense and usability and it must be part of the design" [159].

## PHP

One of the most striking things about the PHP programming language is that beginner programmers can achieve simple goals rather quickly. The problem, on the other hand, is that many programmers are not aware about what is happening on the backend. It's known that security is sacrificed for the sake of convenience. PHP has flexible file handling functions (inclue(), require() and fopen()) [160]. These functions accept local paths and remote files. A lot of vulnerabilities are due to incorrect handling and path names. Another problem is that PHP writes most of the variables into the global scope. This has indeed some advantages, however you can get lost in big scripts. EGPCS (Environment, GET, POST, Cookie and Server) variables are put into the global scope.
One concept you must always remember is that user input is unreliable and you shouldn't trust it quickly; examples for input validation are [161]:

- Partially lost in transmission between server and client
- Corrupted by some in-between process
- Modified by the user in an unexpected manner
- Intentional attempt to gain unauthorized access
- Crash the application

That is why it is extremely important to validate user input before processing. Programming plain PHP is rather boring without an SQL connection. One problem is that SQL queries with unchecked variables are dangerous.

Of course, JavaScript validation and other client-side validations are entirely useless, since they can be easily bypassed. Register Globals (a directive in php.ini to automatically make variables out of environment, GET, POST, cookies, and server data (true/false)) has been criticized a lot, however this is "not security vulnerability; it is a risk and a bad practice [162]". As a result, it has been disabled by default since version 4.1.0. $_GET and $_POST have been used instead. The PHP manual contains a great section especially for security precautions when coding PHP scripts. The manual notes when possible security risks exist and how to prevent them or minimize their side effects. Not validating input to SQL queries ultimately creates vulnerabilities to SQL injection and not validation user input and cookies creates XSS (cross site scripting) vulnerabilities; these are some of the biggest problems in PHP. Also, scattering SQL queries with other PHP codes can create a mess.

### Scala

Scala can be thought of as less secure than Java in a few superficial ways mostly related to visibility of members from outside classes, but in terms of helping you write code that's free of security holes, it should be better if you follow the best practices of immutability, concurrency, etc. While JSP is kind of a domain specific language with a (bad) mixture of HTML, XML and Java, Scala is a general-purpose programming language. JSP is a template language compiles to Java servlets, and allows arbitrary Java code in the snippets. Scala is a general-purpose programming language with which you can do 99.99% of what Java can do, usually in 1/3 to 1/5th the code size. Scala is basically the same as Java in the respect of the common programming practices. The Scala programming language is used by many companies [163] to develop commercial software and production system, such as EDFT, Twitter, Xebia, Xerox, FourSquare, Sony, Siemens, GridGain, AppJet, Reaktor and many others.

### 2.4.2 Web Application Development

**PHP**

PHP is great for web applications. PHP scripts are executed on the server. PHP supports many databases (MySQL, Informix, Oracle, Sybase, Solid, PostgreSQL, Generic ODBC, etc.) [164]. One of the main reasons why PHP is so popular is because it is open source software and is free to download and use. PHP is released under the PHP License [165]. PHP combined with MySQL are cross-platform (you can develop in Windows and serve on a UNIX platform). PHP runs on different platforms (Windows, Linux, UNIX, etc.). PHP is compatible with almost all web servers used today (Apache, IIS, etc.). PHP is easy to learn and runs efficiently on the server side; that explains why many of the new programmers tend to program in PHP and stick to it and further their knowledge in this language. However, with the rise of Web 2.0, programming for Web 2.0 uses mostly Ajax which is more about JavaScript, XML and CSS. And there is also the rise of HTML 5; all those make PHP in a difficult position, as only a server side language, where Ajax (and later HTML 5) will be the key for web applications.

**Scala**

Lift is just one, albeit the most popular, Scala web framework. Play is another one that people seem to like.

Lift is a free web application framework. Lift aims to deliver similar benefits to Ruby on Rails except that Lift applications are written in Scala instead of Ruby. The use of Scala means any existing Java library and Web container can be used in running Lift applications. Lift app-dev is pretty much the same as Java development. Lift programmers use standard Java environments like Eclipse and IDEA.

Lift is written in the Scala programming language [166], which is a modern language for the Java virtual machine. Java 5 or higher version is needed for developing and running Lift projects as well as suitable versions of the Scala libraries and compiler are needed. Lift depends on the Servlet API 2.5, hence you need a suitable Servlet container to run a Lift-based web application, e.g. Jetty 6 or 7 or Tomcat 6. Scala and Lift code can be as brief and expressive as Ruby code. Lift offers developers amazing productivity gains versus traditional Java web frameworks, just as Rails does. On the other hand, Lift code scales much better than Rails code. Lift code is type-safe and the compiler becomes your friend.

The Lift Web Framework is an awesome Ajax and Comet support [167]; it's more powerful and concise than Rails. It's scalable and secure and it's in production from Web 2.0 to SAP.

### 2.4.3 Web services design and composition

Web services are [168] typically application programming interfaces (API) or web APIs that are accessed via Hypertext Transfer Protocol (HTTP) and executed on a remote system hosting the requested services. Web services can be classified in two categories: Big Web Services and RESTful Web Services. Web services allow you to exchange information over HTTP using XML. For example if you want to know the weather for another city, you can write a short script to gather that data in a format you can easily manipulate. From a developer's perspective, it's as if you're calling a local function that returns a value [169].

**PHP**

A web service consists of a server to serve requests to the web service and a client to invoke methods on the web service. The PHP class library provides the SOAP extension to develop SOAP servers and clients and the XML-RPC extension to create XML-RPC servers and clients.

An important advantage of web services is ubiquity across platforms and languages. A PHP script running on Linux can talk to an IIS server on Windows using ASP without any communication problems. When the server switches over to Solaris, Apache, and JSP, everything transitions without a glitch.

SOAP is the most popular web services format [170]. It's a W3C standard for passing messages across the network and calling functions on remote computers. PHP does not come with a bundled SOAP extension. Before you can begin, you need to download and install files that let you easily integrate SOAP into your applications.

It's very simple to use SOAP and WSDL with PHP. These clients allow you to gather information from across the net to use in your scripts. Amazon.com is not the only major company to provide a SOAP front end to its data. Google lets you search their listings up to 1000 times a day. Additionally, XMethods has a large directory of SOAP servers that you can experiment with and use.

**Appendix A.5:** Create a basic web service that provides an XML or JSON response using some PHP and MySQL.

**Scala**

Building Web Services in Lift is somehow easy because of the pattern matching, higher-order functions, Scala's XML support and Lift's built-in support for REST and other web services.

**Appendix A.6:** Building Web Services in Lift.

## 2.4.4 OO-based abstraction

**PHP**

New features appear with each new version released while existing features are improved. PHP's object support is one feature that is being improved on each version. Object oriented support first appeared in PHP 3. PHP 4 made additional improvements, such as the way constructors are handled. With PHP's object support growing, many of the reasons developers might not take an object oriented approach are diminishing. Object-Oriented started to be really interesting with PHP 5.

**Appendix A.7:** An example of creating a very simple abstract class called OOPHPAbstractClass, and OOPHPClassToExtendAnAbstract which extends it.

**Scala**

In Java you have abstract methods, but you can't pass a method as a parameter. You don't have abstract fields, but you can pass a value as a parameter. And similarly you don't have abstract type members, but you can specify a type as a parameter.

Scala team decided to have the same construction principles for all three sorts of members.
So you can have abstract fields as well as value parameters. You can pass methods as parameters, or you can abstract over them. You can specify types as parameters, or you can abstract over them. You can model one in terms of the other. You can express every sort of parameterization as a form of object-oriented abstraction. In a sense, Scala is a more orthogonal and complete language.

### 2.4.5 Reflection

PHP 5 [171] comes with a complete reflection API that adds the ability to reverse-engineer classes, interfaces, functions, methods and extensions. Additionally, the reflection API offers ways to retrieve doc comments for functions, classes and methods.

Some parts of the internal API are missing the necessary code to work with the Reflection extension. For example, an internal PHP class might be missing reflection data for properties. These are considered bugs and should be discovered and fixed.

No external libraries are needed to build this extension and there is no installation needed to use these functions; they are part of the PHP core. This extension has no configuration directives defined in php.ini.

This API takes the language's introspective abilities to a far more mature stage. What's more, it includes some convenient methods that permit developers to dissect both classes and interfaces down to their bare bones, which can be very useful.

**Appendix A.8** Reflection Example from Shell (a Terminal).

**Scala**

Scala's support for reflection is the same in Java, but Scala has richer types which are not fully reflected in bytecode. There's a Scala reflection library in the works.

Scala reflection is built on top of Java reflection. To reassemble a Scala-like view of the program from the damaged view returned by Java reflection, it implements an abstract API, which it shares with the Scala compiler, which can recreate the original Scala view of a program from class files. This API is the backend of the Scala reflection library. The frontend ties the backend and Java reflection into a user-friendly system to do Scala reflection.

The Scala reflection library is in development. For the time being, no usable version is available.

Object-oriented languages usually implement an API supporting meta-level operations such as reflection [172]. However, reflection APIs generally do not follow the three design principle for reflection and meta-programming (encapsulation, stratification and ontological correspondence). Scala does not have any specific API supporting meta-programming. However, since it is compiled into Java byte-code, it is compatible with the complete Java library. Consequently, one can use the Java reflection API in order to access meta-level information about a Scala program. Anyhow, this approach presents important limitations and raises some usability problems.

call object methods with reflection

```
val c = Class.forName("scala.Console")

val m = c.getMethod("println",classOf[Any])

m.invoke(null,"hi")
```

get the reference of the singleton object

```
val c = Class.forName("scala.Console$")

val console = c.getField("MODULE$").get(null).asInstanceOf[Console.type]

console.println("Hi")
```

### 2.4.6 Aspect-orientation

The Aspect Oriented Programming (AOP) concept was created originally by Java developers. They developed a compiler that implements the AOP white box approach. AOP is a methodology meant to implement new aspects in software component using external components, but without altering the code that implements the core functionality.

**PHP**

There are some approaches to facilitate Aspect-Oriented Programming with PHP:

- **PHPAspect** uses a compiler, written in PHP that performs static weaving using source code transformations. A downside of this approach is that advantages that stem from PHP's interpreted nature are lost.

- **Aspect-Oriented PHP** uses a preprocessor for PHP written in Java that is responsible for the weaving of aspect- and base-code. Due to its Java implementation, this approach does not integrate seamlessly with the PHP platform.

- **aspectPHP** is a reimplementation of Aspect-Oriented PHP in C, available as a patch against (not as an extension to) PHP 4.3.10.

- **The AOP Library for PHP** requires manual changes to the base-code. AOP Library for PHP has been implemented to implement Aspect Oriented Programming by executing the code of classes that enable orthogonal aspects at run-time. AOP Library for PHP's package implements a framework that provides a PHP solution that does not rely on a pre-compilation stage. Therefore it can be used right away without the eventual complication of the compiler based AOP implementations.

- **BunnyAspects** is another implementation of AOP inside of pure PHP5. You do not need any extensions and it uses the existing qualities and abilities of the language, which carries its own set of problems. The essence of BunnyAspects is to wrap a BunnyAspect object around our target object. This BunnyAspect object keeps track of what is being woven into it, and then with the magic __call method, intercedes between all of the method calls, and calls before or after advice as needed.

**Scala**

There is usually less of a need for aspects in Scala since the language itself is more powerful and expressive i.e. like traits for composing behavior.

There is no AspectJ-like implementation of AOP for Scala. Scala supports attributes and so we could say that Scala can support AOP via attribute-based meta-programming. Scala also supports mixins, which enable separation of code that crosscuts class hierarchies, but this is only related to the introduce member AspectJ advice. Scala knows closures and compiles those closures to inner classes on the bytecode level and creates multiple bytecode classes for one class. One would have to mentally transform Scala into bytecode then into Java in order to know how to write pointcuts and advices Scala code.

### 2.4.7 Functional programming

The essential approach in functional languages is the definition and application of functions. They allow functions to be treated as values and support higher order functions i.e. functions which can take functions as parameters and which can combine functions to create new functions.

**PHP**

With the rise of JavaScript, and languages like Python and Ruby, functional programming is becoming more mainstream. Fn.php [173] is an attempt to define lots of useful higher-order functions to PHP, and fix some of the things that are inconsistent with the others. Fn.php already supports the things in PHP that already exist, but adds foldr, compose, zip, andf, orf, not, any, every, curry, I, K, S, flip and a new short hand way to define functions with strings. There’s virtually no documentation, and very little in the way of examples or tests.

**Scala**

Scala programs execute on the Java Virtual Machine (JVM) and can interoperate with Java programs and application programmer interfaces (APIs). It is a multiparadigm programming language that natively supports the imperative, object-oriented, and functional styles of programming. Scala functions are [174]:

- First-class values i.e. functions are objects
- Can be higher-order: take functions as arguments or return them as result
- Can be anonymous
- May be curried: take arguments one at a time, allowing partial application
- Are often passed in a closure, with references to free variables they manipulate
- Provide ability to build powerful libraries of higher-order functions

As a hybrid object-functional language, Scala does not require functions to be pure, nor does it require variables to be immutable. It does, however, encourage you to write your code this way whenever possible.

Robert Fischer mentions [175] that “Scala is not a functional programming language. It is a statically typed object oriented language with closures.” Rich Hickey’s mentions [176] that Scala “isn’t really a functional language”

The core of the argument made people against Scala as a functional language goes like this [177]:

- Mutable variables as first-class citizens of the language
- Uncontrolled side-effects (ties in with the first point)
- Mutable collections and other imperative libraries exist on equal footing
- Object-oriented structures (class inheritance, overloading, etc)
- Verbosity

### 2.4.8 Declarative programming

**PHP**

Most modern programming languages, PHP included, are imperative: the programmer describes how the computer is to perform a particular task. In contrast declarative programming languages’ philosophy goes as this: the computer is told what the programmer wants and the how is left to the computer itself. While declarative programming languages are often considered somewhat exotic, there is a declarative language that is seen nearly every day by application developers: SQL. With SQL, the programmer formulates a query and it is left to the database engine’s query analyzer to figure out what combination of disk reads, index lookups and other functions are necessary to satisfy the query.

PHP is not a declarative programming language. However, this does not prevent programmers from adopting a declarative style of programming [178].

Consider the code fragment:

```
require_login();
require_permission('LIST_TRNANSACTIONS');
```

This piece of code states that only logged-in users and only those users with a particular permission can access this page. This is far more readable than the equivalent imperative style which would involve an explicit if statement to test if the user has logged in and if they have the correct position. Or consider the code used to read the query parameters:

```
// Get parameters
$display    = get_parameter('display',  PARAM_TYPE_ENUM, array_keys($display_modes), 'weekly');
$from_date  = get_parameter('from',     PARAM_TYPE_DATE, timestamp_to_mysql_date(0));
$to_date    = get_parameter('to',       PARAM_TYPE_DATE, timestamp_to_mysql_date(time()));
$sort_field = get_parameter('sort',     PARAM_TYPE_ENUM, $fields, $fields[0]);
$order      = get_parameter('order',    PARAM_TYPE_ENUM, $orders, $orders[0]);
$currency   = get_parameter('currency', PARAM_TYPE_ENUM, array_keys($currencies),
$_SESSION['default_currency']);
```

By writing the code in this way, we can see at a glance what type of data is expected in each parameter and the default value to use if any parameter is omitted. This style helps to make the code self-documenting. The declarative style is common in many modern web development frameworks, notably Ruby on Rails.

**Scala**

Scala list comprehensions provide a declarative syntax for abstracting over data that hides less declarative function call and closure creation details. Using operator overloading in Scala, it is possible to also write decent Prolog code [179].

Declarative programming style in Scala makes software product development from proof-of-concept to deployment enjoyable. It is important to note that declarative reading is math definition.

The definition of factorial n, an integer, is, n times factorial n-1

```
def fact(n: Int): Int = n*fact(n-1)
```

The definition of isPrime n, an integer, is, given a range of integers thru n for all X, n modulo x is not zero

```
def isPrime(n: Int) = List.range(2, n) forall (x => n % x !=0)
```

### 2.4.9 Batch scripting

**PHP**

cron is a module that allows you to run commands at predetermined times or intervals. cron is normally available on all Unix and Linux distributions. It is a daemon which allows you to schedule a program or script for a specific time of execution. Using cron, you can automate many tasks; for example, to update the content of a website, generate quota reports, remove expired articles, send out e-mails on a given date and a lot more. An important aspect in cron is that it sends any error output to a specified e-mail address so you can debug problems when they occur.

cron is driven by a crontab, a configuration file that specifies shell commands to run periodically on a given schedule. The crontab files are stored where the lists of jobs and other instructions to the cron daemon are kept. Users can have their own individual crontab files and often there is a system wide crontab file (usually in /etc or a subdirectory of /etc) which only system administrators can edit. Each line of a crontab file represents a job and is composed of a cron expression, followed by a shell command to execute.

A cron expression is a string comprising 6 or 7 fields separated by white space that represents a set of times, normally as a schedule to execute some routine.

The following is a cron configuration to run a PHP script once a day at 11 p.m. [180]:

```
0 23 * * * username /usr/bin/php /users/home/username/myscript.php
```

The first five fields define the times when the script should be executed. Then comes the name of the user who should be used to run the script. The rest of the line is the command line to execute (we need to know where PHP and our scripts are). The time fields are minute, hour, day of the month, month, and the day of the week.

The command:

```
15 * * * * username /usr/bin/php /users/home/username/myscript.php
```

runs the script at the 15-minute mark of every hour.

The command:

```
30 23 * * 6 username /usr/bin/php /users/home/username/myscript.php
```

runs the script at 11:30 p.m. on Saturdays (the day of the week specified as 6).

If your host has cPanel, you may have a cron job interface which will take the commands as an option and allow you to configure visually the times at which it will execute.

**Scala**

A Scala program may also be run as a shell script respectively as a batch command [181]. The bash shell script script.sh containing the following Scala "Hello World" code (and shell preamble)

```
#!/bin/sh
exec scala "$0" "$@"
!#
object HelloWorld {
def main(args: Array[String]) {
println("Hello, world! " + args.toList)
}
}
HelloWorld.main(args)
```

can be run directly from the command shell:

```
> ./script.sh
```

It is important to note that the file script.sh has to have execute access and the search path for the Scala command is specified in the PATH environment variable.

You can use Scala as a replacement for console scripting languages (Batch under windows, Bash/Perl under Linux/Cygwin). The Scala package contains – like Ruby and Python do – its own console to test code snippets and more. So unlike in Java, you can also do real scripting in Scala. The advantages [182]:

- You can re-use Scala-scripts in larger programs later, even in Java projects
- You have access to the complete JDK and you can put other Java/Scala-Libraries into your Classpath to use in a script
- You can do all file operations on a high and object oriented level, much easier than writing Bashscripts or Batch-scripts.

- You can do functional-like programming; you get small and readable scripts
- If the script gets more advanced, you can use the Swing-Wrapper to create a GUI for you script

One downside when using Scala as a scripting language is that every time you start a Scala-script, it lasts few seconds to start. The majority of the taken time is needed for loading the compiler and compiling and start the JVM. However when using the *–savecompiled* flag, then Scala will save a compiled version of your script as a .jar file and load that version instead. However, it is still not as fast as Perl, Python and co.

### 2.4.10 UI prototype design

**PHP**

PHP-GTK is a set of language bindings for PHP which allow GTK+ GUI applications to be written in PHP. PHP-GTK provides an object-oriented interface to GTK+ classes and functions.

PHP-GTK, originally a proof-of-concept inspired by PyGTK to show that it could be done [183]. PHP isn't just for the Web anymore so if you need a GUI application, you can use your favorite language.

PHP-GTK 2 fully utilizes PHP 5's powerful object model support [184], and brings the improved portability of GTK 2.6 as well as its new set of widgets. The project also has some new extensions such as GtkSourceView, which provides a rich source editor widget, alongside some of the old favorites. Documentation for PHP-GTK 2 is filling out rapidly. Several articles and tutorials have been written on the topic, and around half the classes have been fully documented. Scott Mattocks, an active member of the PHP-GTK documentation group, has also written a book (Pro PHP GTK, Apress) on the subject of PHP-GTK programming. GTK+ is platform independent.

**Scala**

GUI applications developed are based on a Scala library that provides access to Java's Swing framework of GUI classes [158]. Scala-swing is a "wrapping" library approach. The Scala library resembles the underlying Swing classes, but hides much of their complexity.

The Scala wrapper is based on the original Java Swing framework. "The Scala wrappers resemble the underlying Swing classes, but try to simplify concepts where possible and make them more uniform. The simplification makes extensive use of the properties of the Scala language. Its "everything is an object" philosophy makes it possible to inherit the main method of a GUI application. Scala's first-class functions and pattern matching make it possible to formulate event handling as the reactions component property".

Below is a simple Swing application in Scala.

```
import scala.swing._
object FirstSwingApp extends SimpleGUIApplication {
def top = new MainFrame {
title = "First Swing App"
contents = new Button {
text = "Click me"}}}
```

## 2.5 Scheme vs. BPEL

### 2.5.1 Default more secure programming practices

The author in [57] describes language safety as a general concept that includes type safety and memory-safety and language implementations enforce the program's intended semantics. This can be achieved by static (e.g. compile-time type checking) and dynamic check (e.g. run-time type checking, array-bounds checking) to "trap nonsensical operations" [58]. Scheme is an example of a safe language [6, 57, 58]. Unchecked or unsafe operations (required in certain scenarios) are explicitly defined by language abstractions [60]. Dynamic type checking is supported by "run-time type tags" to differentiate structures allocated in the heap [58]. Types provide context for operations (the underlying system selects the correct operation) and they limit the semantically valid operations (catching inadvertent logic errors) [6]. It has been observed that nominative type systems (the names in distinct declarations determine the variable type) have a better safety than structural type system (allowing for coincidental structural matches among types) [6]. An example of a low-level memory safety feature is automatic garbage collection, which prevents the occurrence of dangling pointers, memory leaks and double deletes [6]. Racket, a Scheme dialect, implements two kinds of garbage collection: 3m (generational garbage collector, the standard used, less expensive allocation of short-lived objects) and CGC (conservative garbage collector, allows for interaction with C programs, is more expensive and slower) [60]. An example of high-level memory safety capability is stack-inspection access control [57].

Another language safety concern is portability, where the language is completely defined by its manual [58]. The abstractions the language provides are safe: knowledge about details of how the language is implemented, how it manages memory and the order of object construction/destruction is not necessary in order to write a safe program [58]. Scheme's language manual, as an example, specifies the exact behavior of all programs written in the language [58]. Although, Scheme dialects maintain their own language manuals; they are well documented and they comply with the standard Scheme core. They run on Unix, Max OS X, and Windows platform. However, they all implement their own non-standardized libraries in addition to the standard R6RS Libraries.

Language security is important because as observed in [57] many security violations are due to vulnerabilities in the language design and implementation. Safe language features such as access-control lists modifications and access checks, capability-based security (distributing "unforgeable" references to sensitive objects), specification of temporal properties (to allow for communication protocols and resource usage specification and enforcement), control over execution environment (controlled access to local resources: CPU, files etc. to prevent denial-of-service attacks by sandboxing "untrusted" processes) among others [57]. Sandboxing is provided by Racket by sandbox-security-guard (from racket/sandbox) [63]. Some weaknesses are mentioned in Scheme Security Project [64] related to the eval language construct. Security guards in Racket provide explicit filesystem-access and network-access checks [61]. PLT MzScheme also offers several fundamental security mechanisms such as custodians (to manage resource allocation), security guards, inspectors (controlled access to opaque structures), namespaces (control access to Scheme bindings), and thread groups (control the allocation of the CPU) all of which and more can be found in Racket as well [62]. Safe language technologies set the ground for building secure systems. For example, a Web server built in PLT Scheme [57] (currently known as Racket) does not allow buffer overflow attacks. High-level security policies' enforcement is supported by low- and intermediate-level secure language technology [57].

Type systems and type checking play a role at a high-level of abstraction, as well. In [58] the author observes that the interfaces of modules can be seen as types, "providing summary of the facilities" to be exposed to the client modules participating in the composition, or as a contract between "implementers and users". In [59], the author discusses the type safety of XML documents enforced by schemas that define their structure. XML documents can be validated against its schema definition.

In [16], the OASIS Standard of Web Services Business Process Execution Language Version 2.0, requires conformant implementations of WS-BPEL to enforce conservative static analysis of BPEL processes. The analysis, as described by the standard "is to detect any undefined semantics or invalid semantics within a process definition that was not detected during schema validation against the XSD...". Static analysis includes checks performed on the WSDL interface and also on the partners invoked by the process, among many others. BPEL standard provides schemas for all its language constructs. In addition, since every BPEL process to be invoked must be described by a WSDL interface, which it makes available to clients, BPEL processes are themselves typed according to the observation given in [58]. Moreover, BPEL defines WS-Security mechanisms for high-level security protection, such as checking for modified or forged messages while "in transit or while residing at destinations", detecting invalid or expired messages, support timestamps on signed messages, protection against denial-of-service attack. After searching for current BPEL implementations, it was found that they have been implemented using Java language. In [57] Java is described as a safe language and BPEL implementations possibly inherit some of the safe-language properties; however, this writing does not investigate how WS-Security mechanisms or other safety requirements are implemented.

### 2.5.2 Web Application Development

Web applications are programs that can be accessed from a browser. Clients (human) access applications' functionality stored at the server's side via a browser. Businesses are building web applications to provide access to the software they are hosting and maintaining. This allows them to distribute their software products without the need to install the applications at the client's side. The business model of leasing software solutions to clients is known as software-as-service. The advantage of this is that the applications are maintained centrally by application service providers (ASP) and as the software evolves, clients are not directly affected. The disadvantage is the inflexibility of the software to be adjusted to satisfy the clients' needs better, i.e. the software product is accepted "as is". In addition, access to the application is done by browsing or manually initiating different functionality. The human user is "composing" the services provided by the ASP. The difficulties to integrate applications (or make them interoperable) and the incapacity to customize the software to better fit customers' needs led to the development of Web services which built on the software-as-service model to allow the development of complex applications by composing existing [15].

Some of the existing technology that allows us to build web applications are (1) for presentation layer in the browser: Java, JavaScript, DHTML, Flash, and Silverlight; (2) for application layer, server-side: ASP, ASP.NET, CGI, ColdFusion, JSP/Java, PHP, Perl, Python, Ruby on Rails, and Struts2. HTTP is the communication protocol used [18].

**Web Applications: Scheme vs. BPEL**

There are Scheme implementations that allow the programmer to build web applications. An example of this is the Racket implementation of Scheme [65]. An example of a web application implemented using Scheme [65] and BPEL can be found in Appendix A. The application accepts a number from the user and returns that many "Hello" strings. The input and output to the Scheme server come from a client's browser. The BPEL application requires schema to describe the input and output represented in a XML document. The web page was not created in this example; however it would require the use of JSP technology and a servlet, for example, to dispatch the request from the browser to the BPEL process.

Similar web client functionality as Scheme's can be achieved using other technologies such as JSP, Servlets, JavaScript are some of the possibilities. In addition, the Scheme application was modified to add the number obtained from the user and to return its double. The only changes required was a one line code that computes the response. For lack of space this functionality was not implemented in BPEL. The second functionality would have simplified the BPEL application, the while scope will not be needed.

Compared to BPEL, Scheme web server application is programmed at a lower level. It explicitly manipulates TCP connections, threads, creation of HTTP pages to build the response, handling of security issues, are some examples. A lot of research effort is directed in building secure Web servers in Scheme. A collection of papers reporting on the accomplishments can be found in [66]. BPEL operates at a higher level of abstraction. BPEL processes require a Web server on which they can be deployed and run. Although, BPEL is verbose language built on top of XML technology, existing implementations provide graphical IDE which is capable of auto-generating source code from the models. The IDE provide wizards to generate WSDL interfaces from schemas (e.g. DTD, XSD). Schemas can be created using the graphical environment, as well. Low-level security, transport protocol handling are handled by the implementations as defined by the standard. To build a Web application in Scheme requires expertise, experience with the language, and more effort compared to building a BPEL process.

### 2.5.3 Web services design and composition

Web services are applications developed primarily to enable automated interaction between programs. It uses Web services technology to permit the development of complex business processes and transactions as services, which allows the users to customize the business solutions to better fit their needs, something that is deficient in web applications. The programming model of web services is applications built as components are hiding their implementation details behind standardized web service interfaces (WSDL) that allow access to their functionality through standardized communication protocols (e.g. SOAP over HTTP) and exchange data via standardized wire protocols (XML) [15, 31].

Web services can be atomic or composite. Atomic web services provide a simple, cohesive--from the business prospective--functionality. They can be composed into complex business processes (*composite* web services or web service compositions) and expose their functionality using web service interfaces. Customized solutions can be built either by composing web services *statically* (static binding protocol) during the design of the services or *dynamically* (dynamic binding protocol or late binding) at run time depending on the business requirements [15, 32]. Web services can be stateless (do not keep state between distinct operation invocations.) This is inherent to the underlying transportation protocol (HTTP) which is also stateless. In order to process a request, the state needs to be retrieved from persistent data repository [31]. Web services can also be stateful. In long-running transactions where a *session* (*conversation*) spans multiple requests, a composite web service can preserve transient data for each conversation. The *interaction pattern* can be synchronous (request-reply) or asynchronous (one way.) The *messaging style* for *fine-grained* services can be *RPC-style*, where the message contains individual parameters of the operation. For *course-grained* services the messaging style is *document style*, where a whole document is sent (e.g. purchase order) to the receiving service [15].

Web services have roles in the interaction with other Web services. They can be web service providers or web service requestors. They can also fulfill both roles in a given conversation. Web service providers describe their functionality, non-functional properties, and how clients can invoke their services via *WSDL interfaces*. They advertise their capabilities by publishing the WSDL interface to a service registry, where clients search for services to satisfy their business needs. When a web service client finds the best web service provider in the registry, it can bind to it using the information in the web service interface [15, 32].

**Web Services: Scheme vs. BPEL**

WS-BPEL is the OASIS standard built to fulfill the role of a composite web service capable of holding state. A detailed tutorial on how to build a BPEL application is provided in [33]. It details on all the components required to build the process including creating XML Schema, WSDL, BPEL process, composite application (Service Assemblies) and how to test the composite application. Although, there are Scheme language implementations supporting the development of Web servers and applications [34], there is no literature found that discusses how Scheme language can support web service development. A possible reason is that web services are platform and language neutral [15] and any of the existing languages can implement Web service functionality by wrapping it in a WSDL interface. Clients can invoke the functionality without being aware how it was implemented. This independence is achieved using the Web service technology stack described in detail in [15].

## 2.5.4 OO-based abstraction

**Evolution**

Programming models and the languages supporting them which are based on classes and objects are said to be object oriented. The three fundamental concepts of the object–oriented paradigm are *encapsulation*, *inheritance*, and *dynamic method binding* [6]. Object-orientation is an evolutionary paradigm building upon earlier ones. The earliest languages introduced features that abstracted from the underlying platform. It made programs easier to understand. The concepts important to high-level language design are *names* (used to replace low-level concepts such as machine instructions and addresses), *scopes* (supported by *scoping rules* to determine where in the program text the name is visible and can be used), *bindings* (association made between different language features) and *binding time* (when those associations are made, e.g. compile, load, or runtime), *referencing environment* ( the set of active bindings; provides the context of the program at a given point), *aliases* (referring to an object in a given scope by more than one name), *overloading* (referring to more than one object in a given scope by one name depending on the context when the reference is made), and *polymorphism* (an object can have more than one type depending on the execution history and context) [6, 54].

Classes are data abstractions that encapsulate the representation of the data and provide operations for their manipulation. The evolution of data abstractions began with global variables. They were visible in all parts of the program and therefore modified by any executable statement. Program bugs were difficult to find. Variables exist for the whole duration of the program increasing the program's footprint. Local variables were introduced to reduce their visibility only to a segment of the program. This mechanism provides fault containment by limiting the program text where the fault might occur. Variables are allocated memory only for the time when the segment is executing and deallocated after the segment completes. With static variables, the "life" or "extent" of the objects (the period during which they are allocated computer memory) and their visibility are separated. The extent of the static variables is the whole program execution. Their visibility is controlled by the scoping rules [6].

The next stage of the evolution of data abstractions were the introduction of modules. They encapsulate a set of subroutines that share a set of static variables. Modules can be one of two kinds: module as a "manager" and module as a "type." The "manager" kind exports an abstract type (variables or a record) and a set of subroutines. The type can be instantiated by the module's users and passed as a reference to the module's subroutines. The "type" kind allows for creation of instances of the whole module, i.e. the module is the abstract type (including variables and subroutines) [6].

Object-oriented paradigm extends the notion of "module-as-type" by adding inheritance (a form of reusability mechanism distinct from composition) and dynamic method binding which allows the functionality of the actual object in memory to be executed in a context where the base type functionality is expected [6]. Encapsulation mechanism was refined by providing different levels of visibility (e.g. private, protected, and public) to allow for finer control of which names are exported and which are only local to the class [6].

**Programming Languages Variations**

Some languages allow classes to be nested. Nesting is useful in creating object closers. It becomes a mechanism capable of capturing an explicit representation of the referencing environment (a context). At a later point in the program a context can be bound to a subroutine. It allows for the separation of a subroutine and its referencing environment which provides a greater flexibility. The subroutine can execute in a context created earlier [6].

Classes can hold variables or reference to variables that point to objects of classes (reuse by composition.) Programming languages vary in provision of reference model or value model for variables. Value model dictates the implicit creation of automatic variables from the program stack. With the reference model, objects are dynamically allocated on the heap by explicitly initiating their construction. Their destruction can be either explicit via language construct or via a language mechanism such as garbage collection [6]. Languages can very by the type(s) of polymorphism they support. Subtype polymorphism is achieved using inheritance as the underlying mechanism. Implicit parametric polymorphism is realized by allowing variables to bind to any type of object (value) at runtime [6]. Explicit parametric polymorphism (generics) is a language feature that allows types themselves to be variables and instantiation of those type variables (binding of a type variable to a concrete type) is usually done at compile time [55].

**Scheme Language**

Scheme language consists of small set of base constructs such as variables, constants, conditionals, assignments, procedures and functions applications, constructs implementing variable definitions, and procedure parameter specifications and function or macro definitions. Specific Scheme language dialects such as PLT Scheme (currently Racket), DrScheme extend the base language with constructs: modules, generative structure definitions, expressive macro system, and class systems. Classes in PLT Scheme are first class values with lexical scoping rules and single namespace. Mixins and explicit inheritance are implemented in macros. Claims are made that objects, the instances created by the class system are space and time efficient when compared to those of Java object and Smalltalk. Method-invocation efficiency is compared to those of Smalltalk and the interface-based Java method invocations [20].

OO-based abstractions will be discussed in the context of Racket dialect of Scheme [56]. A class expression in Racket is a first-class value. The syntax of a class is: (class superclass-expr decl-or-expr ...). The built-in root class is object%. There is no method constructor, instead there are initialization expressions. In addition, the class hold the declaration of methods and fields. Classes are instantiated with the new form. Initializing arguments need to be given values at the instantiation time and cannot be referenced directly by methods. They are used to initialize fields. Class fields are available to methods. Class instances can be named or anonymous. Racket provides inheritance mechanism. The (super-new) expression is used to initialize the super class. The initializing arguments, field declaration and (super-new) can be placed in any order in the class. However their relative order determines the order in which they will be evaluated when the class is being instantiated. If there are dependencies between them, they need to be ordered. Methods' declaration order is not important, because they are fully defined when the class is instantiated [56].

Interfaces in Racket are used to be able to check whether an object of a class implements a specific set of methods. To declare a class that implements an interface class* form is used. If a class does not implement the methods declared in the interface it extends, a runtime error will occur during its evaluation. An example of a class declaration is shown in Figure 1 . An object as an instance of this class was created and initialized. Its field is updated at runtime and its value displayed [56].

```
#lang racket

(define fish%                    ;the name of the class
  (class object%                 ;extends the buit-in superclass
    (init size)                  ; initialization argument,
                                 ;invoked at object instantiation

    (define current-size size) ; field, initialize with size

    (super-new)                  ; superclass initialization

    (define/public (get-size)   ; public method to return the field
      current-size)

    ;public method grow accepts one argument
    ;updates the field by adding amt to it
    (define/public (grow amt)
      (set! current-size (+ amt current-size)))

    ;public method grow accepts one argument, assumed type is fish%
    ;gets the size of the other-fish and updates with it the
    ;size of this fish%
    (define/public (eat other-fish)
      (grow (send other-fish get-size)))))

  ;creates an object of class fish%, initialized with size 10
   (define charlie (new fish% [size 10]))

Welcome to DrRacket, version 5.0 [3m].
Language: racket; memory limit: 128 MB.
> (send charlie grow 6)
> (send charlie get-size)
16
```

**Figure 1: Class fish% and an instance charlie in Racket [56]**

**BPEL Language**

In [21] the authors are comparing OOP paradigm to Services Oriented Computing paradigm and the similarity and differences in concepts and abstractions. For example, we think of objects as providing functionality (services) and hiding implementation details and so do web services. We describe objects and services in terms of the interfaces they provide to their clients other objects or services. OOP method invocation is contrasted with SOC find-bind-use. The shared execution context of OOP is compared to the multiple contexts of SOC, where web services run and interoperate in heterogeneous environment which is middleware and operating system platform and language transparent to the SOC model. Another difference is the interaction between components in OOP and their distributed counterparts in SOC. OOP is using synchronous method invocation as the communication mechanism by contrast the distributed nature of SOC necessitates asynchronous message passing to achieve flexibility.

Composition is central to SOC. In OOP it is a design activity, i.e. the architecture of the system is static. Many patterns are created to assist the programmer to build better solutions for the business problem. By contrast, in SOC the composition of services is dynamic. Web services are built out of other services that are dynamically discovered and composed. Since the architecture is dynamic, its constituents (the web services), can be selected and replaced based on other properties such QoS [15, 32]. From architectural viewpoint, this is a major shift from designing a static architecture in OOP to providing an infrastructure to enable and support the *dynamic selection* and *composition* of distributed components in SOC. BPEL is one of the languages to fulfill this objective.

The authors in [21] propose thinking of inheritance as a *special form* of composition where the interface (behavior) of the composed object is preserved; as opposed to typical composition where the interface of the enclosed object is wrapped and filtered by the enclosing object. Then comparisons can be drawn with the composition of web services in SOC.

Polymorphism in OOP is the ability of an operation to take different operands (objects of different classes) and provide different implementation or behavior depending on the operands' types. To be able to build a system that uses polymorphism, the minimum requirement is for a language to have a formal type system. According to [21] typing and inheritance is not formalized and the notion of polymorphism is non-existent.

Other OOP abstractions such as encapsulation information hiding (i.e. hiding the implementation details with a stable interface to protect the system from changes) are relevant to SOC as well.

### 2.5.5 Reflection

Reflection is the mechanism through which programs perform computations on themselves by means of inspecting their internal structure and types (introspection) or altering the structure and the behavior of the program (intercession.) This mechanism allows for programs to query the program's symbol table at run time or to inspect their execution stack [6, 7]. Reflection can be useful to provide performance statistics, in debugging tools, for self-optimization, self-modification, and self-activation among others [6, 8].

Languages that support reflective programming need to have the following characteristics: incorporate meta-object facilities that keep information about the program structure, API for observing and modifying the program execution at runtime. Criteria for assessing language's support for reflective programming include: (1) source code discovery and modification at runtime; (2) conversion of a string to a language construct; (3) evaluation of a string as an expression; (4) runtime interpreter creation to give new meaning to a programming construct [11].

**Reflection: Scheme vs. BPEL Language**

Some characteristics of Scheme language is homogeneity of program and data, i.e. Scheme programs and data are manipulated using the same language mechanisms. In addition Scheme's operational semantics can be defined by Scheme interpreter written in Scheme language itself, this is Scheme is capable of self-defining. [6]. Scheme is a dynamic language that can expand itself. Reflection mechanisms are incorporated into existing implementations of Scheme. Examples are provided in Racket implementation of Scheme [51].

Figure 2 demonstrates reflection in Scheme. The code snipped is adapted from [52]. The eval function takes a quoted expression as argument and evaluates it. Because of Scheme's lexical scoping rules and the capability of eval to dynamically evaluate an expression, eval cannot see the local bindings in the context where an evaluation is attempted. To make a set of bindings visible or available for dynamic evaluation, a namespace mechanism is used. Namespaces can be encapsulated within a module. To access the current module, a reflective hook is needed: define-namespace-anchor. Then a namespace can be defined within the current module using namespace-anchor->namespace, as the example in Figure 2 shows.

```
#lang racket

 ;declare a reflection hook on the module
 (define-namespace-anchor a)

 ;reel in the module's namespace
  (define ns (namespace-anchor->namespace a))

  ;;definition is visible in ns
  ;;without "ns" we get compile error ==>[compile: unbound identifier...]
  (define (eval-formula formula)
      (eval `(let ([x 2]
                   [y 3])
               ,formula) ns))

  ;;formula can be read from text file
  ;;and evaluated ar tuntime
  ;;the scope of x and y is defined in "ns"
(eval-formula '(+ x y))
(eval-formula '(+ (* x y) y))
```

```
Welcome to DrRacket, version 5.0 [3m].
Language: racket; memory limit: 128 MB.
5
9
>
```

**Figure 2: Scheme reflection example [52]**

The current programming model of BPEL is based on the assumption that information and service composition will remain static throughout its lifetime. As a consequence the language does not support reflective features. In [22] the authors propose a reflective framework as an extension to BPEL to allow dynamic changes to the composition based on changes in the environment reflected in the data being processed. The authors' goal is through the reflective framework to allow the creation of more adaptive service composition and reliable control flow and data flow correctness [53]. Their proposed framework uses computational reflection mechanism structured in two levels: base-level (or application-level computations) and meta-level keeps information about the base-level. The structures used in the framework are $T_D$ (data table) containing participants declarations and data declarations and PFG (Parallel Flow Graph) representing the control flow with explicit synchronization for parallel programs. The operations the framework provides are *adding* and *deleting*: nodes, synchronization edges (between two parallel branches), data variables, and participants; *replacing* nodes of the same type; *reordering* nodes [53].

### 2.5.6 Aspect-orientation

Experience with creating systems in different application domains showed that using only procedural, object-oriented, functional, or logic programming techniques is insufficient to clearly capture and implement certain design decisions. The problem domain or the operationalization of non-functional requirements, forces some functionality to be scattered across different components. This results in a "tangled" and "scattered" code that is difficult to develop and maintain. Aspect-oriented programming allows this cross-cutting functionality to be identified, isolated, composed, and reused [9].

There are similarities between Aspect-oriented programming and computational reflection and meta-object protocols [9]. Specifically, it is possible to exploit the reflective and meta-object facilities of a language to develop AOP prototype systems, i.e. reflection can be used as a tool [9].

Model transformation is another approach to achieve the goal of AOP [9]. Different source models are "weaved" or in MDE terminology transformed into a target model using model transformation definitions written in a model-transformation language [10].

From this discussion can be seen that non-aspect oriented languages and tools can achieve the goal of AOP. In this report, the criteria for a language to be aspect-oriented are dedicated support for specifying aspect, pointcuts, and advice, an aspect compiler to "weave" aspects with the base language, provision of aspect-oriented programming environment. This implies that the conceptual model of declaring an aspect which specifies both the pointcuts (selection of join points) and advice (adding functionality to the base model) should be expressible in that language.

**Aspect Orientation: Scheme vs. BPEL Language**

In [23] the authors propose the semantics for lambda_AOP advice weaving by extending the Hindley-Milner type inference system to be able to "inject applicable advices into lambda expressions during typing." In [24] PLT Scheme is extended with features for pointcuts and advice for higher-order functions using the powerful macro system of the language to extend it. Also [25] provides an extension to Scheme language and resolve issues of dynamic scope generally needed for AOP in higher-order languages. AO4BPEL is an aspect-oriented extension to BPEL which allows the specification of crosscutting concerns such as logging, persistence, auditing, and security and support for dynamic adaptation of composition at runtime [26]. The authors use the AO4BPEL extension to BPEL to build a transparent security mechanism around the processes [27]. In both Scheme and BPEL aspect orientation is implemented as an extension to the base language.

### 2.5.7 Functional programming

The theoretical roots of functional programming can be traced back to Lambda calculus, in which computation is done by macro-style substitution of arguments in functions. As the author states in [6], any program (regardless of the computation paradigm used) can be seen as a "constructive proof of the proposition that, given any appropriate inputs, there exist outputs that are related to the inputs in a particular, desired way." The difference stems from the way the paradigms express the computation and the relationship among the abstractions. For example, in functional programming computation is expressed as a mathematical function over its inputs without side effects. Typical features of the functional languages are higher-order functions (capable of accepting other functions as parameters); first-class function values (allowing functions to be threaded as any other value in the program, e.g. to be assign to variables, store it in structures, to be passed and returned from a function); implicit parametric polymorphism (the types of all variable are universally quantified over all types [35], i.e. the values they hold determine their type), structured types such as Lists; constructors to build the structured types (also called aggregates); garbage collection (to automatically reclaim memory for all dynamically allocated data in the case where stack allocation is unsafe) [6]. Garbage collection is essential to functional languages since variables are generally allocated space from the heap and have unlimited extend. This is necessary in the case where the referencing environment needs to be saved as a closure to be retrieved later. It is used in the creation of co-routines or when passing functions as parameters to other functions.

The implementation needs to guarantee that memory allocated to variables that will not be accessed later in the program will be released to ensure that the program will not run out of memory [6].

**Functional Programming: Scheme vs. BPEL Language**

Scheme possesses all characteristics of a functional language. In addition to those, it has characteristics that are typical of the Lisp-family languages [6]. These include homogeneity of programs and data both of which can be represented as lists and can be compiled/interpreted at run time. Because of this capability, languages in the Lisp family can express their operational semantics in terms of an interpreter expressed in the given language. Consequently, the given language is capable of self-definition [6].
The program snippet below demonstrates the functional characteristics of Scheme. It provides a Scheme implementation of computing the factorial of a number using tail recursion [36]. The function is evaluated in Racket environment using DrRacket (Figure 3).

```
#lang racket

(define factorial
  (lambda (n)
    (if (= n 0) 1
        (* n (factorial (- n 1))))))
```

```
Welcome to DrRacket, version 5.0 [3m].
Language: racket; memory limit: 128 MB.
> (factorial 5)
120
> 
```

**Figure 3: Factorial function implemented in Scheme using tail recursion and the output of a test run [36]**

An approach to assess whether BPEL can be considered a functional language is to compare the characteristics of functional languages to these of BPEL. The main characteristics such as higher-order functions and first-class function are not available as a language feature. In addition, BPEL is executed for its side-effects and keeps state. The control structures of the language are imperative in style. Some examples are while loop, sequential flow (support for parallel as well), and assignments. However, comparison performed by other authors of BPEL to another functional language(s) was not found in the existing literature.

A while structure is used to implement the factorial function in BPEL. Initially, the input value is assigned to the output and if it is greater than one, the while scope is entered. The value of the input is decremented by one and multiplied by the output and another attempt is made to enter the while scope. The outline of the control structure is the same as the one used for building the web application example (details provided in Appendix A.) The difference comes from the assignments, computation, and predicate termination condition. Because of its size, the source code, sample input, and output are provided in Appendix A.

Algorithms in Scheme can be expressed concisely, without a heavy, supporting infrastructure required by BPEL. The Scheme tail-recursive implementation of the factorial function is purely functional; there are no side effects (i.e. updates of variables.) In this paper, the possibility of BPEL to be recursive is not investigated. The implementation of the factorial algorithm using recursion requires sequence of updates to both input and output variables. BPEL as a language is meant to be stateful. The web services which it composes can be stateless; although their implementation can have side effects by updating an underlying data repository [31].

### 2.5.8 Declarative programming

In literature, declarative programming is defined informally as description of what is to be computed and not how to be computed [1, 4, 5]. In his seminal article [2], Robert Kowalski separates the definition of an algorithm into two concerns: a logic component which he defines to be the knowledge used for solving the problem, and a control component which defines the problem-solving strategies for using the knowledge. The logic part of the algorithm determines its meaning and the control part its efficiency [2]. Declarative programming is then about defining the logic component and the underlying execution engine will implement the control component. Others view declarative programs as theories in some logic and computations as deduction from that theory [1]. Languages based on first-order logic such as pure Prolog, higher-order logic such as Gödel, λ-calculus such as Haskell and Scheme qualify according to this definition. This view is broad and includes logic (a.k.a. relational [3]) and functional programming among other applications such as formal methods, theorem proving, algebraic specification, and program synthesis. According to [1] declarative programming can be subdivided into weak where the programs are theories but complemented possibly with control information by the programmer to improve efficiency; and strong where programs are theories and the system supplies all control. In [4] Michael Hanus has similar view of declarative programming as a style of programming that describes the properties of the problem domain and its solution, rather than the steps of the computation needed to obtain the solution. In [4] he describes the similarity of classes (depending on the underlying formalism) of programming languages that support the declarative paradigm. In [5] description of languages that combine both logic and functional programming is provided.

**Declarative Programming: Scheme vs. BPEL Language**

According the description above of what constitutes a declarative language; Scheme qualifies as such since it is a functional language which is included in the definition of declarative languages [1, 4, 5].

BPEL allows for the specification of operational behavior descriptions; it uses programming language constructs to describe a process behavior. Different formalisms are used to specify system's behavioral model (based on automata theory) and the specification of the expected properties (based on logic, i.e. declarative in nature.) The dynamic interface view of a business process can be described via a protocol (the set of all legal conversations between the process and the outside world.) This is a description of the process's interface behavior. The purpose of providing a declarative specification of the interface protocol is to verify conformance of the concrete process implementation to the specification. BPEL abstract process specifies the business protocol. However, since BPEL is imperative process language, its abstract specification is non-declarative. The authors in [50] extend BPEL with declarative behavioral specification language based on a form of regular expressions.

### 2.5.9 Batch scripting

Generally, programs can be seen as functions that produce a result by manipulating the input in a "well-understood way" [6]. These programs are self-contained: built to fulfill a specific purpose or a goal. In practical systems there is a need to configure and coordinate discrete functionality implemented in programs in a prescribed way to fulfill a higher business goal. This can be achieved by using a general-purpose programming language, since they have all the language features necessary -- such as control flow (sequencing, selection, iteration, and recursion), variables, data types, subroutines, and control and data abstractions -- to automate the coordination between independent programs. However, their purpose is to provide features and security mechanisms (e.g. static type checking) built into the language to support the programmers in creating efficient, maintainable, portable programs [6]. The main requirements of scripting languages, by contrast, are *flexibility, local customization, rapid development, dynamic type checking,* and extensive use of structured types such as *tables, lists, files, and operations such as pattern matching and string manipulation, easy access to system facilities, economy of expression* [6]. These requirements change the emphasis on the feature set provided by the scripting languages.

Batch scripting (as opposed to interactive commands processing) is the process of specifying commands to a command interpreter in a file (batch file) which are consequently executed (interpreted) usually line by line [28]. Their main purpose is to automate a process that needs to be performed on regular bases.

**Batch Scripting: Scheme vs. BPEL Language**

Searching the existing literature provided many examples of the Scheme language being used as a scripting language [37, 38, 39, 40, 41, 42]. There are online tutorials showing how to write scripts within a specific implementation of the language, which is not necessarily compatible with other implementations of Scheme. Some examples of command-line parsing using Racket implementation of Scheme are provided in [41]. A Scheme program can be converted into an executable script in Unix and Mac OS X by replacing the language declaration #lang racket at the top of the definition area with #! /usr/bin/env racket and the file permissions need to be changed using chmod+x <program_filename> on the command line to make the new program executable [38].
Similarly, batch files can be created in Windows [39]. For example Figure 4 and Figure 5 show the batch file and the output produced after it is run in the Command Prompt. The command to execute the script has to be written after <; > a semicolon. The operating system automatically attaches the path to the script as an argument to the command Racket.exe. Racket implementation provides command-line command to parse the arguments supplied to the script. The script Hello1.bat takes a string argument and a flag. If flag is “-v” then a long message is produced, otherwise the input argument is echoed back to the user.

```
 ; @echo off
 ; Racket.exe "%~f0" %*
 ; exit /b
 #lang racket
(define verbose? (make-parameter #f))
 (define greeting
   (command-line
    #:once-each
    [("-v") "Verbose mode" (verbose? #t)]
    #:args
    (str) str))
 (printf "~a~a\n"
         greeting
         (if (verbose?) " to you, too!" ""))
```

**Figure 4: Hello1.bat batch file execute Racket code from Windows .bat batch files [39]**

```
c:\Program Files\Racket>hello1 -v Hi
Hi to you, too!

c:\Program Files\Racket>
```

**Figure 5: Output after running Hello.bat file from Command Prompt**

In [43] the authors describe an implementation of a BPEL application that executes script commands in the current run time environment using embedded Java commands Process p = Runtime.getRuntime().exec(cmdString). In [44] describe a possible way to invoke a BPEL process from a sell script to be able to be monitored and controlled from a Control-M environment. Since BPEL engine requires a Java EE server and needs to be remotely invoked, the virtual machine that is launched by the sell script will be different using the current technology. A proposed solution is to have a WS client as an intermediary between the shell script and the BPEL process which will handle the remote invocation and possible exceptions thrown by the server. In [45] the authors illustrate the requirements for a job control language to describe the complex grid applications and their dependencies, error handling, failure recovery, monitoring, resource release among others. They compare existing languages that support job control or grid applications orchestration. BPEL is one of the languages that satisfy the requirements they set forward. For example, they explain that most of the job orchestration can be implemented using BPEL, specific to grid applications task handlers such as job submission, file transfer can be achieved using “Java Partner Links.” Other requirements—capability for setting of a threshold for failed tasks tolerance or task execution monitoring—cannot be easily satisfied. Although powerful, expressing error handling and inter-task communication requirements would be complex when using BPEL.

As demonstrated in [44], it is possible to execute external processes/commands in BPEL using Java API, it requires extensive knowledge of both BPEL and Java API implementing this capability. In addition the BPEL program footprint is larger than Scheme’s batch script footprint. The reason is that Scheme language has explicit support for batch scripting and this is not the case with BPEL.

### 2.5.10 UI prototype design

User interface development tools need to support the use and management of UI components, organizing and arranging layout capabilities, and functionality for testing the prototype interfaces at minimum. The programming language used to specify (e.g. by programming or scripting) user interactions affect the usability of these tools Different phases of software development require different tool support for user interface prototyping. In the early stages during requirements elicitation the end users need to be closely involved in the UI design. The tools support has to target non-programmers. In addition, the developers are specialists in graphical design and not necessarily expert programmers. During the implementation phase the tools need to support GUI or non-GUI components management and usage to allow for creation of UI prototypes [46]. The set of features or capabilities that UI prototyping tools need to provide is therefore different and depend on the stage of the software development and the application requirements.

**UI Prototype Design: Scheme vs. BPEL Language**

In [47] the author introduces MysterX object-oriented Scheme toolkit which is capable of building interactive applications. This is achieved by using reflective mechanism to integrate and host COM components capable of creating visual interfaces (e.g. Microsoft's Visual Basic, or to create Dynamic HTML using JavaScript or VBScript.) The window components' properties can be set using Scheme language. Scheme is used to create event handlers to respond to user's actions. MysterX combines different technologies and languages to enable the creation of GUI applications. In [29] is given a description of an implementation of Scheme language that allows the programmers to build graphical interfaces. It provides support for Core Windowing Classes, Geometry Management (defining containers and so called containees, defining new types of containers), mouse and keyboard events, windowing class reference (window area, canvas, button, frames, check box and other window widgets) and many more capabilities. Similarly, Racket implementation of Scheme provides its own implementation for GUI design capabilities [48]. In addition it has capabilities to build dynamic Web pages as demonstrated in [65]. BPEL is designed to support composition of web services using machine-to-machine communication; however, it does not support human interactions. Support for human interactions was proposed as an extension to the BPEL language. BPEL4People enables the definition of client applications to support human interactions [30, 49]. Intra-process UI interaction in BPEL can also be achieved via XML-test cases; however those were not meant to be used in production [33]. BPEL is deployed on a server and invoked remotely through its WSDL interface. An external client has to exist [44]. It can be another BPEL process, a browser, or a Web service [15].

# 3 Consolidated Analysis and Synthesis of the Results

| **Criteria / PL** | **1. Default more secure programming practices** | **2. Web applications development** | **3. Web services design and composition** | **4. OO-based abstraction** | **5. Reflection** |
|---|---|---|---|---|---|
| **C++** | C++ is close to the operating system, especially when it relates to pointers and memory access, and this fact makes C/C++ unsafe | C++support creating web applications using Wt C++ Toolkit . | C++ web services use XML remote procedure protocol (XML-RPC) or Simple Object Access Protocol (SOAP) as a transport layer to send a request and receive a response. | C++ supports object oriented based programming, it provides multiple inheritance and exception handling as well as interface definition. Static dynamic binding | C++ doesn't support run-time reflection, but does support compile-time reflection. |
| **JavaScript** | JavaScript is dynamically typed and considered unsafe. | JavaScript is imperative scripting language that mainly used for client side internet applications. | JavaScript can call a web services as services as being the client using, JAX-WS is a Java API for XML-Based Web Services is a framework helps to do web service. | JavaScript doesn't support inheritance or virtual functions. | JavaScript is build on top of Java, ExtendScript is a construct facilitates a reflection behavior. |
| **PHP** | Dynamically and weakly typed language, Garbage collection, Many functions defaulted to | Extremely Popular, Efficient, Fast, Free, Open-Source, Many books, The PHP Manual, | Provides the SOAP extension to develop SOAP servers and clients and the XML-RPC | Object-Oriented, PDO for uniform access to any database, Criticized | Supports, Complete reflection API, reverse engineer classes, interfaces, |
| | true, Popular | Easy to learn, Supports many web servers and DBMS, Cross-Platform, Portable | extension to create XML-RPC servers and clients , Simple to use | | functions, methods and extensions. |
| **Scala** | Static and strongly typed which uses type inference to deduce the type, Garbage collection, Used by many companies, Less secure than Java in a few superficial ways | Several web frameworks for web development, Code is brief and expressive, Type-safe, Ajax and Comet support, Scalable | Frameworks (Lift, etc.), Supports. Pattern matching, Higher-order functions, XML support, Lift's built-in support for REST and other web services | Object-Oriented, more orthogonal and complete language | Same as Java, Has richer types, which are not fully reflected in bytecode, Scala reflection library is in development |
| **AspectJ** | Not really type-safe, static typing, memory management by creating new Aspect | Several cross-cutting concern like monitor & detect app. Failure, error handling & repair, authentication, verification and sessions management | Complicated development and deployment compare to C#, can be used to monitor multiple web service, service call, services failure and repair | An aspect has very similarities like class, supports the basic principles like abstraction, encapsulation, inheritance (partial in some cases) | Has very powerful, but controlled mechanism for reflection through Reflection API |
| **C#** | Type-safe, Static Typing, dynamic typing available by declaration, Garbage collection | Flexible, convenient and faster development and deployment | Simple and faster development, lots of enhancement for better web | Primarily language from object-oriented paradigm | Supports through Reflection namespace |
| | | | application development | | |
| **Haskell** | Thread safe, static typing, type casting is done by explicit conversion function, Module system is used for access control, Sub-typing is not available, internal Garbage Collection | The monad Transformers libraries are used to add application specific functionality. | HXML toolbox using generic data model | This style is achieved by class types | Static Reflection using Derive tool |
| **Java** | Robust, statically typed, implicit type casting, Access control is available, sub-typing is possible, Automatic garbage Collection | Java API's ; such as Servlet, JNDI etc | Java Specification Requests submitted to Java Community Process | Object oriented | Static & Dynamic Reflection |
| **Scheme** | Dynamically and strongly typed language, implicit type casting, Access control is available, sub-typing is possible, Automatic garbage Collection | The core language is extended with libraries to allow web app development | The core language is extended with libraries to allow web services development | The core language is extended with libraries to allow the development in OO style with classes, objects, inheritance, polymorphism etc. | Powerful reflection mechanism, homoiconic language capable of self-defining, program and data treated uniformly, dynamic language and program expansion; |
| | | | | | dynamic structure inspection |
| **BPEL** | Statically, strongly typed, security requirements part of the standard, Java implementations of the language | Web app via web clients invoking BPEL web services | BPEL was built with the goal to support web services composition | BPEL is not OO language and currently does not support OO paradigm | Currently does not support it; however, there is ongoing research to extend the language |

| **Criteria / PL** | **6. Aspect-orientation** | **7. Functional programming** | **8. Declarative programming** | **9. Batch scripting** | **10. UI prototype design** |
|---|---|---|---|---|---|
| **C++** | Aspects in AspectC++ implement crosscutting concerns in a modular way. C++ have the feature to weave advice code into destination code | FC++ is a rich library that supports functional programming in C++. | Pro*C mainly used for dealing with databases, it allows the developer to connect into a database and extract or manipulate SQL statements. Pro*C has some declarative statements mainly used to execute SQL statements. | Batch scripting can be done using C++ standard library to execute system commands [255]. | UI prototyping can be achieved in C++ via libraries (e.g. Win32 API [256], gtk+ [257]) |
| **JavaScript** | JavaScript have the feature to weave advice code into destination code | JavaScript support standard functions such as map reduce and select. | In JavaScript declarative programming can be achieved using *Functional* [258] library. | JavaScript allows for execution of system command as shown in [259]. | JavaScript can be used to create web-based GUI running on the client side [260]. |
| **PHP** | Supported via several extension (The AOP Libarary for PHP and others), Performance issues | Fn.php is a basic attempt bring functional programming to PHP | Not Supported, Programmers can adopt a declarative style of programming, Code self-documented | Supports scripting and batch processing via plugins and cron | Supports via GTK+, Utilizes PHP 5's object model support, Platform independent, Full documentations, A book |
| **Scala** | No AspectJ-like implementation of AOP for Scala, Supports mixins, which enable separation of code that crosscuts class hierarchies | Multi-paradigm supports the functional style. Not considered so by some, Statically typed object oriented language with closures | Supports, Functional paradigm, Declarative, Declarative reading is Math definition | Supports, Real Scripting, Access to complete JDK, do file operations on a high and object oriented level | Supports, based on a Scala library, provides access to Java's Swing framework of GUI classes, Alternative frameworks |
| **AspectJ** | Specially designed for Aspect-oriented programming | Doesn't have any direct support or extension to achieve functional programming. Can be achieved from the base language | Doesn't have any direct support or extension to achieve declarative style. Can be achieved from the base language | Possible to monitor, automate, schedule etc. through the API of the base language by creating aspect at certain jointpoints | Policy, rules, regulation updates etc. can be notified by implementing AOP, where the basic UI designing can be performed through the base language (Java) |
| **C#** | Supported through other implementation like Spring.Net, not by Microsoft | Functional programming can be achieved through the implementation of lambda expression and own namespace | Supported through its built in Library of LinQ and Regex. | External commands can be executed through System.Diagnostic namespace, also support for internal command and automation | Extensive supports of tools & control from the IDE, considered one of the best IDE for UI design |
| **Haskell** | Embedding this style provides TRex | Pure functional | It is declarative. | Can be achieved but in complex way by System.cmd API. | Can be done by portable and native library |
| **Java** | Implementation is AspectJ | Mimics Functional style but not pure | This facility can be integrated with JSetL library | Process Builder Class Provides this facility | Provides lightweight and heavyweight - components |
| **Scheme** | Recent research and development to extend implementation of Scheme with macros to allow specifications of advice, poitcuts etc. supporting AOP | Supported via first-class functions, lang. constructs to allow side-effect free programming, based on Lambda calculus, non-pure functional | Declarative, by virtue of being functional | Concise expression of batch scripts, widely used for this purpose | The core language is extended with libraries to support UI/GUI development |
| **BPEL** | Only via extension to the language: AO4BPEL, much ongoing research | Not a functional language, all its language constructs are imperative (structured programming), there is no notion of | It is primarily imperative, there no language constructs to support declarative programming, BPEL abstract process has | Possible via Java language code inserted into BPEL script, done as a prove of concept | No native support, only via the language BPEL4People to allow for human interaction with the workflow |
| | | function | the potential to be expressed declaratively, ongoing research for declarative extensions | | |

### 3.1 Secure Programming Practices

As per our studies, Java, C++, C#, Haskell and BPEL are type-safe languages and more specifically support static typing. JavaSript is dynamically typed, means it checks data types ar execution time. Scheme, on the other hand, is dynamically, yet strongly typed language. According to our findings, Scala is static and strongly typed which uses type inference to deduce the type. AspectJ is observed as not a type-safe language in some researches. PHP is a dynamically and weakly typed language. Java, C#, Haskell, BPEL and Scheme are the most secure programming languages although they provide it in a different manner. In addition, automated garbage collection used for safe memory management is provided by Haskell, Java, Scheme, Scala, and PHP. BPEL's implementations that we found are built on top of Java and they use all the capabilities of the language.

### 3.2 Web Application Development

Most of the languages provide capabilities to build web applications. PHP is the most popular according to our research. Java builds high quality web applications using embedded APIs and Scala has several web frameworks for web development. Scheme has programming constructs that allow building a secure web server and web applications by extending the core language with macros. Haskell relies on external libraries for building those applications and C# has extensive supports for web development through its Visual Studio IDE. While BPEL can expose WSDL interface and browsers can invoke the services provided by it, AspectJ provides a better way to implement the cross-cutting concern without hampering the existing code. We have found AspectJ functionality has been exposed as a web application. C++ supports web development by hiding many network communications along with web browsers. JavaScripts mainly used for web applications, by running a script at user side, which help to distribute computations and reduce network bandwidth.

### 3.3 Web Service Design and Composition

All the languages can expose their functionality via WSDL interfaces. However, for composing web services, BPEL is the language that has been built for this purpose. C#, using the .NET framework provides a very good support for interoperability through Web services. C++ uses XML-RPC helps for web service messaging and to establishes the communication across different platforms. JavaScript uses JAX-WS API for XML-Based Web Services to do web service across different platforms.

## 3.4 OO-based Abstraction

C#, Java, Scala and PHP are object-orientated languages. Scheme is a multi-paradigm language and supports the object-oriented paradigm. AspectJ's program implementation also supports some properties of the object-oriented language. In Haskell, with the help of type classes, we can use this feature. C++ is object oriented based programming style support Class, Instance, Method, Message passing, Inheritance, Abstraction, Encapsulation, and (Subtype) polymorphism. JavaScript support simple object-based paradigm, it doesn't support inheritance or virtual functions (polymorphism).

## 3.5 Reflection

In our analysis and research, we have found that Reflection is supported by most of the languages. For example, Java supports both static and dynamic reflection. But, in Haskell we need a separate tool (Derive) to avail the support of reflection. Scheme is homoiconic language, reflection is supported natively. The programs and the language itself can be expanded and compiled at run time. PHP comes with a complete reflection API; we can reverse engineered classes, interfaces, functions, methods, and extensions. In C#, reflection is supported through its namespace construct, whereas AspectJ offers an alternative way to access the static and dynamic context through its reflection API. BPEL doesn't currently support reflection; however, there is ongoing research to expand the language to support the reflection mechanism. Scala's support for reflection is same as Java, but has richer types, which are not fully reflected in bytecode. C++ support compile time reflection but it doesn't support run-time reflection due to the unrecognizing detailed type information which be hidden by C++ compiler, however, run-time reflection in C++ is limited to some features such as monitoring expression types and querying the type name. JavaScript supports reflection object, which is a construct that provides the program with reflected-objects' contents.

## 3.6 Aspect-oriented programming

In our comparative analysis, we observed that AspectJ is the only language designed for AOP. Most of the other languages require extensions or libraries for AOP support. For example, AspectJ is one of the extensions of Java to provide AOP, BPEL uses AO4BPEL to provide support for the Aspect-oriented paradigm. Haskell achieves AOP via GHC compiler. There are some approaches to facilitate AOP in PHP (e.g. PHPAspect, Aspect-Oriented PHP, and others); however, most of these extensions have performance issues. There is no AspectJ-like implementation of AOP for Scala; however, Scala supports mixins, which enable separation of code that crosscuts class hierarchies. For C#, there are also some supporting extensions for AOP; the most updated and developed one is Spring.Net. Scheme supports aspect-orientation via a powerful macro system, which extends the core language. AspectC++ is one of the extension of C++ that implement crosscutting concerns in a modular way, it supports code pointcuts and name pointcuts. JavaScript and C++, have the feature to weave advice code into destination code, this helps to change the behavior of the program without refractor the original code. The advice code is written in a separate file, and woven into the original code at execution time.

## 3.7 Functional Programming

Haskell is a pure functional language which has no side effects. This is in contrast with Scheme which is a multi-paradigm language including functional; however, it allows side effects. Java can have the features of functional programming with side effects and BPEL does not support the functional paradigm according to our research. Fn.php is an attempt to define lots of useful higher-order functions to PHP; however, it's very basic. Scala is a multi-paradigm programming language that supports the functional style; yet some don't consider it a really functional language (it is a statically typed object oriented language with closures). C# is also a multi-paradigm programming language which supports functional programming. AspectJ (the extension of Java), on the other hand, does not support the functional programming paradigm, but its subset Java does, according to our studies. FC++ is an extension to C++ language, it is a rich library that supports functional programming in C++. JavaScript support some functional programming such as map reduce and select, in javascript block does not have scope, only function have scope, which helps to maintain some states, which is good for functional paradigm.

### 3.8 Declarative programming

All languages that support the functional paradigm are considered declarative according to our references. These include Haskell, Scheme, Scala, and C#. However, among those languages, Haskell is the only pure functional language. PHP is not a declarative programming language; however, you can adopt a declarative style of programming. Our studies show that AspectJ and BPEL do not support the declarative paradigm. C# and Java have libraries that support declarative programming. CPP is an imperative, object based and functional programming paradigm according to our understanding, Pro*C is an extension to CPP language used for dealing with databases in declarative style. *Functional* library can be used in JavaScript to achieve declarative programming.

### 3.9 Batch Scripting

Batch scripting is supported by most of our languages; however, Scheme, Scala, and Haskell allow expressing scripts concisely. We have found that PHP also supports scripting and batch processing via plugins and cron. On the other hand, Java and C# support it via their own libraries. BPEL and AspectJ allow writing scripts using Java. C++Script is C++ library, it uses a dynamic programming and can be compiled with C++ compiler, it is considered one of the script languages. JavaScript allows for execution of system commands.

### 3.10 UI prototype design

C# has very extensive support for UI prototype design using its IDE, where as AspectJ can provide some support for monitoring user interface threading, synchronization, notification etc. But the basic designing for core concerns can be achieved through the base language. Java, PHP, and Scheme allow creating interfaces via a toolkit. In order to support human interaction, BPEL was extended with the BPEL4People language. For the development of GUI applications, Scala provides a library that has access to Java's Swing framework of GUI classes. Haskell also uses different libraries which provide toolkits for interface development. UI prototyping can be achieved in C++ via libraries (e.g. Win32 API, gtk+). JavaScript can be used to create web-based GUI running on the client side.

# 4 Conclusion

We have made a comprehensive research and survey of ten languages that support most of the popular paradigms based on specified criteria. We have found that, C++, JavaScript, PHP, and AspectJ are not considered safe languages. For example, C++ allows pointers to access memory, while in AspectJ, which is statically-typed, a binding between a pointcut and an advice can give rise to type errors at runtime. Furthermore, JavaScript and PHP are weakly, dynamically typed languages and also considered unsafe. On the other hand, C#, Scala, Haskell, Java, Scheme, and BPEL are safe languages. Among these, Scheme is the only one which is dynamically typed.

Web application can be developed using most of the languages we have studied. JavaScript is the only language that specially designed for client-side application development. Among these which are used for server-side web application development, PHP, C#, and Java are well suited for it. BPEL is specifically designed and standardized language for web service composition. We have found that most of the languages are capable to implement web services, whereas JavaScript can invoke web services as a client.

From the research we have done, we have found that BPEL and JavaScript do not support the Object-oriented paradigm. They don't have constructs for inheritance. AspectJ does support some basic principles of OO paradigm. All other languages have OO capabilities either natively or via libraries.

Scheme, Java, C#, AspectJ, and PHP have powerful reflection mechanism. Haskell supports static reflection via a tool. C++ has only compile-time reflection, while Scala reflection library is in development phase. JavaScript also accomplishes reflection via tools like ExtendScript.

AspectJ is the only language built specifically for Aspect-orientation. JavaScrpit can achieve partial AOP. It has the feature to weave advice code into destination code. All of the other languages can have the AOP properties via libraries or extensions.

C++, JavaScript, PHP, and C# can obtain functional style via either libraries or extensions. On the other hand, Scheme, Haskell, and Scala are considered as functional languages, among them Haskell is the only one which is pure functional. No implementation of functional programming has been observed for AspecJ, but its base language Java can mimic some functional styles. BPEL doesn't support the functional paradigm, but some research is ongoing to achieve this. All functional languages are considered as declarative by definition.

Although most of the languages we have researched support batch scripting, some cannot accomplish it in a concise way. AspectJ and BPEL can also support batch scripting using Java API from its base language.

Support for web-based GUI is provided by JavaScript. C# uses ASP .Net web form with its server side controls to built web GUI with the support from its IDE. Additionally, Java applets are good example of applications running at the client-side browsers. BPEL allows for human interaction with the workflow via another language called BPEL4People. Scheme, PHP and Java Servlet technology create HTML web pages that are sent through HTTP to the client to be displayed in the browser. All studied languages support command prompt UI prototyping. Non web-based GUI is also supported via libraries or toolkits.

In addition to our base research work, we have also studied the popularity of our programming languages. According to TIOBE Programming Community Index for August 2010 (http://www.tiobe.com/), we have found that C# and Scheme increased in popularity compared to August 2009. On the other hand, C++, Java, and PHP remain in the same position as they were last year. In addition, according to the same source, the ranking of the most popular languages to the least are as follows: Java, C++, PHP, C#, Scheme, Haskell, and Scala. As per, TIOBE's Categories of Programming Languages, object-oriented statically typed languages have been the most popular for more than four years. In fact, statically-typed languages increased in popularity by 2.5% since last year. Functional and logical programming languages rank after object-oriented and procedural languages; however, they have increased in popularity by 0.2%. According to (http://www.langpop.com/), the following rankings: Powell's Books, Google Code, Ohloh, and Slashdot placed Java on top position. Moreover, some rankings like Yahoo! Search Results, Discussion Site Results, and Lambda The Ultimate show that C++ ranks first.

# Appendix A

## A.1 Installing Racket

The Scheme dialect used in this paper for testing is Racket, formally known as PLT Scheme. It can be downloaded from (http://racket-lang.org/). Racket can be run from command line Racket or from the graphical environment DrRacket. The programs related to writing web applications need to be saved in the same directory where Racket is installed (http://docs.racket-lang.org/more/).

To test the installation, create a small program in a text editor or DrRacket like the following:

```
#lang racket

 (define (test-it)
   'Hello-World)
```

Save the program as "first-one.rkt" and run from command prompt using commands:

(enter! "first-one.rkt") and after loading the files enter command (test-it). See output below:

```
> (enter! "first-one.rkt")
 [loading C:\Program Files\Racket\first-one.rkt]
 [loading C:\Program Files\Racket\collects\racket\lang\compiled\reader_rkt.zo]
 [loading C:\Program Files\Racket\collects\syntax\compiled\module-reader_rkt.zo]

 [loading C:\Program Files\Racket\collects\syntax\compiled\readerr_rkt.zo]
> (test-it)
'Hello-World
> _
```

## A.2 Web Server in Scheme

The following is the implementation of a server in Scheme ran on Racket. It is taken in whole from [34] and commented. Also, it has been slightly modified to produce the sum of the number entered.

```
#lang racket


;library used to parse URL and format HTML responses
(require xml net/url)


;Web server implemented as a function that
;accepts IP port-no from client
(define (serve port-no)

  (define main-cust (make-custodian))
  (parameterize ([current-custodian main-cust])

  ;server accepts TCP connection through listener
  ;#t is to avoid waiting for time-out for port-no
    (define listener (tcp-listen port-no 5 #t))
    (define (loop)  ;loop to accept connections from listener
      (accept-and-handle listener)
      (loop))
    (thread loop))
  (lambda ()
    (custodian-shutdown-all main-cust))
  ); end of (define (serve port-no)
```

**Figure 6: Startup server and accept port no from user**

## A.2.1 Running the Scheme Server

To start the server first open the Racket command prompt and then enter (enter! pgmName.rkt). On the new prompt enter (define stop(serve 8081)) after which the server will be running. Open a browser window and enter: http://localhost:8081/many. A web page ready to accept a number will be displayed. Enter a number and press the submit button. A response page will be displayed with as many “Hello”(s) as the number entered. Code fragment 1 on Figure 6: (define (serve port-no) function listens for client requests and invokes (accept-and-handle listener) function in its own thread to process the request.

```
;accept-and-handle function definition
;param listener
(define (accept-and-handle listener)
  (define cust (make-custodian))
  (custodian-limit-memory cust (* 50 1024 1024))
  (parameterize ([current-custodian cust])

     ;accept tcp connection which creates two streams from/to client
    (define-values (in out) (tcp-accept listener))
    (thread (lambda () ;to handle multiple requests
                (handle in out) ;create new thread to handle each request
                (close-input-port in)
                (close-output-port out))))
  ;; Watcher thread:
  (thread (lambda ()
              (sleep 10)
              (custodian-shutdown-all cust))))
```

**Figure 7: (accept-and-handle listener) function definition**

The function invokes (handle (in out)…) function and passes the input and output streams (Figure 8). The function invokes the (define (dispatch str-path) function with the path as a string (Figure 9).

```
;handle function definition
;param: in, out: streams
(define (handle in out)

  (define req  ;Extract request
    ;; Match the first line to extract the request:
    (regexp-match #rx"^GET (.+) HTTP/[0-9]+\\.[0-9]+"
                  (read-line in)));end of (define req ...
  (when req
    ;; Discard the rest of the header (up to blank line):
    (regexp-match #rx"(\r\n|^)\r\n" in)


  ;dispatch function takes a requested URL and
  ;produces a result to send back to the client

    ;; Dispatch:
    (let ([xexpr (dispatch (list-ref req 1))])
      ;; Send reply:
      (display "HTTP/1.0 200 Okay\r\n" out)
      (display "Server: k\r\nContent-Type: text/html\r\n\r\n" out)
      (display (xexpr->string xexpr) out)
      );end of (let ([xexpr (dispatch (list-ref req 1))]) ...
    );end of (when req ...
  ); end of function (define (handle in out)
```

**Figure 8: (handle (in out)…) function definition**

```
;Dispatch function definition
;The dispatch function consults a hash table
;that maps an initial path element, like "foo", to a handler function:

(define (dispatch str-path)
  ;; Parse the request as a URL:
  (define url (string->url str-path))
  ;; Extract the path part:
  (define path (map path/param-path (url-path url)))
  ;; Find a handler based on the path's first element:
  (define h (hash-ref dispatch-table (car path) #f))
  (if h
      ;; Call a handler:
      (h (url-query url))
      ;; No handler found:
      `(html (head (title "Error"))
             (body
              (font ((color "red"))
                    "Unknown page: "
                    ,str-path))));end of `(html (head (title "Error"))
  );end of (define (dispatch str-path)



(define dispatch-table (make-hash));path => handler function
```

**Figure 9: (dispatch str-path) function definition**

Dispatch table is declared as shown on Figure 9 to map the string path of the request/response to a function request/response handler. It is implemented with a hash data structure. If a handler is not found an error response is generated to the client otherwise the appropriate handler function is retrieved from dispatch table and invoked.

```
;; New helper function: builds and HTML page
;; for a form that has a "number" field and
;; a "hidden" field:

(define (build-request-page label next-url hidden)
  `(html
    (head (title "Enter a Number to Add"))
    (body ([bgcolor "white"])
          (form ([action ,next-url] [method "get"])
                ,label
                (input ([type "text"] [name "number"]
                        [value ""]))
                (input ([type "hidden"] [name "hidden"]
                        [value ,hidden]))
                (input ([type "submit"] [name "enter"]
                        [value "Enter"]))))))
  ;many function definition to handle the request from browser
(define (many query)
  ;; Create a page containing the form: ...and
      ;search dispatch-table for reply handler and call function
  (build-request-page "Number of greetings:" "/reply" ""))

;reply function definition to handle the reply to browser
(define (reply query)
  ;; Extract number from query and sore it in n
  (define n (string->number (cdr (assq 'number query))))

  `(html (body ,@(for/list ([i (in-range n)])
                 " hello")));concatenate n * hello(s)

;`(html (body "The sum is " ,(number->string (+ n n)))) ;double n
  ); end of (define (reply query) ...
```

**Figure 10: (define (many query)...); (define (reply query)...); (build-request-page label next-url hidden) function definitions**

The (define (many query)...) is invoked to handle the request by calling a helper function (build-request-page label next-url hidden) (Figure 10) to build the request form and sends the form to another url-path: “/reply”. The user responds and the “/reply” handler is retrieved from the dispatch table and invoked. It builds an html page with hello(s) corresponding to the number received. There is an alternative response that has been commented but (at the bottom of the code snippet) which adds the number to itself and displays the sum on the browser with an appropriate message. To run this code no other changes to the rest of the program need to be done. It was not part of the original code.

```
(hash-set! dispatch-table "many" many) ;many => many handler function
(hash-set! dispatch-table "reply" reply);reply => reply handler
```

**Figure 11: Building the dispatch-table**

The dispatch table is built using the Scheme's imperative hash-set! special form. The key is the url-path and the value is the function that handles the request/response.

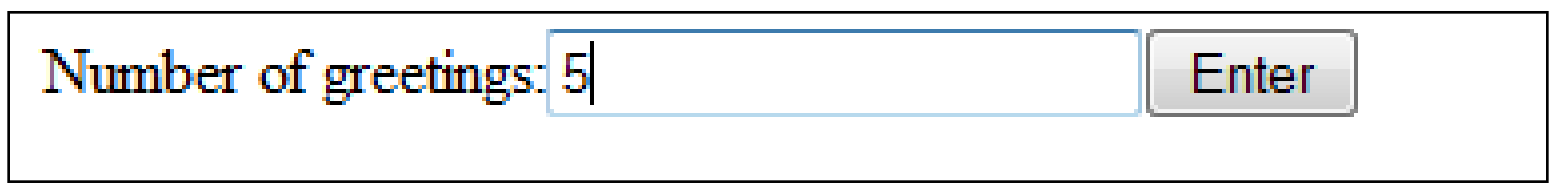

**Figure 12: Request form displayed on browser window**

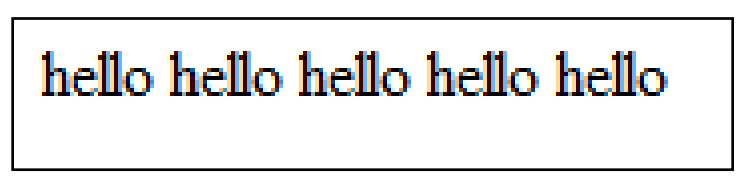

**Figure 13: Response produced by the server**

The request form is displayed on Figure 12 and the response on Figure 13.

## A.3 Web Server in BPEL

A similar behavior can be achieved using BPEL as the language. Sine, the primary goal of BPEL is to support automated interactions between web services communicating using SOAP or another transport protocol over HTTP, with XML as the wire protocol (or data exchange protocol) through well-defined standard interfaces, the basic structure of the basic invocation differs from the one of Scheme. For example, the input and output is represented using XML document which conforms to the XML grammar rules and the structure of the data is defined in XSD Schema which validates the data. In Scheme the data was passed to the server in the request sent from the browser. To invoke the correct functionality Scheme uses dispatch-table. BPEL requests are wired directly to the control statements that handle the request. BPEL can be seen as single function with possibly many nested scopes, event and message handlers, exception handlers, correlations grouped together. Part of the computation is performed within BPEL module and part of it is delegated to other web services by invoking their interfaces.

The base structure of the functionality that implements this simple web application has been obtained from the existing SynchronousSample which comes with NetBeans 6.5. An online tutorial of how to obtain and use it can be found online(http://netbeans.org/kb/61/soa/synchsample.html). The

SynchronousSample.wsdl document that describes the interface of BPEL and SynchronousSample.xsd defining the structure of the input and output messages (the data) remain unmodified. The data consists of a single xsd.string type input and output fields, wrapped in messages described in WSDL. The SynchronousSample.bpel was entirely modified to implement the functionality. The application is implemented using primarily the Design view which automatically generates equivalent BPEL code from the visual model and the Design view Palette to obtain the programming constructs that BPEL supports. The application consists of nested sequence, while and another sequence scopes. The first assignment within the outer sequence scope is used to initialize the output. The predicate expression of the while scope compares the input to zero. If it is not zero, control enters the while scope. Within the true path of the while scope there is a nested sequence within which there are two assignments. One is used to decrement the input used later in the predicate of while. The other is concatenating a string "Hello" to the output. The output is wrapped in an XML document. Test cases can be generated using the integrated JUnit included in the NetBeans IDE. The input and output for one test case and the generated source code can be seen in Figure 15, Figure 16, and Figure 17, respectively.

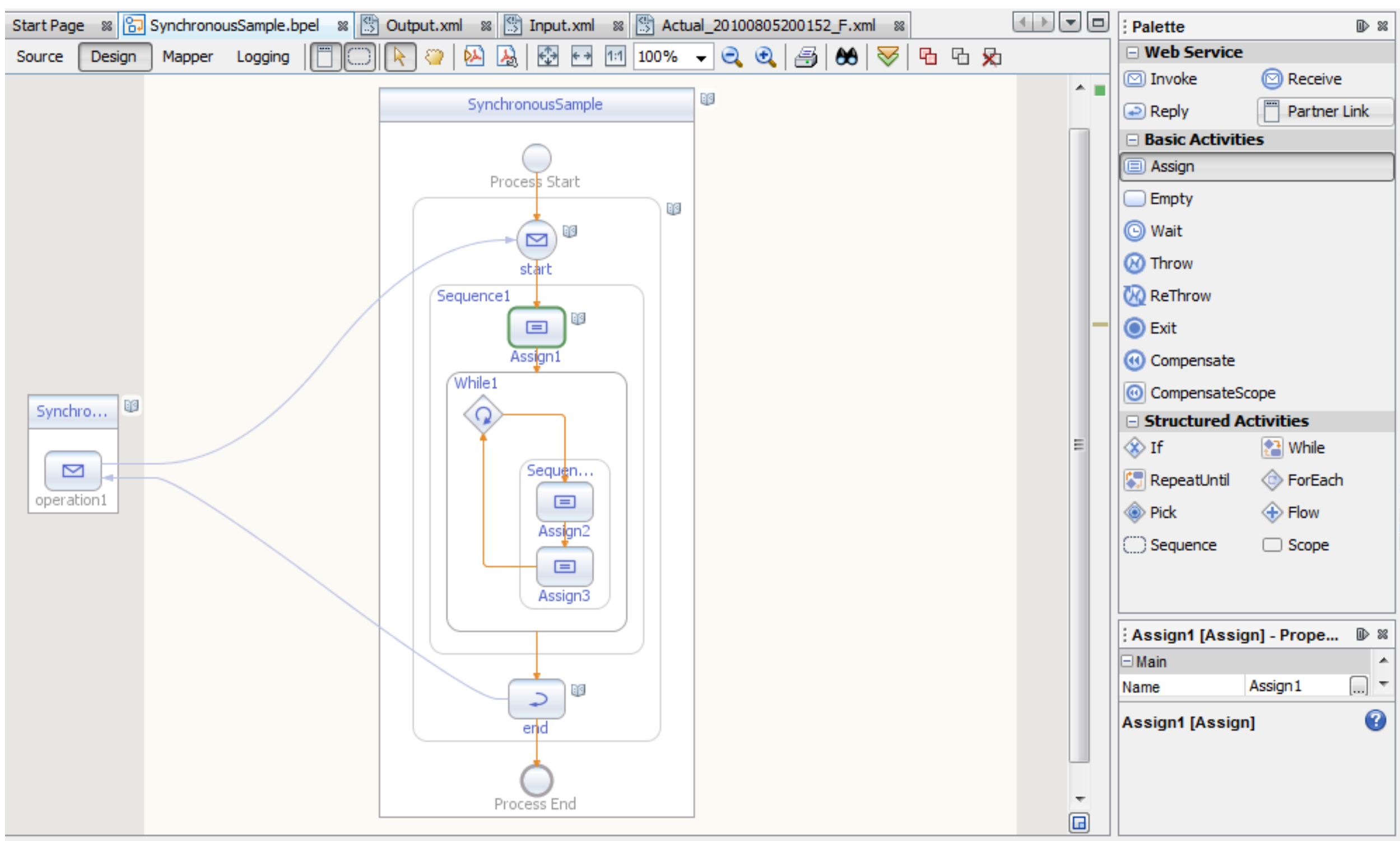

**Figure 14: Graphical representation of server application implemented using BPEL**

```xml
<soapenv:Envelope xsi:schemaLocation="http://schemas.xmlsoap.org/soap/envelope/
http://schemas.xmlsoap.org/soap/envelope/"
xmlns:xsi="http://www.w3.org/2001/XMLSchema-instance"
xmlns:xsd="http://www.w3.org/2001/XMLSchema"
xmlns:soapenv="http://schemas.xmlsoap.org/soap/envelope/"
xmlns:syn="http://xml.netbeans.org/schema/SynchronousSample">
  <soapenv:Body>
    <syn:typeA>
      <syn:paramA>5</syn:paramA>
    </syn:typeA>
  </soapenv:Body>
</soapenv:Envelope>
```

**Figure 15: Sample input represented in XML**

```xml
<?xml version="1.0" encoding="UTF-8"?>
<SOAP-ENV:Envelope xmlns:SOAP-ENV="http://schemas.xmlsoap.org/soap/envelope/"
xmlns:xsd="http://www.w3.org/2001/XMLSchema"
xmlns:xsi="http://www.w3.org/2001/XMLSchema-instance"
xsi:schemaLocation="http://schemas.xmlsoap.org/soap/envelope/
http://schemas.xmlsoap.org/soap/envelope/">
  <SOAP-ENV:Body>
    <ns1:typeA xmlns:msgns="http://localhost/SynchronousSample/SynchronousSample"
    xmlns:ns1="http://xml.netbeans.org/schema/SynchronousSample">
      <ns1:paramA>Hello Hello Hello Hello Hello </ns1:paramA>
    </ns1:typeA>
  </SOAP-ENV:Body>
</SOAP-ENV:Envelope>

```

**Figure 16: Sample output represented in XML**

```
</receive>
<sequence name="Sequence1">
    <assign name="Assign1">
        <copy>
            <from>''</from>
            <to>$outputVar.resultType/ns2:paramA</to>
        </copy>
    </assign>
    <while name="While1">
        <condition>0 != $inputVar.inputType/ns2:paramA</condition>
        <sequence name="Sequence2">
            <assign name="Assign2">
                <copy>
                    <from>$inputVar.inputType/ns2:paramA - 1</from>
                    <to>$inputVar.inputType/ns2:paramA</to>
                </copy>
            </assign>
            <assign name="Assign3">
                <copy>
                    <from>concat($outputVar.resultType/ns2:paramA, 'Hello ')</from>
                    <to>$outputVar.resultType/ns2:paramA</to>
                </copy>
            </assign>
        </sequence>
    </while>
</sequence>
<reply
```

**Figure 17: Code fragment implementing the functionality, generate by BPEL engine**

# A.4 Factorial Function Implemented in BPEL

The generated source code, test input and the corresponding output are shown on Figure 18, Figure 19, and Figure 20.

```xml
</receive>
<sequence name="Sequence1">
    <assign name="Assign1">
        <copy>
            <from>$inputVar.inputType/ns2:paramA</from>
            <to>$outputVar.resultType/ns2:paramA</to>
        </copy>
    </assign>
    <while name="While1">
        <condition>1 != $inputVar.inputType/ns2:paramA</condition>
        <sequence name="Sequence2">
            <assign name="Assign2">
                <copy>
                    <from>$inputVar.inputType/ns2:paramA - 1</from>
                    <to>$inputVar.inputType/ns2:paramA</to>
                </copy>
            </assign>
            <assign name="Assign3">
                <copy>
                    <from>$outputVar.resultType/ns2:paramA * $inputVar.inputType/ns2:paramA</from>
                    <to>$outputVar.resultType/ns2:paramA</to>
                </copy>
            </assign>
        </sequence>
    </while>
</sequence>
<reply
```

**Figure 18: Generated source code of factorial function**

```xml
<soapenv:Envelope xsi:schemaLocation="http://schemas.xmlsoap.org/soap/envelope/
http://schemas.xmlsoap.org/soap/envelope/"
xmlns:xsi="http://www.w3.org/2001/XMLSchema-instance"
xmlns:xsd="http://www.w3.org/2001/XMLSchema"
xmlns:soapenv="http://schemas.xmlsoap.org/soap/envelope/"
xmlns:syn="http://xml.netbeans.org/schema/SynchronousSample">
  <soapenv:Body>
    <syn:typeA>
      <syn:paramA>5</syn:paramA>
    </syn:typeA>
  </soapenv:Body>
</soapenv:Envelope>
```

**Figure 19: Input test of factorial function implemented in BPEL**

```xml
<?xml version="1.0" encoding="UTF-8"?>
<SOAP-ENV:Envelope xmlns:SOAP-ENV="http://schemas.xmlsoap.org/soap/envelope/">
  <SOAP-ENV:Body>
    <ns1:typeA xmlns:msgns="http://localhost/SynchronousSample/SynchronousSample"
    xmlns:ns1="http://xml.netbeans.org/schema/SynchronousSample">
      <ns1:paramA xmlns:soapenv="http://schemas.xmlsoap.org/soap/envelope/"
      xmlns:syn="http://xml.netbeans.org/schema/SynchronousSample"
      xmlns:xsd="http://www.w3.org/2001/XMLSchema"
      xmlns:xsi="http://www.w3.org/2001/XMLSchema-instance">120</ns1:paramA>
    </ns1:typeA>
  </SOAP-ENV:Body>
</SOAP-ENV:Envelope>
```

**Figure 20: Output from testing factorial function implemented in BPEL with input of 5**

## A.5 PHP – Creating web service

Here is how to create a basic web service that provides an XML or JSON response using some PHP and MySQL. Author: **David Walsh [187]**.

The PHP / MySQL

```php
/* require the user as the parameter */
if(isset($_GET['user']) && intval($_GET['user'])) {

/* soak in the passed variable or set our own */
$number_of_posts = isset($_GET['num']) ? intval($_GET['num']) : 10; //10 is the default
$format = strtolower($_GET['format']) == 'json' ? 'json' : 'xml'; //xml is the default
$user_id = intval($_GET['user']); //no default

/* connect to the db */
$link = mysql_connect('localhost','username','password') or die('Cannot connect to the DB');
mysql_select_db('db_name',$link) or die('Cannot select the DB');

/* grab the posts from the db */
$query = "SELECT post_title, guid FROM wp_posts WHERE post_author = $user_id AND post_status =
'publish' ORDER BY ID DESC LIMIT $number_of_posts";
$result = mysql_query($query,$link) or die('Errant query:  '.$query);

/* create one master array of the records */
$posts = array();
if(mysql_num_rows($result)) {
while($post = mysql_fetch_assoc($result)) {
$posts[] = array('post'=>$post);}}

/* output in necessary format */
if($format == 'json') {
header('Content-type: application/json');
echo json_encode(array('posts'=>$posts));
} else {
```

```
header('Content-type: text/xml');
echo '<posts>';
foreach($posts as $index => $post) {
if(is_array($post)) {
foreach($post as $key => $value) {
echo '<',$key,'>';
if(is_array($value)) {
foreach($value as $tag => $val) {
echo '<',$tag,'>',htmlentities($val),'</',$tag,'>';}}
echo '</',$key,'>';}}}
echo '</posts>';          }
@mysql_close($link); } /* disconnect from the db */
```

Take the following sample URL for example

```
http://mydomain.com/web-service.php?user=2&num=10
```

The XML Output

```
<posts><post>
<post_title>SSLmatic SSL Certificate Giveaway Winners</post_title>
<guid>http://davidwalsh.name/?p=2304</guid>
</post>
<post>
<post_title>MooTools FileManager</post_title>
<guid>http://davidwalsh.name/?p=2288</guid>
</post>
<post>
<post_title>PHPTVDB: Using PHP to Retrieve TV Show Information</post_title>
<guid>http://davidwalsh.name/?p=2266</guid>
</post>
<post>
<post_title>David Walsh: The Lost MooTools Plugins</post_title>
<guid>http://davidwalsh.name/?p=2258</guid>
</post>
<post>
<post_title>Create Short URLs Using U.Nu</post_title>
<guid>http://davidwalsh.name/?p=2218</guid>
</post>
<post>
<post_title>Create Bit.ly Short URLs Using PHP</post_title>
<guid>http://davidwalsh.name/?p=2194</guid>
</post>
<post>
<post_title>Represent Your Repositories Using the GitHub Badge!</post_title>
<guid>http://davidwalsh.name/?p=2178</guid>
</post>
<post>
<post_title>ZebraTable</post_title>
<guid>http://davidwalsh.name/?page_id=2172</guid>
```

```
</post>
<post>
<post_title>MooTools Zebra Table Plugin</post_title>
<guid>http://davidwalsh.name/?p=2168</guid>
</post>
<post>
<post_title>SSLmatic: Quality, Cheap SSL Certificates and Giveaway!</post_title>
<guid>http://davidwalsh.name/?p=2158</guid>
</post></posts>
```

The JSON Output

```
{"posts":[{"post":{"post_title":"SSLmatic SSL Certificate Giveaway
Winners","guid":"http:\/\/davidwalsh.name\/?p=2304"}},{"post":{"post_title":"MooTools
FileManager","guid":"http:\/\/davidwalsh.name\/?p=2288"}},{"post":{"post_title":"PHPTVDB: Using
PHP to Retrieve TV Show
Information","guid":"http:\/\/davidwalsh.name\/?p=2266"}},{"post":{"post_title":"David Walsh: The
Lost MooTools Plugins","guid":"http:\/\/davidwalsh.name\/?p=2258"}},{"post":{"post_title":"Create
Short URLs Using U.Nu","guid":"http:\/\/davidwalsh.name\/?p=2218"}},{"post":{"post_title":"Create
Bit.ly Short URLs Using
PHP","guid":"http:\/\/davidwalsh.name\/?p=2194"}},{"post":{"post_title":"Represent Your Repositories
Using the GitHub
Badge!","guid":"http:\/\/davidwalsh.name\/?p=2178"}},{"post":{"post_title":"ZebraTable","guid":"http:\
/\/davidwalsh.name\/?page_id=2172"}},{"post":{"post_title":"MooTools Zebra Table
Plugin","guid":"http:\/\/davidwalsh.name\/?p=2168"}},{"post":{"post_title":"SSLmatic: Quality, Cheap
SSL Certificates and Giveaway!","guid":"http:\/\/davidwalsh.name\/?p=2158"}}]}
```

## A.6 Scala - Building Web Services in Lift

Here is how to build Web Services in Lift. Author: **Steve Jenson** [188].

Start with a simple class

```
package com.hellolift.api

import net.liftweb.http._
import com.hellolift.model._
import scala.xml._

class BlogAPI(val request: RequestState)
 extends SimpleController {

  def index: ResponseIt = {
    // To be filled in. }

  def get(id: String): ResponseIt = {
    // To be filled in. }
```

```
    def create: ResponseIt = {
      // To be filled in. }

    def delete(id: String): ResponseIt = {
      // To be filled in. }
  }
```

GET an existing Item: GET /api/$itemid/ Route a request to the proper method via pattern matching

```
def index: ResponseIt {
  request match {
    case RequestState(ParsePath("api" :: itemid :: Nil,
      _, _), GetRequest, _, _) => get(itemid)
  }}

def get(id: String): ResponseIt = {
  val it = Item.find(By(Item.id, Helpers.toLong(id)),
                     By(Item.author, user)
  it match {
    case Full(item) => AtomResponse(item.toAtom())
    case Empty => NotFoundResponse()
  }}
```

Outputting XML: scala.xml makes it really easy to output valid XML. The following is in our Item model object.

```
def toAtom = {
  val id = "http://example.com/api/" + this.id
  val formatter = new
    SimpleDateFormat("yyyy-MM-dd'T'HH:mm:ss'Z'")
  val updated = formatter.format(this.lastedited.is)

  <entry xmlns="http://www.w3.org/2005/Atom">
    <id>{id}</id>
    <updated>{updated}</updated>
    <author>
      <name>{name}</name>
    </author>
    <content type="xhtml">
      <div xmlns="http://www.w3.org/1999/xhtml">
        {body}
      </div>
    </content>
  </entry>
}
```

## A.7 PHP Abstract Class Basics

The following is an example of creating a simple abstract class “OOPHPAbstractClass”, and “OOPHPClassToExtendAnAbstract” which extends it. To successfully extend OOPHPAbstractClass, OOPHPClassToExtendAnAbstract must have getName() and setName functions(). [189] [190] [191] [192]

OOPHPClassToExtendAnAbstract.php

```
//copyright Lawrence Truett and FluffyCat.com 2007, all rights reserved

  //this class "extends" OOPHPAbstractClass

  include_once('OOPHPAbstractClass.php');

  class OOPHPClassToExtendAnAbstract extends OOPHPAbstractClass {

    private $instanceName;

    //OOPHPAbstractClass has the abstract function getName,
    //  so we must implement it here.
    public function getName() {
      return $this->instanceName;
    }

    //OOPHPAbstractClass has the abstract function setName,
    //  so we must implement it here.
    public function setName($nameIn) {
      $this->instanceName = $nameIn;
    } }
```

OOPHPAbstractClass.php

```
//copyright Lawrence Truett and FluffyCat.com 2007, all rights reserved

  //OOPHPAbstractClass - a simple OO PHP Abstract Class
  //  this defines two functions, getName() and setName($nameIn)
  //  which any class extending this must have

  abstract class OOPHPAbstractClass {

   abstract public function getName();

   abstract public function setName($nameIn);  }
```

testOOPHPAbstract.php

```
//copyright Lawrence Truett and FluffyCat.com 2007, all rights reserved

  include_once('OOPHPClassToExtendAnAbstract.php');

  define('BR', '<'.'BR'.'>');

  echo 'BEGIN TESTING PHP ABSTRACT CLASSES'.BR;
  echo BR;

  echo 'test 1 - create a class which extends an abstract'.BR;
  $classOne = new OOPHPClassToExtendAnAbstract();
  echo BR;
  $classOne->setName("Harold");
  echo $classOne->getName();
  echo BR.BR;

  echo 'END TESTING PHP ABSTRACT CLASSES'.BR;
```

Output of testOOPHPAbstract.php

```
BEGIN TESTING PHP ABSTRACT CLASSES


test 1 - create a class which extends an abstract


Harold


END TESTING PHP ABSTRACT CLASSES
```

# A.8 PHP - Reflection

Reflection Example from Shell (a Terminal) [193]

```
$ php --rf strlen
$ php --rc finfo
$ php --re json
$ php --ri dom
```

The above example will output something similar to:

```
Function [ <internal:Core> function strlen ] {

  - Parameters [1] {
```

```
    Parameter #0 [ <required> $str ]
  }
}

Class [ <internal:fileinfo> class finfo ] {

  - Constants [0] {
  }

  - Static properties [0] {
  }

  - Static methods [0] {
  }

  - Properties [0] {
  }

  - Methods [4] {
    Method [ <internal:fileinfo, ctor> public method finfo ] {

      - Parameters [2] {
        Parameter #0 [ <optional> $options ]
        Parameter #1 [ <optional> $arg ]
      }    }

    Method [ <internal:fileinfo> public method set_flags ] {

      - Parameters [1] {
        Parameter #0 [ <required> $options ]
      }    }

    Method [ <internal:fileinfo> public method file ] {

      - Parameters [3] {
        Parameter #0 [ <required> $filename ]
        Parameter #1 [ <optional> $options ]
        Parameter #2 [ <optional> $context ]
      }    }

    Method [ <internal:fileinfo> public method buffer ] {

      - Parameters [3] {
        Parameter #0 [ <required> $string ]
        Parameter #1 [ <optional> $options ]
        Parameter #2 [ <optional> $context ]
      }    } }}

Extension [ <persistent> extension #23 json version 1.2.1 ] {

  - Constants [10] {
```

```
   Constant [ integer JSON_HEX_TAG ] { 1 }
   Constant [ integer JSON_HEX_AMP ] { 2 }
   Constant [ integer JSON_HEX_APOS ] { 4 }
   Constant [ integer JSON_HEX_QUOT ] { 8 }
   Constant [ integer JSON_FORCE_OBJECT ] { 16 }
   Constant [ integer JSON_ERROR_NONE ] { 0 }
   Constant [ integer JSON_ERROR_DEPTH ] { 1 }
   Constant [ integer JSON_ERROR_STATE_MISMATCH ] { 2 }
   Constant [ integer JSON_ERROR_CTRL_CHAR ] { 3 }
   Constant [ integer JSON_ERROR_SYNTAX ] { 4 }
  }

  - Functions {
   Function [ <internal:json> function json_encode ] {

     - Parameters [2] {
       Parameter #0 [ <required> $value ]
       Parameter #1 [ <optional> $options ]
     }   }
   Function [ <internal:json> function json_decode ] {

     - Parameters [3] {
       Parameter #0 [ <required> $json ]
       Parameter #1 [ <optional> $assoc ]
       Parameter #2 [ <optional> $depth ]
     }   }
   Function [ <internal:json> function json_last_error ] {

     - Parameters [0] {
     }   } }}

dom

DOM/XML => enabled
DOM/XML API Version => 20031129
libxml Version => 2.7.3
HTML Support => enabled
XPath Support => enabled
XPointer Support => enabled
Schema Support => enabled
RelaxNG Support => enabled
```